\documentclass[11pt]{article}
\usepackage{tikz-qtree}
\usepackage{geometry}
\usepackage{amssymb,amsmath,amsfonts,eurosym,ulem,graphicx,caption,color,setspace,sectsty,comment,natbib,pdflscape, array, soul, xcolor}
\usepackage[hidelinks]{hyperref}
\usepackage{algorithm, algpseudocode}
\usepackage[bottom]{footmisc}
\usepackage{tabularx}
\usepackage{hyperref}
\usepackage{graphicx}
\usepackage{booktabs}
\usepackage{bm}
\usepackage{float}
\usepackage{multirow}
\usepackage{caption}
\usepackage{subcaption}
\usepackage{scalerel,stackengine}
\usepackage{color}
\usepackage{amssymb}
\usepackage[bottom]{footmisc}
\usepackage{algorithm, algpseudocode}
\usepackage{algcompatible}
\usepackage{amsmath}
\usepackage{amsthm}
\usepackage{epsfig}
\usepackage{booktabs}
\usepackage{graphicx}
\usepackage{graphics}
\usepackage{float}
\usepackage{multirow}
\usepackage{color}
\usepackage{caption}
\usepackage{enumitem}
\usepackage{subcaption}
\usepackage{fullpage}
\usepackage{tikz-qtree}
\usetikzlibrary{positioning}
\usepackage{makeidx}
\usepackage{xspace}
\usepackage{wrapfig}
\usepackage{tikz}
\usepackage{setspace}
\usepackage{array,makecell}
\usepackage{multirow}
\usepackage{natbib}
\allowdisplaybreaks
\stackMath
\newcommand{\mc}{\multicolumn{1}{c}}
\newcommand\reallywidehat[1]{%
	\savestack{\tmpbox}{\stretchto{%
			\scaleto{%
				\scalerel*[\widthof{\ensuremath{#1}}]{\kern-.6pt\bigwedge\kern-.6pt}%
				{\rule[-\textheight/2]{1ex}{\textheight}}
			}{\textheight}%
		}{0.5ex}}%
	\stackon[1pt]{#1}{\tmpbox}%
}
\makeatletter
\newcommand*{\indep}{%
	\mathbin{%
		\mathpalette{\@indep}{}%
	}%
}
\newcommand*{\nindep}{%
	\mathbin{
		\mathpalette{\@indep}{\not}
	}%
}
\newcommand*{\@indep}[2]{%
	\sbox0{$#1\perp\m@th$}
	\sbox2{$#1=$}
	\sbox4{$#1\vcenter{}$}
	\rlap{\copy0}
	\dimen@=\dimexpr\ht2-\ht4-.2pt\relax
	\kern\dimen@
	{#2}%
	\kern\dimen@
	\copy0 
} 
\makeatother

\definecolor{non-photoblue}{rgb}{0.64, 0.87, 0.93}  
\definecolor{bubbles}{rgb}{0.90, 1.0, 1.0}  
\definecolor{whitish}{rgb}{0.93, 1.0, 1.0}  
\definecolor{ceruleanblue}{rgb}{0.16, 0.57, 0.75}  
\definecolor{tuftsblue}{rgb}{0.28, 0.57, 0.81}
\definecolor{teagreen}{rgb}{0.75, 0.94, 0.75}  
\definecolor{tea}{rgb}{0.82, 0.99, 0.80}  
\definecolor{olivedrab(web)(olivedrab3)}{rgb}{0.30, 0.70, 0.14}  
\definecolor{olivedrab(web)}{rgb}{0.30, 0.56, 0.14}
\definecolor{darkolivegreen}{rgb}{0.33, 0.42, 0.18}  
\definecolor{pistachio}{rgb}{0.58, 0.77, 0.45}  
\definecolor{officegreen}{rgb}{0.0, 0.5, 0.0}

\newcommand\independent{\protect\mathpalette{\protect\independenT}{\perp}}
\def\independenT#1#2{\mathrel{\rlap{$#1#2$}\mkern2mu{#1#2}}}
\newcolumntype{L}[1]{>{\raggedright\let\newline\\arraybackslash\hspace{0pt}}m{#1}}
\newcolumntype{C}[1]{>{\centering\let\newline\\arraybackslash\hspace{0pt}}m{#1}}
\newcolumntype{R}[1]{>{\raggedleft\let\newline\\arraybackslash\hspace{0pt}}m{#1}}

\newcommand{\bX}{\mathbf{X}}

\newcommand{\1}{\mathbbm{1}}

\definecolor{ao(english)}{rgb}{0.0, 0.5, 0.0}
\newcommand\redsout{\bgroup\markoverwith{\textcolor{red}{\rule[0.5ex]{2pt}{2pt}}}\ULon}
\newcommand\greensout{\bgroup\markoverwith{\textcolor{ao(english)}{\rule[0.5ex]{2pt}{1pt}}}\ULon}
 \makeatletter
\let\@fnsymbol\@alph
\makeatother

\begin{document}
\setstcolor{red}
\begin{titlepage}
\title{{ Heterogeneous causal effects with imperfect compliance:\\
a \redsout{novel} Bayesian machine learning approach}}\footnotetext{We are grateful to participants at Harvard University, KU Leuven, IMT School for Advanced Studies and Flemish Ministry of Education seminars and at the Atlantic Causal Inference Conference (ACIC 2019). We want to thank  Bart Baesens, Tatiana Celadin, Giovanna D'Inverno, Francesca Dominici, Avi Feller, Laura Forastiere, Steven Groenez, Michael William Johnson, Alessandra Mattei, Fabrizia Mealli, Anna Mergoni, Rachel Nethery, Mike Smet, Melissa Tuytens, Davide Viviano, Xiao Wu and Paolo Zacchia for the useful comments provided. Falco J. Bargagli-Stoffi acknowledges funding from the Alfred P. Sloan Foundation Grant for the development of ``Causal Inference with Complex Treatment Regimes: Design, Identification, Estimation, and Heterogeneity" and funding from the 2021 Harvard Data Science Initiative Postdoctoral Research Fund Award. Kristof De Witte acknowledges funding from Steunpunt SONO and KU Leuven (C24/18/005).}
\author{Falco J. Bargagli-Stoffi\thanks{\scriptsize Corresponding author. Harvard University, United States of America. Department of Biostatistics, Harvard School of Public Health, 655 Huntington Ave, Boston, MA 02115, United States. Mail to: \href{mailto:fbargaglistoffi@hsph.harvard.edu}{fbargaglistoffi@hsph.harvard.edu}.} \and Kristof De Witte\thanks{\scriptsize KU Leuven, Leuven, Belgium and Maastricht University, Maastricht, The Netherlands. LEER - Leuven Economics of Education Research, Faculty of Economics and Business, KU Leuven, Naamsestraat 69 - 3000 Leuven, Belgium. UNU-Merit, Maastricht University, Minderbroedersberg 4 - 6211 LK Maastricht, The Netherlands.} \and Giorgio Gnecco\thanks{\scriptsize IMT School for Advanced Studies, Lucca, Italy. Laboratory for the Analysis of Complex Economic Systems, IMT School for Advanced Studies, piazza San Francesco 19 - 55100 Lucca, Italy.}}
\date{}
\maketitle
\vspace{-0.5cm}
\begin{abstract}
\noindent{This paper introduces an innovative Bayesian machine learning algorithm to draw interpretable inference on heterogeneous causal effects in the presence of imperfect compliance (e.g., under an irregular assignment mechanism). We show, through Monte Carlo simulations, that the proposed Bayesian Causal Forest with Instrumental Variable (BCF-IV) methodology outperforms other machine learning techniques tailored for causal inference in discovering and estimating the heterogeneous causal effects {\color{blue} while controlling for the familywise error rate (or -- less stringently -- for the false discovery rate) at leaves' level}. BCF-IV sheds a light on the heterogeneity of causal effects in instrumental variable scenarios and, in turn, provides the policy-makers with a relevant tool for targeted policies. Its empirical  application evaluates  the  effects  of  additional  funding  on  students'  performances. The results indicate that BCF-IV could be used to enhance the effectiveness of school funding on students’ performance.} \\
\vspace{0in}\\
\noindent\textbf{Keywords:} causal inference; instrumental variable; heterogeneous effects; interpretability; machine learning; school funding; students' performance\\
\vspace{0in}
\end{abstract}
\setcounter{page}{0}
\thispagestyle{empty}
\end{titlepage}

\section{Introduction}

\subsection{Motivation}
Starting from the year 2002, the Flemish Ministry of Education promoted the ``\textit{Equal Educational Opportunities}" program (henceforth referred as EEO) to ensure equal educational opportunities to all students \citep{oecd2017}. The EEO program  provides additional funding for secondary schools with a higher share of disadvantaged students. Proceeding from the seminal contributions of \cite{coleman1966equality} and \cite{hanushek2003failure} to recent contributions by \cite{jackson2015effects} and \cite{jackson2018does}, the question on whether or not an increase in school spending affects students' performances has been central in the social science literature. However, the bulk of studies on this topic have focused on average treatment effects \citep{hanushek2016handbook, hanushek2017school, dewitte2018disadvantaged}, often disregarding potential heterogeneities in how different subgroups of students and schools may be differentially affected by additional resources. While we acknowledge that the average treatment effect (ATE) is the most common starting point for any impact evaluation analysis, just focusing on it may mask potentially meaningful \redsout{and interesting} heterogeneities in the causal effects. \redsout{Insights on the heterogeneous effects are utterly relevant for developing cost-efficient targeted policies. As policy-makers are always facing financial constraints, they may want their policies to target just subsets of the population that really benefit from it (or, in turn, do not target those that do not benefit). This is particularly true for education policies. Human capital plays a fundamental role in modern societies and it is deemed important to understand which the characteristics of schools and students that get the highest return on their performance from additional school transfers are. Moreover, the evaluation of educational policies is a promising field for the discovery of heterogeneous causal effects and, in turn, targeted policies for at least two factors: (i) in the education context, there is a clear source of heterogeneity given by the disparate profiles of schools and students; and (ii) it is possible to gather large (administrative) datasets. Thus, the research question that we investigate in this paper is the following: \textit{what kinds of students most or least benefit from additional EEO school funding?}}

From a methodological point of view, the evaluation of the heterogeneous effects of additional school resources poses \redsout{a number of} major challenges\redsout{. The gold standard for assessing the causal effects of an intervention is to conduct a Randomized Control Trial (RCT) by randomly assigning the treatment to the study population. However,}{\color{blue}:} (i) school funding is generally not randomly assigned to schools {\color{blue} but its assignment correlates with schools' characteristics and potential anticipations of the expected benefits (i.e., \textit{confounding} issue)} and; (ii) some schools may not comply with funds' requirements and opt out even when eligible {\color{blue} (i.e., \textit{imperfect compliance} issue)}. When funding is not randomized but its assignment is based on the realized value of a variable, usually called the \textit{forcing} (or \textit{running}) variable, researchers can compare schools with very close values of the forcing variable -- namely around the point where the discontinuity is observed -- but with different levels of treatment. This comparison between units with different treatment levels, just above and below a given value of the forcing variable, is referred to as Regression Discontinuity (RD) design \citep{thistlethwaite1960regression,cook2008waiting}. RD designs are considered to be \textit{quasi-experimental} set ups and lead to valid inference on causal effects of the treatment at the threshold. In the case of our application on EEO data, the forcing variable used to evaluate which schools are eligible for additional funding is the share of disadvantaged students. Schools above an exogenously set threshold of 10\% on the share of disadvantaged students are eligible to receive the funding, schools below are not. However, even when a quasi-experiment is set up, one cannot force individuals (or schools) to comply with the treatment assigned. This is the case of studies in which a certain number of units (e.g., individuals, groups of individuals, schools, companies and so on) are randomly assigned to receive a treatment (e.g., a drug, a training course, additional school funding, and so on), but not all the units that are assigned to receive it are actually treated (and vice versa for units assigned to the control branch). This issue is widely known as \textit{imperfect compliance} problem \citep{angrist1996identification, balke1997bounds}. \redsout{Imperfect compliance may arise in all those scenarios where the receipt of the treatment require the units to take, or subject themselves to, a particular action (e.g., taking a drug, entering a job training program, and so on). For instance, according to the EEO guidelines,} {\color{blue} For instance, in the case of our application,} eligible schools need to comply with a minimal amount of additional teaching requirement to access the funding. In these cases, researchers can make use of a secondary treatment or \textit{instrument} (i.e., being eligible for additional funding) to isolate the causal effects of the primary treatment (i.e., actually receiving the funding) on the outcome of interest. 
These designs are referred to as instrumental variable (IV) settings \citep{angrist1996identification, angrist2001instrumental}. In scenarios where the instrument is not randomized, but assigned based on threshold, and there is room for imperfect compliance between the treatment assigned and the treatment actually received, we are in the presence of a so-called \textit{fuzzy RD} design \citep{trochim1984research, hahn2001identification}. Hence, fuzzy RD and IV methodologies are two sides of the same coin, and both techniques are tailored to draw causal inference in imperfect compliance settings \cite[for further discussion on the similarities between fuzzy RD and IV see][]{lee2010regression}.\redsout{However, both these methodologies do not provide any insight on which are the subgroups with heterogeneous causal effects.} {\color{blue} However, both these methodologies are not built to data-drivenly discover the subgroup with heterogeneous causal effects.}

\subsection{\sout{Related works}}
\redsout{A growing literature seeks to apply supervised machine learning techniques to the problem of estimating heterogeneous treatment effects. In their seminal contributions, \cite{hill2011bayesian} and \cite{foster2011subgroup} propose to directly apply machine learning to estimate the unit level causal effect as a function of the units' attributes. In other papers, machine learning techniques are adapted to estimate the heterogeneous causal effects using decision-tree based algorithms \citep{su2012facilitating, athey2016recursive}, ensemble of trees \citep{wager2018estimation, athey2019generalized, lechner2019modified}, Bayesian ensemble of trees \citep{hahn2020bayesian, starling2019targeted}, or doubly robust approaches \citep{semenova2020debiased, knaus2020double, fan2020estimation, zimmert2019nonparametric, kennedy2020optimal}. When the structure of the heterogeneity can be assumed to be relatively simple, one can also adopt alternative approaches such as targeted maximum likelihood \citep{van2006targeted}, Bayesian algorithms \citep{green2012modeling, wang2017bayesian}, LASSO-based approaches \citep{imai2013estimating}, or meta-learners \citep{kunzel2019metalearners}.}

{\color{blue} In recent years, various algorithms have been proposed to discover and estimate heterogeneous effects \cite[see][for a critical review]{Dominici2021From}.}
However, {\color{blue}most of} these techniques are tailored for drawing causal inference in settings where the treatment is randomly assigned to the units and do not address imperfect compliance issues.
Nonetheless, in the real world, the implementation of policies or interventions often results in imperfect compliance which makes the policy evaluation complicated. Recently, some machine learning techniques have been proposed to discover heterogeneous effects while dealing with imperfect compliance using tailored reworks of tree-based algorithms \citep{bargagli2020causal, bargagli2018estimating, wang2018instrumental, johnson2019detecting}, ensemble of trees algorithms \citep{athey2019generalized} or deep learning methodologies \citep{hartford2017deep}. However, these methods exhibit four principal limitations: (i) random forest-based algorithms for causal inference require large samples to converge to a good asymptotic behaviour for the estimation of causal effects, as shown in \cite{hahn2019atlantic} and \cite{wendling2018comparing}; (ii) deep learning-based algorithms require computationally expensive explorations of the space of possible hyper-parameters \redsout{(i.e., the deep neural network's learning rate, cost-function, regularization parameter, and number of hidden layers)} configurations \citep{zhang2017sensitivity}, are extremely sensitive to tuning parameters \citep{novak2018sensitivity} and are often not able to tolerate significant sources of unmeasured variation \citep{prosperi2020causal}; (iii) results obtained from ensemble methods, such as forest-based algorithms and deep neural networks, are hard to interpret by human experts because their non-linear parametrizations of the covariate space are complex to probe \citep{lee2020causal}; (iv) more interpretable algorithms are based on single learning that {\color{blue} performs} worse as compared to multiple learning {\color{blue} algorithms} (i.e., ensemble methods).\footnote{ \redsout{Ensemble methods have extensively been shown to outperform single learning algorithms in prediction tasks \citep{van2007super}.}} \redsout{Moreover, in machine learning applications, inference and uncertainty quantification are often of secondary importance after predictive performance. However, in policy decision settings it is crucial to know the credibility and variance of the counterfactual predictions \citep{athey2019machine, dominici2020controlled}. }

\subsection{Contributions}
To address and accommodate the shortcomings introduced in the previous Sections, this paper innovates the literature in both a methodological and an empirical perspective.  

First, we develop a machine learning algorithm tailored to draw causal inference in the presence of imperfect compliance. These situations, where the assignment depends on the observed and unobserved potential outcomes, are also referred to as \textit{irregular assignment mechanisms} in the causal inference literature \citep{imbens2015causal}. \redsout{In particular, we propose to modify a machine learning technique, namely, Bayesian Causal Forests, developed for causal inference goals \citep{hahn2020bayesian} to fit an IV setting.} The proposed method, Bayesian Instrumental Variable Causal Forest (BCF-IV), is an ensemble  semi-parametric  Bayesian  regression model  that  directly  builds  on the  Bayesian Additive Regression Trees (BART) algorithm \citep{chipman2010bart}. BCF-IV is tailored to discover and estimate the heterogeneous effects on the subgroup of units that comply with the treatment assignment (see Section \ref{subsec:BCF-IV}), {\color{blue} the so-called \textit{compliers}}. In this sense, the estimated effects can be seen as \textit{doubly local effects} (namely, subgroup effects for the compliers subpopulation). {\color{blue} In particular, BCF-IV discovers the heterogeneity in the form of an interpretable tree structure where each node of the tree corresponds to a discovered subgroup. For each leaf (final node) of the generated tree, BCF-IV allows to perform multiple hypotheses tests adjustments to control for Type I error rate (familywise error rate), or false discovery rate.}

We evaluate the fit of the proposed algorithm by comparing it with two alternative machine learning methods explicitly developed to draw causal inference on the heterogeneous effects in the presence of irregular assignment mechanisms: namely, the Generalized Random Forests (GRF) algorithm \citep{athey2019generalized} and the Honest Causal Trees with Instrumental Variables (HCT-IV) algorithm \citep{bargagli2020causal}. Using Monte Carlo simulations, we compare and evaluate the algorithms with respect to \redsout{two} {\color{blue} three} critical dimensions: (i) the ability of the algorithms to correctly detect the subgroups with heterogeneous causal effects; \redsout{and} (ii) the overall performance in terms of estimation accuracy for the conditional causal effects within the correctly discovered subgroups {\color{blue} and; (iii) the false positive rate}. These dimensions are consistent with recent evaluations of various machine learning methods for causal inference that highlight the excellence of Bayesian algorithms for causal inference \cite[see, e.g., ][]{hahn2019atlantic, wendling2018comparing}. We show that for each dimension, BCF-IV outperforms both GRF and HCT-IV in small samples and converges to an optimal large sample behaviour. 

Second, in an empirical application, BCF-IV is used to evaluate the EEO policy. In particular, we employ BCF-IV to evaluate the heterogeneity using a unique administrative dataset on the universe of Flemish pupils in the first stage of education in the school year 2010/2011 (135,682 students). As the additional funding is allocated to schools based on being above or below an exogenously set threshold regarding the proportion of disadvantaged students, this provides us with a quasi-experimental identification strategy. There is also a second, exogenously set, eligibility criterion stating that schools have to generate a minimum number of teaching hours. This source of imperfect compliance, given by the fact that not all the schools fulfill both the criteria, enables us to exploit a fuzzy regression discontinuity design to draw causal effects. We focus on the effects of additional funding on \redsout{two outcomes: (i)} students' performance: namely if a student gets the most favorable outcome (\textit{A-certificate}). \redsout{; and (ii) students' progresses to the following year without retention in their grades.} The results of our empirical application suggest that, although the effects of additional funding on the overall population of students are found to be not statistically significant, there is appreciable heterogeneity in the causal effects {\color{blue} and consistent sources of effect variation. The results are discussed in the application's Section.}\redsout{younger and less senior principals (less than 55 years old and with less than 30 years of experience) seem to play an important role in boosting the effects of the additional funding and, in turn, improving the performance of students. These results can advise policy-makers in multiple ways: the heterogeneous drivers could, on the one hand, help them enhancing the policy effectiveness by targeting just the schools with the highest shares of pupils that benefit the most from additional funding. On the other hand, policy-makers could investigate more in depth the reason why some schools do not benefit from the policy and ultimately provide additional tools to these schools to enhance the policy outcomes.}

The methodology proposed in this paper can be more widely applied to evaluations of the heterogeneous impact of an intervention in the presence of an irregular assignment mechanism in social and biomedical sciences. The remainder of this paper is organized as follows: in Section \ref{framework} we provide a general overview on causal inference and the applied machine learning frameworks, and we introduce our algorithm. In Section \ref{simul} we compare the performance of our algorithm with the performance of other methods already established in the literature. In Section \ref{application} we depict the usage of our algorithm in an educational scenario to evaluate the heterogeneous causal effects of additional funding to schools. Section \ref{conclusion} discusses the results and highlights the further applications of heterogeneous causal effects discovery and targeted policies in education as well as in social and biomedical sciences. {\color{blue} \texttt{R} code for the BCF-IV function can be found at \texttt{https://github.com/fbargaglistoffi/BCF-IV}.}

\section{Bayesian Instrumental Variable Causal Forest} \label{framework}

\subsection{{Notation}}
This paper contributes to the causal inference literature by establishing a novel machine learning approach for the estimation of conditional causal effects in the presence of an irregular assignment mechanism. 

We follow the standard notation of the Rubin's causal model \citep{rubin1974estimating, rubin1978bayesian, imbens2015causal}. Given a set of $N$ units, indexed by $i=1,...,N$, we denote with $Y_i$ a generic outcome variable, with $W_i$ a binary treatment indicator, with $\mathbf{X}$ a $N\times P$ matrix of $P$ control variables, and with $\bX_i$ the $i$-th $P$-dimensional row vector of covariates. Given the Stable Unit Treatment Value Assumption (SUTVA), that excludes interference between the treatment assigned to one unit and the potential outcomes of another \citep{rubin1986comment}, we can postulate the existence of a pair of potential outcomes: $Y_i(W_i)$. 
Specifically, the potential outcome for a unit $i$ if assigned to the treatment is $Y_i(W_i=1) = Y_i(1)$, and the potential outcome if assigned to the control is $Y_i(W_i=0) = Y_i(0)$.
We cannot observe for the same unit both the potential outcomes at the same time. However, we observe the potential outcome that corresponds to the assigned treatment: $Y_i^{obs} = Y_i(1) W_i + Y_i(0)(1-W_i)$.
\par In order to draw proper causal inference in observational studies researchers need to assume \textit{strong ignorability} to hold. This assumption states that:
\begin{equation}
    Y_i(W_i) \independent  W_i \mid \bX_i,
\end{equation} 
and
\begin{equation}
    0 < Pr(W_i=1\mid  \bX_i = x) < 1 \:\: \forall \: x \in \mathcal{X},
\end{equation}
where $\mathcal{X}$ is the features space.
The first assumption (\textit{unconfoundedness}) rules out the presence of unmeasured confounders while the second condition (\textit{common support}) needs to be invoked to be able to estimate the unbiased treatment effect on all the support of the covariates space. If these two conditions hold, we are in the presence of the so-called \textit{regular assignment mechanism} \citep{imbens2015causal}. In such a scenario the Average Treatment Effect (ATE) can be expressed as:
\begin{equation}
    \tau = \mathbb{E}\left[Y_i^{obs}\mid W_i=1\right] - \mathbb{E}\left[Y_i^{obs}\mid W_i=0\right],
\end{equation}
and one can define, following \cite{athey2016recursive}, the Conditional Average Treatment Effect (CATE) as:
\begin{equation}
    \tau(x) = \mathbb{E}\left[Y_i^{obs}\mid W_i=1, \bX_i=x\right] - \mathbb{E}\left[Y_i^{obs}\mid W_i=0, \bX_i=x\right].
\end{equation}
CATE is central for targeted policies as it enables the researcher to investigate the heterogeneity in causal effects. For instance, we may be interested in assessing how the effects of an intervention vary within different sub-populations. 
\par In observational studies, the assignment to the treatment may be different from the reception of the treatment. In these scenarios, where one allows for non-compliance between the treatment assigned and the treatment received, one can assume that the assignment is unconfounded, wherein the receipt is confounded \citep{angrist1996identification}. 
In such cases, one can rely on an instrumental variable (IV), $Z_i$, to draw proper causal inference.\footnote{Throughout the paper we assume the IV to be binary but, one could relax this assumption. However, as there are currently only a few studies that develop machine learning algorithms for the estimation of heterogeneous causal effects with a continuous treatment variable \cite[see][]{woody2020estimating}, we leave the investigation of these algorithms to further research.} $Z_i$ can be thought as a randomized assignment to the treatment, that  affects the receipt of the treatment $W_i$, without directly affecting the outcome $Y_i$ (\textit{exclusion restriction}). Thus, one can then express the treatment received as a function of the treatment assigned: $W_i(Z_i)$.

{\color{blue} In the following we assume the classical four IV assumptions \citep{angrist1996identification} -- \textit{monotonicity, existence of compliers, unconfoundedness of the IV, exclusion restriction} -- to hold. These assumptions need to hold to interpret the estimands proposed below as causal, and, in particular, researchers should devote attention to the necessary monotonicity assumption. Monotonicity states that for each subject, the level of the treatment that a subject would take if given a level of the IV is a monotonic increasing function of the level of the IV \citep{angrist1996identification}. Hence, this assumption, rules out the presence of so-called \textit{defiers}: i.e., no unit who would be always defying from the treatment assigned, taking the treatment when assigned to the control and viceversa. Under the four IV assumptions, IV can then be used to identify the causal effect in the subset of units that comply with the treatment assigned (i.e., the compliers). We refer the reader to the Supplementary Material for a detailed discussion of the four assumptions and how they are assumed to hold in our application reported in Section \ref{application}.}
If the assumptions hold, one can get the causal effect of the treatment on the sub-population of compliers, the so-called Complier Average Causal Effect (CACE), that is:

    \begin{equation} \label{cace} \small
   \tau^{cace} = \frac{\mathbb{E}\left[Y_i\mid Z_i = 1\right]-\mathbb{E}\left[Y_i\mid Z_i = 0\right]}{\mathbb{E}\left[W_i\mid Z_i =
    	1\right]-\mathbb{E}\left[W_i\mid Z_i = 0\right]}={ITT_Y\over\pi_C},
    \end{equation}
where the numerator represents the average effect of the instrument, also referred to as Intention-To-Treat effect, and the denominator represents the overall proportion of units that comply with the treatment assignment, also referred to as proportion of compliers \citep{angrist1996identification}.
CACE is also sometimes referred as Local Average Treatment Effects \cite[see][]{angrist2008mostly} and represents the estimate of causal effect of the assignment to treatment on the principal outcome, $Y_i$, for the subpopulation of compliers \citep{imbens2015causal}.
In this paper we consider the following conditional version of CATE.
The conditional CACE, $\tau^{cace}(x)$, can be thought as the CACE for a sub-population of observations defined by a vector of characteristics $x$:
    \begin{equation} \label{ITT_C} \small
   \tau^{cace}(x) =  \frac{\mathbb{E}\left[Y_i\mid Z_i = 1, \bX_i=x\right]-\mathbb{E}\left[Y_i\mid Z_i = 0, \bX_i=x\right]}{\mathbb{E}\left[W_i\mid Z_i =
    	1, \bX_i=x\right]-\mathbb{E}\left[W_i\mid Z_i = 0, \bX_i=x\right]}= {ITT_Y(x)\over\pi_C(x)},
    \end{equation}
where the numerator is the conditional intention to treat Intention-To-Treat effect, and the denominator the conditional proportion of compliers \citep{angrist1996identification}.

\subsection{Bayesian Instrumental Variable Causal Forest}\label{subsec:BCF-IV}

\redsout{In recent years, various algorithms have been proposed to estimate conditional causal effects (i.e, CATE and $\tau^{cace}(x)$).  Most algorithms focus on the estimation of CATE \cite[see][]{hill2011bayesian, su2012facilitating, green2012modeling, athey2016recursive, hahn2020bayesian, wager2018estimation, lee2021discovering, lechner2019modified} while just a few focus on the estimation of $\tau^{cace}(x)$ \cite[see][]{athey2019generalized, hartford2017deep, wang2018instrumental, bargagli2020causal}.} In this paper, we propose an algorithm for the estimation of CATE in an irregular assignment mechanism scenario. In particular, we adapt the Bayesian Causal Forest (BCF) algorithm \citep{hahn2020bayesian} for such a task. BCF was originally proposed for regular assignment mechanisms. This algorithm extends to a causal inference setting the Bayesian Additive Regression Trees (BART) algorithm \citep{chipman2010bart}, which in turn builds on the seminal Classification and Regression Trees (CART) algorithm \citep{friedman1984classification}.\footnote{ \redsout{\cite{ chipman2010bart} highlight how BART is different from other ensemble methods such as the Random Forest algorithm \citep{breiman2001random} or boosted regression trees \citep{friedman2001greedy}.}} CART is a widely used algorithm for the construction of binary trees where each node is splitted into only two branches \cite[see][for more details]{zhang2010recursive}.
{\color{blue} The first node of the tree is called the root, its final nodes are referred to as leaves.}

\redsout{Figure \ref{fig:CART} illustrates how the binary partitioning works in practice in a simple case with just two regressors $x_1 \in [0,1]$ and $x_2 \in [0,1]$.}

The accuracy of the predictions of binary trees can be dramatically improved by iteratively constructing the trees {\color{blue}, and aggregating their results}. BART, as well as BCF, are sum-of-trees ensemble algorithms, and their estimation approach relies on a fully Bayesian probability model \citep{kapelner2013bartmachine}. In particular, the BART model can be expressed as:
\begin{equation}\label{tree}  \centering 
    Y_i = f(\bX_i) + \epsilon_i \approx \mathcal{T}_1(\bX_i) + ... + \mathcal{T}_q(\bX_i) + \epsilon_i, \:\:\:\:\:\:\:\:\: \epsilon_i \sim \mathcal{N} (0, \sigma^2),
\end{equation}
where each of the $q$ distinct binary trees is denoted by $\mathcal{T}$, where $\mathcal{T}$ represents the entire tree: its structure, its nodes and its leaves (terminal nodes). The Bayesian component of the algorithm is incorporated in a set of three different priors on: (i) the structure of the trees (this prior is aimed at limiting the complexity of any single tree $\mathcal{T}$ and works as a regularization device); (ii) the probability distribution of data in the nodes (this prior is aimed at shrinking the node predictions towards the center of the distribution of the response variable $Y_i$);  (iii) the error variance $\sigma^2$ (which bounds away $\sigma^2$ from very small values that would lead the algorithm to overfit the training data).\footnote{\redsout{The choice of the priors, and the derivation of the posterior distributions, is discussed in depth by \cite{chipman2010bart} and \cite{kapelner2013bartmachine}. Namely, (i) the prior on the probability that a node will split at depth $k$ is $\beta(1+k)^{-\eta}$ where $\beta \in (0,1), \eta \in [0, \infty)$ (these hyper-parameters are generally chosen to be $\eta=2$ and $\beta = 0.95$); (ii) the prior on the probability distribution in the nodes is a normal distribution with zero mean: $\mathcal{N}(0, \sigma^2_q)$ where $\sigma_q = \sigma_0/\sqrt{q}$ and $\sigma_0$ can be used to calibrate the plausible range of the regression function; (iii) the prior on the error variance is $\sigma^2 \sim InvGamma(v/2, v\lambda/2)$ where $\lambda$ is determined from the data in a way that the BART will improve 90\% of the times the RMSE of an OLS model.}} The aim of these priors is to ``regularize" the algorithm, preventing single trees to dominate the overall fit of the model \citep{kapelner2013bartmachine}. {\color{blue} For further details on the priors specifications we refer to \cite{chipman2010bart}.}

\redsout{Moreover, BART allows the researcher to tune the variables' importance by departing from the original formulation of the Random Forest algorithm where each variable is equally likely to be chosen from a discrete uniform distribution (i.e., with probability $1 \over p$) to build a single tree learner. These Bayesian tools give researchers the possibility to tune the algorithm with prior knowledge. BART has shown particular flexibility and an excellent performance  -- with the need of no or little hyper-parameter tuning --  in both prediction tasks \citep{murray2017log,linero2018bayesian1, linero2018bayesian2, hernandez2018bayesian,starling2018functional} and in causal inference tasks \citep{hill2011bayesian, hahn2020bayesian,logan2019decision,starling2019targeted, nethery2019estimating}.}
 
The BCF algorithm proposed by \cite{hahn2020bayesian} is a semi-parametric Bayesian regression model that directly builds on BART. It, however, introduces some significant changes in order to estimate heterogeneous treatment effects in regular assignment mechanisms (even in the presence of strong confounding).
The principal novelties of this model are the expression of the conditional mean of the response variable as a sum of two functions and the introduction, in the BART model specification for causal inference, of an estimate of the propensity score, $E[W_i=1\mid \bX_i=x]=\pi(x)$, in order to improve the estimation of heterogeneous treatment effects.\footnote{It is important to highlight that the propensity score is not used to estimate the causal effects but to moderate the distortive effects in treatment heterogeneity discovery due to strong confounding. Moreover, since BCF includes the entire predictors' vector, $\mathbf{X}_i$, even if the propensity score is misspecified or poorly estimated, the model allows for the possibility that the response remains correctly specified \citep{hahn2020bayesian}. In the Supplementary Material, we show that even if the estimate $\hat{\pi}(x)$ of the propensity score is incorrectly specified the results are still widely robust.} {\color{blue} Results from empirical Monte Carlo simulations studies have shown an excellent performance of BART and BCF in causal inference tasks \citep{dorie2019automated, hahn2019atlantic, wendling2018comparing}. } \redsout{As depicted from the results of the Atlantic Causal Inference Conference (ACIC) competition in 2016 and 2017, reported by  \cite{hahn2019atlantic}, it was observed that BCF performs dramatically better than other machine learning algorithms for causal inference in the presence of randomized and regular assignment mechanisms.}

\redsout{As argued in the introduction of this paper, the bulk of methodologies developed thus far for the discovery and estimation of heterogeneous effects in the presence of imperfect compliance potentially suffers a series of limitations. In particular, ensemble methodologies such as generalized random forest or deep learning algorithms for IV, despite the success in accurately estimating the conditional CACE, offer little guidance about which covariates or, even further, subpopulations (i.e., subsets of the features space defined by multiple covariates) bring about treatment effect heterogeneity. Outputs/results obtained from existing methods are hard to interpret by human experts because parametrizations of the covariate space are complicated \citep{lee2020causal}. In this regard, tree based algorithms such as causal trees \citep{athey2016recursive} provide enhanced interpretability, but are not suited for imperfect compliance scenarios. Recently, \cite{bargagli2020causal} and \cite{wang2018instrumental} have proposed to rework causal trees to make them tailored for irregular assignment mechanisms showing how these methods outperform causal trees when a suitable IV is available.\footnote{\color{blue} These authors compare the ability of causal trees \citep{athey2016recursive} and their reworked versions tailored for imperfect compliance, with respect to their ability to precisely discover and estimate heterogeneous effects. They find that methodologies directly tailored for imperfect compliance perform consistently better in the presence of unmeasured confounding when a suitable IV is available.} However, these methodologies are based on single learning algorithms, which are usually less precise and less stable (hence, less reproducible) than ensemble learning algorithms \citep{strobl2009introduction}. These reasons drive the need for a new algorithm tailored for causal inference on heterogeneous effects in the presence of irregular mechanisms. 

In the following Sections,} 

Here, we introduce the Bayesian Causal Forest with Instrumental Variable (BCF-IV) algorithm, which is tailored to data-drivenly discover and estimate heterogeneous causal effects in an interpretable way. Heterogeneous effects analyses are typically conducted for subgroups defined \textit{a priori} to avoid potential cherry-picking problems connected to reporting the causal effects only for subgroups with extremely high or low treatment effects \citep{cook2004subgroup}. However, defining a priori these groups has two main shortcomings: (i) it requires a fairly good understanding of the treatment effect; (ii) researchers may miss unexpected subgroups. To overcome these limitations, we propose a data-driven approach to discover heterogeneous effects by using \textit{honest} sample splitting \citep{athey2016recursive}.\footnote{ \redsout{The idea of sample splitting is not new in statistics and the earliest references can be tracked back to \cite{stone1974cross} and \cite{cox1975note}.}} Following a honest sample splitting approach we divide the data into two subsamples: one to build the tree for the discovery of the heterogeneous effects (\textit{discover subsample}: $\mathcal{I}^{dis}$) and another for making inference (\textit{inference subsample}: $\mathcal{I}^{inf}$).
Accordingly to \cite{athey2016recursive}, for both the simulations and empirical application, half of the sample is assigned to $\mathcal{I}^{dis}$ and the other half to $\mathcal{I}^{inf}$. \greensout{In the \texttt{R} function implementing BCF-IV, we leave to the researcher the choice of the ratio between the numbers of discovery and inference subsamples.}\footnote{As highlighted by \cite{lee2020causal} different proportions of units could be assigned to the discovery and inference subsamples. \redsout{However, while the estimation's precision could benefit from different splitting ratios, it is not clear how this could affect the subgroups discovery. We leave this to further research.} {\color{ao(english)} In the \texttt{R} function implementing BCF-IV, we leave to the researcher the choice of the ratio between discovery and inference subsamples.}} As inference is a separated task from model selection, honest sample-splitting enables an honest inference for effect modification.\footnote{Alternatively, honest inference could be obtained by techniques based on multiple testing and controlling for family wise error rate such as the ones proposed by \cite{hsu2015strong} and \cite{johnson2019detecting}.} Algorithm (1) provides a general overview on the proposed BCF-IV algorithm.

\vspace{0.1cm}

\begin{center}
\fbox{\parbox{\textwidth}{\textbf{Algorithm 1} Overview of Bayesian Causal Forest with Instrumental Variable (BCF-IV) 
\hline \vspace{0.2cm}
		
		\footnotesize
		{\bf Inputs}: $N$ units $i$ ($X_i,Z_i,W_i,Y_i$), where $X_i$ is the feature vector, $Z_i$ is treatment assignment (instrumental variable), $W_i$ is the treatment receipt, and $Y_i$ is the observed response.\\
        {\bf Outputs}: (1) a tree structure discovering the heterogeneity in the causal effects, and (2) estimates of the Complier Average Causal Effects within its leaves.
		\vspace{-0.1cm}
		
		\begin{itemize}
		    \item The Honest Splitting Step:
		    \begin{enumerate}
		        \item  Randomly split  the total sample into a discovery ($\mathcal{I}^{dis}$) and an inference subsample $\mathcal{I}^{inf}$.
		    \end{enumerate} 
		    \item The Discovery Step (performed on $\mathcal{I}^{dis}$) (Section~\ref{hetero}):
		\begin{enumerate}  \setcounter{enumi}{1}
			\item Estimation of the Conditional CACE:
			
			\begin{enumerate}
			    \item Estimate the conditional Intention-To-Treat: $\widehat{ITT}(x)$;
			
			    \item Estimate the conditional proportion of compliers: $\hat{\pi}_{C}(x)$;
			
			    \item Estimate the conditional CACE, $\hat{\tau}^{cace}(x)$, using the estimated values from (a) and (b) as in \eqref{cace_bcf} (for a continuous outcome) or in \eqref{eq:cace_bart} (for a binary outcome).
			\end{enumerate}
			
			\item Heterogeneous subpopulations discovery:
			
			\begin{enumerate}[label=(\alph*),start=4]
			    \item Discover  the  heterogeneous  effects  by  fitting  a  decision  tree  using  the  data $(\hat{\tau}^{cace}(x), \bX_i)$.
			\end{enumerate}

		\end{enumerate}
		\end{itemize}
		
		\begin{itemize}
		    \item The Inference Step (performed on $\mathcal{I}^{inf}$) (Section~\ref{subsec:estimation}):
		\begin{enumerate} \setcounter{enumi}{3}
			\item Estimate the $\hat{\tau}^{cace}(x)$ for all the discovered subpopulations {\color{blue} (i.e., nodes and leaves)} in the tree discovered in (d);
			\item {\color{blue} Perform multiple hypotheses tests adjustments of the p-values to control for familywise error rate or -- less stringently --  for the false discovery rate;}
			\item Run weak-instrument tests within every node and discard those nodes where a weak-instrument issue is detected.
		\end{enumerate}
		\end{itemize}
}}    
\end{center}

\subsubsection{Discovery Step: Discovering Heterogeneity in the Conditional CACE} \label{hetero}

After dividing the total sample into a discovery and an inference subsample, we perform the discovery of the heterogeneity in the conditional CACE on $\mathcal{I}^{dis}$.
In particular, the BCF-IV algorithm starts from modifying (\ref{tree}) to adapt it for the estimation of the conditional Intention-To-Treat, by including the IV, $Z_i$:
\begin{equation} \label{iv-tree}
    Y_i = f(Z_i, \bX_i) + \epsilon_i \approx \mathcal{T}_1(Z_i, \bX_i) + ... + \mathcal{T}_q(Z_i, \bX_i) + \epsilon_i, \:\:\:\:\:\:\:\:\:\:\:\: \epsilon_i \sim \mathcal{N} (0, \sigma^2),
\end{equation}
where, for simplicity, we assume the error to be a mean zero additive noise as in \cite{hill2011bayesian}, \cite{hahn2020bayesian}, and \cite{logan2019decision}. The conditional expected value can of $Y_i$ be expressed as:
\begin{equation} \label{mean}
    \mathbb{E}[Y_i\mid  Z_i=z, \bX_i=x] = g(z, x),
\end{equation}
and in turn the conditional intention-to-treat, $ITT_{Y}(x)$, is:
\begin{equation}
    ITT_{Y}(x) = \mathbb{E}[Y_i\mid Z_i = 1, \bX_i=x]-\mathbb{E}[Y_i\mid Z_i = 0, \bX_i=x] = g(1, x) - g(0,x).
\end{equation}
 Then, adapting to an irregular assignment mechanism the model proposed by \cite{hahn2020bayesian}, we adopt the following functional form for (\ref{mean}):
\begin{equation} \label{iv-bart}
    \mathbb{E}[Y_i\mid Z_i=z, \bX_i=x] = \mu(\pi(x),x) + ITT_{Y}(x)  z
\end{equation}
where $\pi(x)$ is the propensity score for the IV: $\pi(x) = E[Z_i=1\mid \bX_i=x]$.
The expression of $\mathbb{E}[Y_i\mid Z_i=z, \bX_i=x]$ as a sum of two functions is central: the first component of the sum, $\mu(\pi(x), x)$, directly models the impact of the control variables on the conditional mean of the response (the component that is independent from the treatment effects) while the second component $ITT_{Y}(x)z$ models directly the intention-to-treat effect as a nonlinear function of the observed characteristics (this second {\color{blue} component} captures the heterogeneity in the intention-to-treat). Both the functions $\mu$ and $ITT_Y$ are given independent priors. These priors are chosen in line with \cite{hahn2020bayesian} to be for the first component the same priors of \cite{chipman2010bart}. However, for the second component the priors are changed in a way that allows for less deep, hence  simpler trees.\footnote{\redsout{The depth penalty parameters are set to be $\eta=3$ and $\beta=0.25$ (instead of $\eta=2$ and $\beta=0.95$).}}

\redsout{The expression of $\mathbb{E}[Y_i\mid Z_i=z, \bX_i=x]$ as a sum of two functions has a double effect: (i) on the one hand, it allows the algorithm to learn which component in the heterogeneity of the conditional mean of the outcome is driven by a direct effect of the control variables and which component is the true heterogeneity in the effects of the assignment to the treatment $Z_i$ on $Y_i$; (ii) on the other hand, it allows the predictions of the treatment effect driven by the BART to be modelled directly and separately with respect to the impact of the control variables \citep{hahn2020bayesian}.}

The estimated propensity score is not used for the estimation of the effects but is included, as an additional covariate, in the first component of (\ref{iv-bart}) to mitigate possible problems connected to \textit{regularization induced confounding} (RIC)\footnote{RIC is analyzed in depth in \cite{hahn2018regularization}. RIC issues rise when the ML algorithm used for regularizing the coefficient does not shrink to zero some coefficients due to a nonzero correlation between $Z_i$ and $\bX_i$ resulting in an additional degree of bias that is not under the researcher's control.} and \textit{targeted selection}.\footnote{Targeted selection refers to settings where the treatment (or in an IV scenario the assignment to the treatment) is assigned based on an ex-ante prediction of the outcome conditional on some characteristics $\bX_i$. We refer to \cite{hahn2020bayesian} for a discussion of targeted selection problems.} Moreover, in scenarios where the IV is not randomized ex-ante, the inclusion of the estimated propensity score, $\hat{\pi}(x)$, leads to an improvement in the discovery of the heterogeneity in the causal effect \citep{hahn2019atlantic}.  \redsout{Furthermore, it is important to highlight that choosing a misspecified definition of $\hat{\pi}(x)$ does not impact in a significant way the quality of the results as shown in the Supplementary Material. This is due to the fact that this first step of our algorithm is not about directly estimating the conditional CACE but is tailored to discover the heterogeneity in $ITT_{Y}(x)$.}

Using a similar rework of BART, one can estimate the conditional proportion of compliers, ${\pi}_{C}(x)$. In particular, we propose to rework \eqref{tree} as follows:
\begin{equation} \label{iv-pic}
    W_i = l(Z_i, \bX_i) + \epsilon_i \approx \mathcal{T}_1(Z_i, \bX_i) + ... + \mathcal{T}_q(Z_i, \bX_i) + \psi_i, \:\:\:\:\:\:\:\:\:\:\:\: \psi_i \sim \mathcal{N} (0, \psi^2),
\end{equation}
where, we assume again, the error $\psi_i$ to be a mean zero additive noise. The conditional expected value of $W_i$ can be expressed as:
\begin{equation}
    \mathbb{E}\left[W_i\mid Z_i = z, \bX_i=x\right] = \delta(z,x),
\end{equation}
and the conditional proportion of compliers is:
\begin{equation} \label{iv-bart-pic}
    \mathbb{E}\left[W_i\mid Z_i = 1, \bX_i=x\right]-\mathbb{E}\left[W_i\mid Z_i = 0, \bX_i=x\right]=\delta(1,x)-\delta(0,x),
\end{equation}
where $\delta(z,x)$ can be estimated, in the case of a binary $Z_i$, using the BART methodology for causal effects estimation proposed by \cite{hill2011bayesian} and implemented in R in the \texttt{bartCause} package \citep{dorie2020package}.

Finally, the conditional CACE can be expressed, reworking \eqref{cace}, as the ratio between \eqref{iv-bart} and \eqref{iv-bart-pic}:
\begin{equation} \label{cace_bcf} \small
   \tau^{cace}(x) = \frac{ \mathbb{E}[Y_i\mid Z_i=z, \bX_i=x]}{ \mathbb{E}[W_i\mid Z_i=z, \bX_i=x]} =\frac{\mu(\pi(x), x) + ITT_{Y}(x)  z}{\delta(1,x)-\delta(0,x)}.
\end{equation}

In the case of a binary outcome, the proposed methodology is implemented using again, instead of BCF, a suitable rework of the causal BART algorithm proposed by \cite{hill2011bayesian}. In this case, the conditional CACE can be expressed as:
\begin{equation}\label{eq:cace_bart}
    \tau^{cace}(x) = \frac{ \mathbb{E}[Y_i\mid Z_i=z, \bX_i=x]}{ \mathbb{E}[W_i\mid Z_i=z, \bX_i=x]} = \frac{g(1, x) - g(0,x)}{\delta(1, x) - \delta(0,x)}, 
\end{equation}
where both $g(\cdot,\cdot)$ and $\delta(\cdot,\cdot)$ are estimated using BART using as an outcome the output $Y_i$ and $W_i$ respectively. 

Once one estimated the conditional CACE as in \eqref{cace_bcf}, one can build a simple binary tree, using a CART model \citep{breiman1984classification}, on the fitted values  ($\hat{\tau}^{cace}(x)$) to discover, in an interpretable manner, the drivers of the heterogeneity {\color{blue} (fit-the-fit approach). Indeed, as argued by \cite{lee2021discovering} and \cite{lee2020causal}, the subgroups discovered from a fit-the-fit approach using CART,} are ideal to guarantee high levels of interpretability, where interpretability can be defined as the degree to which a human can understand the cause of a decision or consistently predict the results of the model \citep{miller2018explanation, kim2016examples}. The CART algorithm used for the detection of heterogeneous subgroups is implemented in \texttt{R} using the \texttt{rpart} package version 4.1-15 \citep{therneau2015package}. The maximal depth of the tree can be set by the researcher. In our empirical example and in the simulations, the maximal depth is set to two to maintain a small level of complexity (and, in turn, enhance interpretability) and guarantee enough observations within each node \cite[see][for further a discussion on tree depth, interpretability and subgroups sizes]{lee2021discovering}.

{\color{blue}  Alternatively, one could directly fit the CART on the estimated unit level ITT in \eqref{iv-bart}. We call this alternative methodology BCF-ITT. BCF-ITT could be potentially helpful in scenarios where (i) one can imagine the heterogeneity to be driven by small values of $\pi_{C}(x)$, or (ii) one is solely interested in the detection of the heterogeneity in the ITT. BCF-ITT may suffer limitations when the ITT is relatively homogeneous while the proportion of compliers in various subgroups is not. In the simulations in the Supplementary Material, we discuss further the comparison between BCF-IV and BCF-ITT and how BCF-IV dominates BCF-ITT in the detection of the heterogeneity in CACE.}

\subsubsection{Estimation of Conditional CACE}\label{subsec:estimation}
Once the heterogeneous patterns in the intention-to-treat (ITT) are learned from the algorithm, one can estimate the conditional CACE, $\tau^{cace}(x)$ on the inference subsample $\mathcal{I}^{inf}$. To do so, one can suitably estimate the method of moments in Equation (\ref{ITT_C}) within all the different sub-populations that were detected in the previous step {\color{blue} (see the Supplementary Material).}
 
\greensout{The conditional CACE can be estimated in a generic sub-sample (i.e., for each $\bX_i \in {\mathbb{X}}_j$, where ${\mathbb{X}}_j$ is a generic node of the discovered tree, like a non-terminal node or a leaf) as:
\begin{equation}\label{eq:conditionalCACEestimate}
    \hat{\tau}^{cace}(\bX_i) = \frac{\reallywidehat{ITT}_{Y}(\bX_i)}{\hat{\pi}_C(\bX_i)},
\end{equation}
where $\hat{\pi}_{C}(\bX_i)$ is estimated as:
\begin{equation}
\hat{\pi}_{C}(\bX_i)=\frac{1}{N_{1,j}}\sum_{l: X_l \in \mathbb{X}_j}  W_l   Z_l -\frac{1}{N_{0,j}}\sum_{l: X_l \in \mathbb{X}_j} W_l  (1-Z_l),
\end{equation} 
and $\reallywidehat{ITT}_{Y}(\bX_i)$ as:
    \begin{equation} \label{IV_propensity}
    \reallywidehat{ITT}_Y(\bX_i)=\frac{1}{N_{1,j}}\sum_{l: X_l \in \mathbb{X}_j}  Y_l^{obs}   {Z_l } - \frac{1}{N_{0,j}}\sum_{l: X_l \in \mathbb{X}_j}  Y_l^{obs}   (1-Z_l),
 \end{equation}
where $N_{k,j}$ (where $k \in \{0,1\})$ is the number of observations with $Z_l=k$ in the sub-sample of observations with  $X_l \in \mathbb{X}_j$: where $N_{1,j}=\sum_{l: X_l \in \mathbb{X}_j} Z_l$ and $N_{0,j}=\sum_{l: X_l \in \mathbb{X}_j}(1- Z_l)$. It is worth highlighting that, since the supervised machine learning technique is used in the discovery phase and not in the estimation phase, the estimators that are proposed here could be used in a more ``traditional way", in settings where the subgroups are defined ex-ante by the researcher.}

In the case of a binary instrument ($Z_i \in \{0,1\}$) and a binary treatment variable ($W_i \in \{0,1\}$), \cite{angrist1996identification} and \cite{imbens2015causal} revealed that the population versions of the method-of-moments estimator correspond to a Two Stage Least Squares (henceforth referred as 2SLS) estimator of $\tau^{cace}$, in the cases where the four IV assumptions can be assumed to hold.
Hence, since this case is analogous to our setting, 
one can apply the 2SLS method in every node $\mathbb{X}_j$ of the {\color{blue} discovered} tree, $\mathcal{T}$, for the estimation of the effect on the compliers population, as it is presented by \cite{imbens1997estimating}.

The two simultaneous equations of the 2SLS estimator are, in the population,
\begin{align}
Y_i^{obs}&=\alpha + \tau^{cace}   W_i + \epsilon_i,\label{IV_reg1} \\ 
W_i&=\pi_0 + \pi_C   Z_i + \eta_i,  \label{IV_reg2}
\end{align} \normalsize
where $\mathbb{E}(\epsilon_i)=\mathbb{E}(\eta_i)=0$, and $\mathbb{E}(Z_i \eta_i)=0$.\footnote{The latter comes from the fact that (\ref{IV_reg1}) is assumed to represent the linear projection of $W_i$ onto $Z_i$.} In the econometric terminology, the explanatory variable $W_i$ is $endogenous$, while the IV variable $Z_i$ is $exogenous$.  \eqref{IV_reg1} is referred to as the \textit{outcome equation} and \eqref{IV_reg2} is referred to as the \textit{treatment equation} \citep{lee2010regression}.

We can express the 2SLS equations, conditional on a subpopulation of a node $\mathbb{X}_j$, as
\begin{align} \label{IV_{X_j}_reg}
Y_{i,\mathbb{X}_j}^{obs}&=\alpha_{\mathbb{X}_j} + \tau^{cace}_{\mathbb{X}_j}   W_{i,\mathbb{X}_j} + \epsilon_{i,\mathbb{X}_j}, \\ \label{IV_{X_j}_reg1}
W_{i,\mathbb{X}_j}&=\pi_{0,\mathbb{X}_j} + \pi_{C,\mathbb{X}_j}   Z_{i,\mathbb{X}_j} + \eta_{i,\mathbb{X}_j},
\end{align} \normalsize
where $\mathbb{E}(\epsilon_{i,\mathbb{X}_j})=\mathbb{E}(\eta_{i,\mathbb{X}_j})=0$, and $\mathbb{E}(Z_{i,\mathbb{X}_j} \eta_{i,\mathbb{X}_j})=0$.

Moreover, the following reduced equation (obtained plugging (\ref{IV_{X_j}_reg1}) into (\ref{IV_{X_j}_reg})) holds:
\begin{align} \label{IV_reg_red}
Y_{i,\mathbb{X}_j}^{obs}&=\left(\alpha_{\mathbb{X}_j} + \tau^{cace}_{\mathbb{X}_j}   \pi_{0,\mathbb{X}_j}\right) + \left(\tau^{cace}_{\mathbb{X}_j}   \pi_{C,\mathbb{X}_j}\right)    Z_{i,\mathbb{X}_j} + \left(\epsilon_{i,\mathbb{X}_j} +  \tau^{cace}_{\mathbb{X}_j}   \eta_{i,\mathbb{X}_j}  \right) \nonumber \\
&=\bar{\alpha}_{\mathbb{X}_j} + \gamma_{\mathbb{X}_j}    Z_{i,\mathbb{X}_j} + \psi_{i,\mathbb{X}_j} .
\end{align} \normalsize 
In the case of a single instrument, the logic of IV regression is that one can estimate the respective parameters $\pi_{C,\mathbb{X}_j}$ and $\gamma_{\mathbb{X}_j}=\tau^{cace}_{\mathbb{X}_j}   \pi_{C,\mathbb{X}_j}$ of the regressions (\ref{IV_{X_j}_reg1}) and (\ref{IV_reg_red}) above by least squares, when the observations in each node are independent and identically distributed, then obtaining an estimate of the parameter $\tau^{cace}_{\mathbb{X}_j}$ in (\ref{IV_{X_j}_reg}). In particular, for every element $\bX_i$ of a node $\mathbb{X}_j$, one can estimate  $\tau^{CACE}(\bX_i)=\tau^{cace}_{\mathbb{X}_j}$ through 2SLS, as the following ratio \citep{imbens2015causal}:
    \begin{equation} \label{eq:52} 
    \hat{\tau}^{CACE}(\bX_i) \equiv \hat{\tau}^{2SLS}_{\mathbb{X}_j}=\frac{\hat{\gamma}_{\mathbb{X}_j}}{\hat{\pi}_{C,\mathbb{X}_j}}.
    \end{equation}
{\color{blue} The theoretical properties of these conditional estimators are reported in the Supplementary Material.}
 
Within each subgroup $\mathbb{X}_j$, we evaluate whether $\hat{\tau}^{2SLS}_{\mathbb{X}_j}$ is significantly different from zero, and we run weak-instrument tests for the estimated effects and we discard those nodes where a weak-instrument issue is detected. This is done to avoid problems connected to the discovery of heterogeneous effects driven by small proportion of compliers in the subgroup (aka potential weak-instrument issue). 

{\color{blue} Furthermore, for each leaf of the generated tree, we implement a set of multiple hypotheses tests adjustments to control for Type I error rate (aka familywise error rate -- e.g., the probability of making a Type I error among a specified group, or \textit{family}, of tests), or false discovery rate (e.g., the expected proportion of false discoveries amongst the rejected hypotheses). Indeed, while the main focus of the BCF-IV algorithm is to discover heterogeneous effects in the form of an interpretable tree structure and then estimate the subgroup effects, it may be of interest to assess the significance of the discovered heterogeneity at leaves-level while controlling for potential spurious heterogeneity discovery. In particular, the adjustment methods that are implemented in BCF-IV are the Bonferroni correction and the corrections proposed by \cite{holm1979simple}, \cite{hochberg1988sharper}, \cite{hommel1988stagewise}, \cite{benjamini1995controlling} and \cite{benjamini2001control}. The first four methods are designed to control for familywise error rate, while the last two corrections control for the false discovery rate. The false discovery rate is a less stringent condition than the family-wise error rate, so the latter two methods are more powerful than the others. By default, BCF-IV implements Holm's method, which is more stringent, dominates the Bonferroni correction, and is valid under arbitrary assumptions, while Hochberg's and Hommel's methods are valid when the hypothesis tests are independent or when they are non-negatively associated \citep{sarkar1997simes}. However, depending on the specific setting of application and the specific research interest, any of the other methods can be implemented instead.}

\section{Monte Carlo Simulations} \label{simul}

The overall goal of this Section is to provide insights on how discovering and estimating the causal effects in the presence of imperfect compliance is a complex task that depends on multiple factors such as the structure and magnitude of heterogeneity in the causal effects, the size of the available data, the strength of the IV used and the heterogeneity of compliance rate among different subgroups.
To do so, a set of Monte Carlo simulations' scenarios is designed to analyse the performance of the proposed methodologies as a function of: (1) the structure and magnitude of the heterogeneous effects within the different subgroups; (2) the size of the data used in the analysis. {\color{blue} {In the Supplementary Material, we provide a number of robustness checks of the Monte Carlo simulation. In particular, we focus on what happens to the fit of the three algorithms when one introduces:
(3) the different compliance rates, ranging from a strong instrument (high compliance) to a weak one (low compliance); and (4) the variation in the compliance rates among different subgroups; (5) confounding in the generation of the IV; (6) covariance in the covariates matrix; and (7) misspecification in the propensity score. The performance of BCF-IV does not significantly deteriorate, as compared to the baseline models introduced in the following simulations.}}

The following simulations were designed to closely mimic the empirical example discussed later in Section 4 in order to furnish a useful guidance on the reliability of the empirical results. In particular, the generated Monte Carlo data reproduce an imperfect compliance setting under the assumption of one-sided non-compliance (i.e., units that are not assigned to the treatment will never be able to get it) with potential heterogeneity in the causal effects and in the compliance rate. Sample sizes (4,000) and compliance rates (0.75) are simulated to be slightly smaller than the ones in the empirical application on EEO data reported later in Section \ref{application} in order to provide conservative guidance on the algorithms' performances.\footnote{The smallest learning sample for the EEO has 4,300 observations and an overall compliance rate of roughly 0.80.}

To evaluate the performance of the BCF-IV and BCF-ITT algorithms, we contrast them with other two methods tailored for drawing causal inference in irregular assignment mechanism scenarios: the Honest Causal Trees with Instrumental Variable (HCT-IV) algorithm \citep{bargagli2020causal} and the Generalized Random Forests (GRF) algorithm \citep{athey2019generalized}. Both these algorithms were shown to outperform other causal machine learning methodologies in irregular assignment mechanisms. {\color{blue} Since the foremost focus of this paper is on discovery of the heterogeneity in the effects in terms of an interpretable tree structure and the estimation of this heterogeneity in all the detected subgroups, we compare and evaluate the algorithms on three critical dimensions: (i) the ability of the algorithms to correctly detect the subgroups (at leaves level) with heterogeneous causal effects, (ii) the capacity of BCF-IV to control for false positive discovery at leaves level through p-value adjustment, and (iii) the overall performance in terms of estimation accuracy for the conditional causal effects within the correctly discovered subgroups at leaves level (namely, the ones with heterogeneous effects).} With respect to the first dimension, we compare BCF-IV and BCF-ITT with HCT-IV as these three methods are able to identify, in an interpretable manner, the groups with highest levels of heterogeneity in the causal effect. In this case, we leave out the comparison with GRF as this technique is not providing interpretable guidance on the subgroups with highest levels of heterogeneity. For the last dimension, we compare the estimation ability of BCF-IV with the one of GRF.\footnote{ Since BCF-IV and BCF-ITT are based on the same estimator we implement just BCF-IV for this contrast.}

We start by generating the set of potential outcomes $Y_i(Z_i)$, potential treatments $W_i(Z_i)$, and covariates $\mathbf\bX_i$ for each unit $i$, where the observed treatment and outcome are based on the value of the instrument $Z_i$. For each data-generating process, we generate the covariate matrix $\bX$ with 10 binary covariates from $X_{i1}$ to $X_{i10}$  where each covariate is sampled from a binomial distribution with success probability equal to 0.5: $X_{ip}  \sim \textit{Binom}(0.5)$. The binary instrument $Z_i$ is also drawn from a binomial distribution, $Z_i  \sim \textit{Binom}(0.5)$. Given one-sided non-compliance, for units not assigned to the treatment their potential treatment is always 0, namely $W_i(0)=0$. The potential treatment is a Bernoulli trial with a success rate (or compliance rate) of $\pi_{X_{i1},X_{i2}}$. This means that the compliance rate depends on the two covariates $X_{i1}$ and $X_{i2}$ mimicking a setting where people assigned to the treatment may decide to actually receive it or not based on the values of some observed covariates. The potential outcomes of units having not received the instrument are sampled from a standard normal distribution, $Y_i(0) \sim \mathcal{N}(0,1)$, while the potential outcomes of units having received the instrument are a function of the treatment value and $\tau^{cace}(\mathbf\bX_i)$: $Y_i(1) = Y_i(0) + W_i(1)\tau^{cace}(\mathbf\bX_i)$. $\tau^{cace}(\mathbf\bX_i)$ indicates the heterogeneous conditional CACE, which is a function of the set of covariates levels of each unit. To simplify, we make $\tau^{cace}(\mathbf\bX_i)$ dependent on $X_{i1}$ and $X_{i2}$ and we generate two scenarios:
\begin{enumerate}
    \item {\em strong heterogeneity scenario}: 
    \begin{equation*}
        \tau^{cace}(\mathbf\bX_i) =\begin{cases}
        \,\,k \quad & \text{if}\quad \mathbf\bX_i \in \ell_{1}=\{\mathbf\bX_i: X_{i1}=0,X_{i2}=0\};\\
        \!\!-k\quad & \text{if}\quad \mathbf\bX_i \in \ell_{2}=\{\mathbf\bX_i: X_{i1}=1,X_{i2}=1\};\\
        \,\, 0 \quad & \text{otherwise};
    \end{cases}
    \end{equation*} 
    \item {\em slight heterogeneity scenario}: 
    \begin{equation*}
        \tau^{cace}(\mathbf\bX_i) =\begin{cases}
        \,\,\,\,\,\,\,\,\,\,k \quad & \text{if}\quad \mathbf\bX_i \in \ell_{1}=\{\mathbf\bX_i: X_{i1}=0,X_{i2}=0\};\\
        \,\,\,\,\,\,\,\!\!-k \quad & \text{if}\quad \mathbf\bX_i \in \ell_{2}=\{\mathbf\bX_i: X_{i1}=1,X_{i2}=1\};\\
        \,\, 0.5k\quad & \text{if}\quad \mathbf\bX_i \in \ell_{3}=\{\mathbf\bX_i: X_{i1}=1,X_{i2}=0\};\\
        \!\! -0.5k \quad & \text{if}\quad \mathbf\bX_i \in \ell_{4}=\{\mathbf\bX_i: X_{i1}=0,X_{i2}=1\};\\
    \end{cases}
    \end{equation*} 
where $k$ is a positive number and, to simplify the notation, we refer to $\ell_1$ for $X_{i1}=0,X_{i2}=0$, $\ell_2$ for $X_{i1}=1,X_{i2}=1$, $\ell_3$ for $X_{i1}=1,X_{i2}=0$, and $\ell_4$ for $X_{i1}=0,X_{i2}=1$.
\end{enumerate}
Half of the learning sample was assigned to the discovery subsample and the other half to the inference subsample.

{ \color{blue}The performance of the BCF-IV algorithm is evaluated using a number of goodness-of-fit measures, averaged over $M=500$ generated data sets. With respect to the first two dimensions -- the ability to correctly discover the subgroups with heterogeneous effects at leaves level and the capacity of BCF-IV to control for false positive discovery at leaves level -- we evaluated the performance of the algorithms using a suitable rework of common performance measures from the classification literature. With respect to the first dimension, we assess the True Positive Rate (TPR) as the rate at which the algorithm is able to correctly detect the subgroups with true heterogeneity at leaves level ($\ell_1$ and $\ell_2$ in the case of strong heterogeneity, and $\ell_1, \ell_2, \ell_3$ and $\ell_4$ in the case of slight heterogeneity). TPR ranges between zero (worst performance) and one (best performance) and is computed as the ratio $\frac{TP}{TP + FN}$ where TP stands for the number of True Positive, and FN stands for the number of False Negatives. Regarding the second dimension, the performance of the algorithm to control for false positive discovery is evaluated using the False Positive Rate (FPR). FPR ranges from zero (best performance) to one (worst performance) and is the ratio  $\frac{FP}{FP + TN}$, where FP stands for the number of False Positives and TN for the number of True Negatives. We refer to Table \ref{tab:hsd} for additional details. The evaluation is performed on leaves level. The intuition is that if the algorithm is able to detect/discard heterogeneous subgroup at the final level (read leaves) it will also be able to detect it at any intermediate level (read nodes).\footnote{\color{blue} We want to highlight that, while the TPR is evaluated comparing BCF-IV (and BCF-ITT) with HCT-IV, FPR is evaluated just for BCF-IV. Indeed, as HCT-IV is not able to correct for false positive (or false discovery) rate the advantage of BCF-IV are manifest in this scenario. Hence -- to build the fairest comparison -- we employed p-values correction just in the evaluation of the FPR for BCF-IV. The significance level is set to 0.05.} Finally, the overall performance in terms of estimation accuracy for the conditional effects within the correctly discovered subgroups is evaluated using a number of performance measures for all the correctly discovered heterogeneous subgroups at leaves level. In particular, we measured for the heterogeneous subgroups the Monte Carlo estimated bias, the Monte Carlo Mean Squared Error (MSE), and the Monte Carlo coverage. These measures are introduced in more detail in the Supplementary Material.}

\begin{table}[H]
    \centering
   \begin{tabular}{ *{5}{|c} | }
  \cline{3-4}
  \mc{} & & \multicolumn{2}{c|}{True Condition} & \mc{} \\
  \cline{3-4}
  \mc{} & & \makecell{HES} & 
    \makecell{Not HES}   \\
  \cline{1-4}
  \multirow{4}{*}{\makecell{\rotatebox[origin=c]{90}{Prediction}}} & 
    \makecell{Predicted \\ as HES} & 
    \makecell{True positive \\  $TP$ \\ \textit{(Correct)}} & 
    \makecell{False positive \\  $FP$ \\ \textit{(Incorrect)}}  \\
  \cline{2-4}
  & 
    \makecell{Not Predicted  \\ as HES} & 
    \makecell{False negative \\  $FN$ \\ \textit{(Incorrect)}} &
    \makecell{True negative \\  $TN$ \\ \textit{(Correct)}}  \\
\cline{1-4}
\end{tabular}
    \caption{Classification for heterogeneous subgroups detection, where HES refers to Heterogeneous Effect Subgroup at leaves level.}
    \label{tab:hsd}
\end{table}

\redsout{In the next Subsections we will depict and comment the results from a set of simulated data generating processes. In particular, in Subsection \ref{subsec:sim1} we will compare different structures (\textit{strong heterogeneity} vs \textit{slight} heterogeneity) and magnitudes of the heterogeneity, and different sizes of the learning dataset. In Subsection \ref{subsec:sim2} we assess the performance of various algorithms in the presence of mild and weak instruments. Finally, in Subsection \ref{subsec:sim3}, we introduce variations in the compliance rates among different subgroups.}


Building on the data generating process described above, we introduce variations in the effect size of the heterogeneous causal effects, in the structure of the heterogeneity and in the sample sizes. These features are important for various reasons. Firstly, as the size of the effect within the heterogeneous subgroups and its difference between various such subgroups are growing the algorithms should depict a higher ability to correctly discover these subgroups. Secondly, it is important to assess the ability of the algorithms to discover the correct subgroups in the presence of various structures of the heterogeneity. Finally, both the discovery and the estimation ability of the algorithms are expected to vary based on the size of the learning sample.

\begin{figure}[H]
    \centering
    \includegraphics[width=0.74\linewidth]{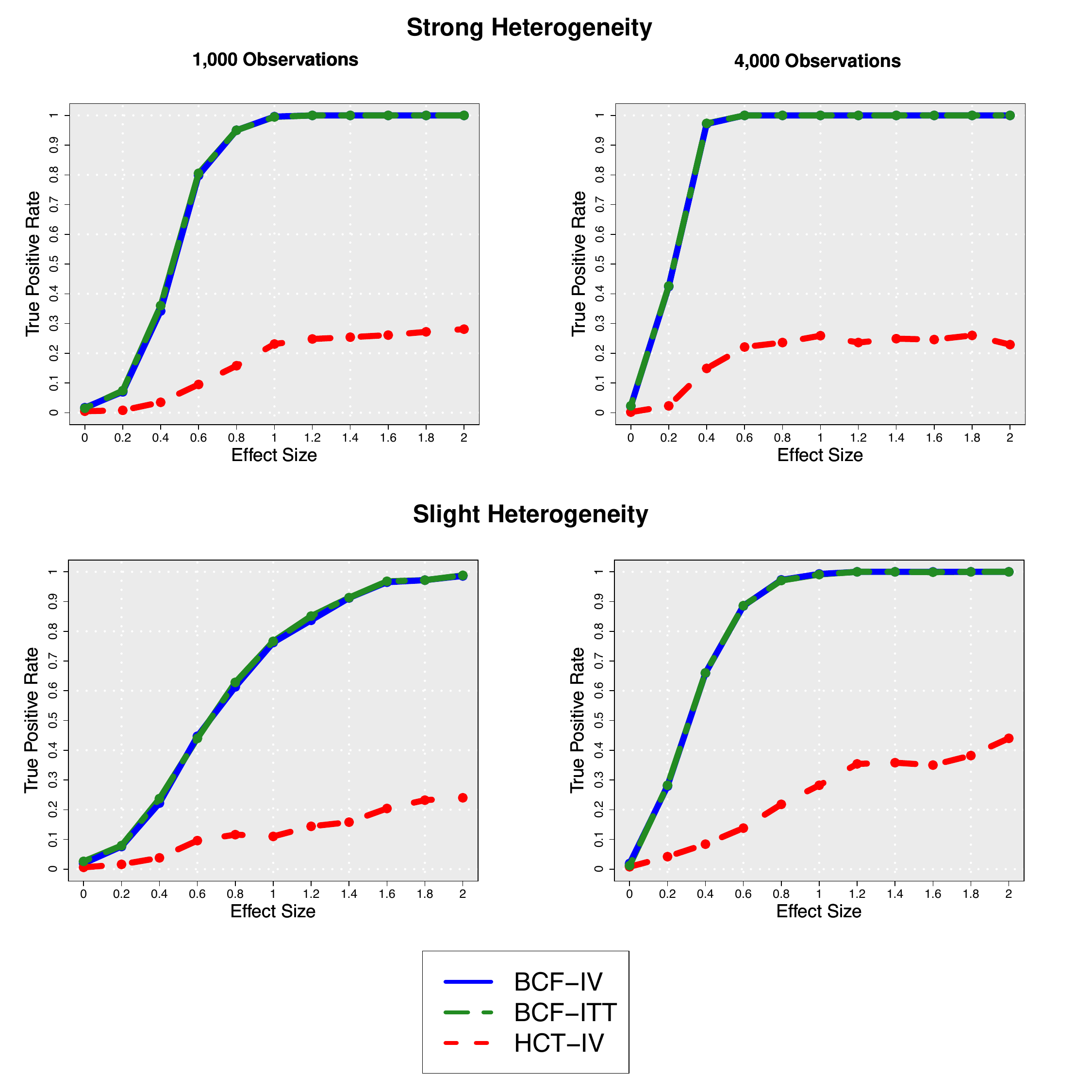}
    \caption{TPR for the \textit{strong heterogeneity} and the \textit{slight heterogeneity} scenarios.}
    \label{fig:rules_mtx1}
\end{figure}

\begin{figure}[H]
    \centering
    \includegraphics[width=0.74\linewidth]{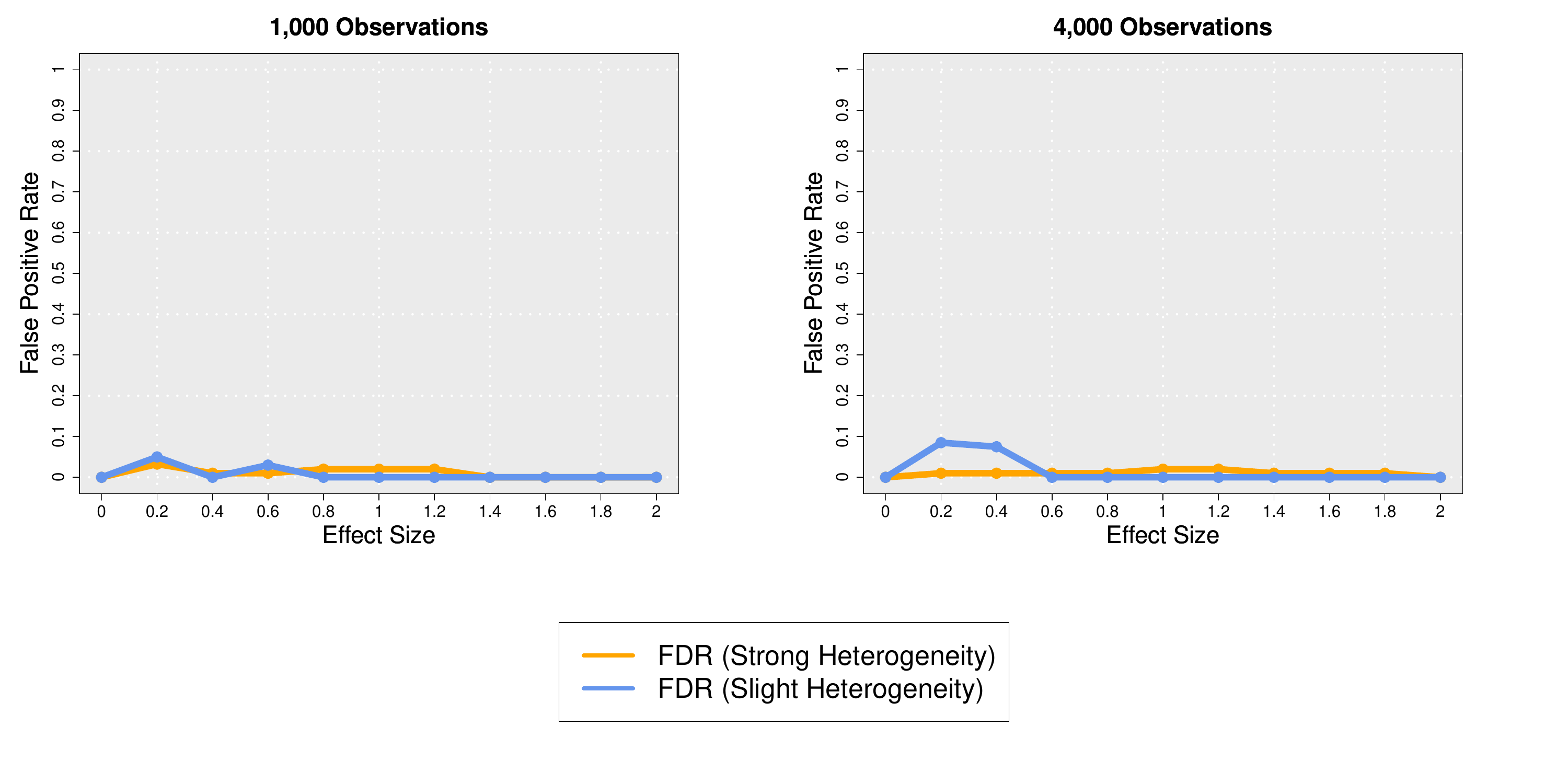}
    \caption{FDR for the \textit{strong heterogeneity} and the \textit{slight heterogeneity} scenarios.}
    \label{fig:rules_mtx2}
\end{figure}

In the proposed simulation setting, we {\color{blue} consider} three varying factors: (i) the effect size $k$ from 0 (no heterogeneity) to 2 (greater level of  heterogeneity); (ii) the structure of the heterogeneity: \textit{slight heterogeneity scenario} vs \textit{strong heterogeneity scenario} and; (iii) the sample size with $N=1,000$, or $4,000$. {\color{blue} These sample sizes refer to the size of data used for either the discovery or the inference step.} The compliance rate is set to be 0.75 (strong instrument scenario) and the covariance between the covariates to be zero. {\color{blue}The results for TPR are depicted in Figures \ref{fig:rules_mtx1}. Figure \ref{fig:rules_mtx2} reports the results for the FDR.} The results for the overall performance in terms of estimation accuracy within the subgroups are shown in Tables \ref{tab:sims_0.75_1000} and \ref{tab:sim_0.75_4000}.
{\color{blue}Figure \ref{fig:rules_mtx1} depicts the results for the TPR in the \textit{strong} and \textit{slight heterogeneity} scenario both with 1,000 and 4,000 observations. We compare the results for the subgroups identified with BCF-IV (blue line) with the ones of BCF-ITT (green line) and HCT-IV (red line). In both scenarios, BCF-IV and BCF-ITT are able to converge to the highest possible TPR as the effect size grows larger. This convergence is faster as the sample sizes increases.  Moreover, their performance is very similar, and they always outperform HCT-IV. Figure \ref{fig:rules_mtx2} depicts the results for the FDR for BCF-IV in the same scenarios. These results complement the information from the previous figure on TPR. While the former results focus on the ability of BCF-IV to correctly discover the heterogeneous subgroups, the latter results evaluate the ability of the algorithm to control for false discoveries (FPR). In both \textit{strong} and \textit{slight heterogeneity} scenarios, BCF-IV is able to keep a FPR very close to zero by discarding false discoveries through p-values adjustments controlling for the familywise error rate.}

Tables \ref{tab:sims_0.75_1000} and \ref{tab:sim_0.75_4000} show the results for the estimated conditional effects within the subgroups with heterogeneous effects for the case of \textit{strong heterogeneity} with 1,000 and 4,000 data points respectively.\footnote{The estimation results for the case of \textit{slight heterogeneity} mimic the results for the case of \textit{strong heterogeneity}, as these two designs do not differently affect the ability of the algorithms to precisely estimate the conditional effects.} Also with respect to the estimation of the conditional CACEs, BCF-IV seems to perform in a very good way irrespective of the effect size. Indeed, the Monte Carlo estimated MSE and bias are smaller than the ones of GRF, and the Monte Carlo coverage approaches the value of 0.95.

\begin{table}[H]
\centering \scriptsize
\begin{tabular}{ccccccc} 
\textit{Effect Size} & MSE($\hat{\tau}^{cace}_{l_1}$) & Bias($\hat{\tau}^{cace}_{l_1}$) & Coverage($\hat{\tau}^{cace}_{l_1}$) & \multicolumn{1}{l}{MSE($\hat{\tau}^{cace}_{l_2}$)} & \multicolumn{1}{l}{Bias($\hat{\tau}^{cace}_{l_2}$)} & \multicolumn{1}{l}{Coverage($\hat{\tau}^{cace}_{l_2}$)} \\ \toprule
\multicolumn{1}{l}{} & \multicolumn{6}{c}{BCF-IV}\\ \toprule
0                    & 0.030                   & 0.138                    & 0.952                        & 0.031                                       & 0.140                                        & 0.938                                            \\
0.1                  & 0.021                   & 0.020                    & 0.974                        & 0.019                                       & 0.018                                        & 0.980                                            \\
0.2                  & 0.028                   & -0.004                   & 0.960                        & 0.027                                       & -0.014                                       & 0.968                                            \\
0.3                  & 0.027                   & 0.008                    & 0.956                        & 0.031                                       & 0.010                                        & 0.938                                            \\
0.4                  & 0.028                   & -0.013                   & 0.962                        & 0.031                                       & -0.022                                       & 0.950                                            \\
0.5                  & 0.029                   & 0.007                    & 0.954                        & 0.026                                       & -0.014                                       & 0.964                                            \\
0.6                  & 0.026                   & -0.004                   & 0.960                        & 0.031                                       & -0.016                                       & 0.944                                            \\
0.7                  & 0.028                   & 0.003                    & 0.958                        & 0.032                                       & -0.012                                       & 0.944                                            \\
0.8                  & 0.030                   & -0.003                   & 0.944                        & 0.029                                       & -0.002                                       & 0.952                                            \\
0.9                  & 0.031                   & 0.010                    & 0.936                        & 0.029                                       & 0.002                                        & 0.944                                            \\
1                    & 0.028                   & 0.004                    & 0.952                        & 0.029                                       & 0.004                                        & 0.950                                            \\ \midrule
\multicolumn{1}{l}{} & \multicolumn{6}{c}{GRF}                                                                                                                                                                                                           \\ \toprule
0                    & 0.018                   & 0.107                    & 0.998                        & 0.017                                       & 0.103                                        & 0.998                                            \\
0.1                  & 0.015                   & -0.072                   & 1.000                        & 0.015                                       & -0.075                                       & 1.000                                            \\
0.2                  & 0.070                   & -0.235                   & 0.962                        & 0.066                                       & -0.222                                       & 0.954                                            \\
0.3                  & 0.135                   & -0.328                   & 0.732                        & 0.136                                       & -0.329                                       & 0.708                                            \\
0.4                  & 0.197                   & -0.398                   & 0.646                        & 0.197                                       & -0.399                                       & 0.648                                            \\
0.5                  & 0.235                   & -0.427                   & 0.652                        & 0.234                                       & -0.431                                       & 0.658                                            \\
0.6                  & 0.249                   & -0.439                   & 0.684                        & 0.264                                       & -0.445                                       & 0.672                                            \\
0.7                  & 0.232                   & -0.394                   & 0.752                        & 0.234                                       & -0.408                                       & 0.752                                            \\
0.8                  & 0.216                   & -0.367                   & 0.798                        & 0.224                                       & -0.384                                       & 0.782                                            \\
0.9                  & 0.208                   & -0.344                   & 0.838                        & 0.219                                       & -0.357                                       & 0.824                                            \\
1                    & 0.186                   & -0.318                   & 0.862                        & 0.181                                       & -0.309                                       & 0.876                                            \\ \bottomrule
\end{tabular}%
\caption{Simulation results for 1,000 data points and strong instrument.}
\label{tab:sims_0.75_1000}
\end{table}

\begin{table}[H]
\centering \scriptsize
\begin{tabular}{ccccccc} 
\textit{Effect Size} & MSE($\hat{\tau}^{cace}_{l_1}$) & Bias($\hat{\tau}^{cace}_{l_1}$) & Coverage($\hat{\tau}^{cace}_{l_1}$) & \multicolumn{1}{l}{MSE($\hat{\tau}^{cace}_{l_2}$)} & \multicolumn{1}{l}{Bias($\hat{\tau}^{cace}_{l_2}$)} & \multicolumn{1}{l}{Coverage($\hat{\tau}^{cace}_{l_2}$)} \\ \toprule
\multicolumn{1}{l}{} & \multicolumn{6}{c}{BCF-IV}\\ \toprule
0                    & 0.007                   & 0.068                    & 0.950                        & 0.008                                       & 0.071                                        & 0.944                                            \\
0.1                  & 0.007                   & 0.004                    & 0.960                        & 0.007                                       & 0.007                                        & 0.944                                            \\
0.2                  & 0.007                   & -0.004                   & 0.950                        & 0.007                                       & -0.003                                       & 0.942                                            \\
0.3                  & 0.007                   & 0.001                    & 0.940                        & 0.007                                       & -0.003                                       & 0.960                                            \\
0.4                  & 0.007                   & 0.002                    & 0.962                        & 0.007                                       & 0.002                                        & 0.940                                            \\
0.5                  & 0.007                   & -0.008                   & 0.958                        & 0.007                                       & 0.003                                        & 0.956                                            \\
0.6                  & 0.007                   & -0.006                   & 0.940                        & 0.008                                       & -0.004                                       & 0.938                                            \\
0.7                  & 0.006                   & 0.002                    & 0.968                        & 0.007                                       & -0.004                                       & 0.948                                            \\
0.8                  & 0.008                   & 0.006                    & 0.942                        & 0.007                                       & -0.001                                       & 0.952                                            \\
0.9                  & 0.007                   & 0.001                    & 0.960                        & 0.008                                       & 0.002                                        & 0.942                                            \\
1                    & 0.008                   & -0.005                   & 0.938                        & 0.006                                       & 0.007                                        & 0.966                                            \\ \midrule
\multicolumn{1}{l}{} & \multicolumn{6}{c}{GRF}                                                                                                                                                                                                           \\ \toprule
0                    & 0.006                   & 0.063                    & 1.000                        & 0.006                                       & 0.063                                        & 1.000                                            \\
0.1                  & 0.013                   & -0.080                   & 1.000                        & 0.012                                       & -0.079                                       & 1.000                                            \\
0.2                  & 0.032                   & -0.143                   & 0.968                        & 0.030                                       & -0.135                                       & 0.976                                            \\
0.3                  & 0.033                   & -0.126                   & 0.972                        & 0.033                                       & -0.129                                       & 0.980                                            \\
0.4                  & 0.031                   & -0.102                   & 0.992                        & 0.028                                       & -0.097                                       & 0.978                                            \\
0.5                  & 0.030                   & -0.093                   & 0.984                        & 0.025                                       & -0.078                                       & 0.986                                            \\
0.6                  & 0.023                   & -0.051                   & 0.994                        & 0.025                                       & -0.055                                       & 0.996                                            \\
0.7                  & 0.017                   & -0.026                   & 1.000                        & 0.019                                       & -0.032                                       & 0.998                                            \\
0.8                  & 0.018                   & -0.014                   & 1.000                        & 0.016                                       & -0.015                                       & 0.998                                            \\
0.9                  & 0.017                   & -0.011                   & 1.000                        & 0.019                                       & -0.005                                       & 1.000                                            \\
1                    & 0.016                   & -0.008                   & 0.996                        & 0.014                                       & -0.004                                       & 0.998                                            \\ \bottomrule
\end{tabular}%
\caption{Simulation results for 4,000 data points and strong instrument.}
\label{tab:sim_0.75_4000}
\end{table}

\section{Heterogeneous Causal Effects of Education Funding} \label{application}

There is a wide consensus that education positively influences labor market outcomes \cite[see the review by][]{Psacharopoulos2018}. Moreover, the question on whether or not school spending affects students' performances has been central in the economic literature \citep{coleman1966equality, hanushek2003failure, jackson2015effects, jackson2018does}. Students' performance can be driven by multiple factors connected with students' characteristics and environmental characteristics. However, to the best of our knowledge, this is the first paper to study the  heterogeneous impact of additional school funding on students' performance using machine learning techniques tailored for causal inference.
In this Section we apply the BCF-IV algorithm to evaluate the impact and estimate the heterogeneity in the effects of additional funding to schools with disadvantaged students on students' performance. First, we describe the data used for this application. Next, we depict the identification strategy. Finally, we describe the results obtained and their relevance in the economics of education literature.

\subsection{Data} \label{data}

The EEO program, promoted by the Flemish Ministry of Education to encourage ``Equal Educational Opportunities",  provides additional funding for secondary schools with a significant share of disadvantaged students. Owing to the funding schools can hire additional teachers and increase the number of teaching hours.  Pupils are considered to be disadvantaged on the basis of five different indicators: (i) the pupil lives outside the family; (ii) the pupil does not speak Dutch as a native language; (iii) the mother of the pupil does not have a secondary education degree; (iv) the pupil receives educational grant guaranteed for low income families; and (v) one of the parents is part of the travelling population. In order for a school to be eligible for the EEO funding, it needs to satisfy two conditions: the first condition is that the share of students with at least one of the five characteristics has to exceed an exogenously set threshold; to avoid fragmentation of resources, the second condition requires that the additional resources should be at least larger than six teaching hours a week. The exogenous threshold is, for students in the first two years of secondary education (first stage students), a minimum share of 10\% disadvantaged students.
\par The Flemish Ministry of Education provided us with data on the universe of pupils in the first stage of education in the school year 2010/2011 (135,682 students). In particular, we have data on student level characteristics and school level characteristics. The student level characteristics cover the gender of the pupil (\textit{gender}), the grade retention in primary school (\textit{retention}) and the inclusion of the pupil in the special needs student population in primary school (which serves as a proxy of student's low cognitive skills). The school level characteristics include both the teacher characteristics, such as the teachers' age, seniority and education, in addition to principal characteristics, such as the principals' age and seniority. Teacher and principal seniority measures the level of experience of the teachers and principals, respectively. These variables assume values in the range of 1 to 7, where the teachers (and principals) with a seniority level of 1 are the least experienced (0-5 years of experience) and teachers (and principals) with a seniority level of 7 are the most experienced (more than 30 years of experience).\footnote{Teachers and principals' seniority classes are the following: class 1: between 0 and 5 years of experience; class 2: between 6 and 10; class 3: between 11 and 15; class 4: between 16 and 20; class 5: between 20 and 25; class 6: between 26 and 30; class 7: more than 30.} Similarly, the ages of teacher and principal are reported as categorical variables that range from 1 to 8, where teachers/principals in the first category are the youngest (less than 30 years old) and teachers/principals in the last category are the oldest (more than 60 years old).\footnote{Teachers and principals' age classes are the following: class 1: less than 30 years old; class 2: between 30 and 34; class 3: between 35 and 39; class 4: between 40 and 44; class 5: between 45 and 49; class 6: between 50 and 54; class 7: between 55 and 60; class 8: more than 60.} { Teachers' education records whether or not the teacher holds a pedagogical training (in the following we will refer to it as ``teacher training")}. All these variables are aggregated at school level in the form of averages (for age and seniority) and shares (for teachers' education) and assigned to each student with respect to the school where he/she is enrolled.

{\color{blue}The outcome variable is a dummy variables defined as follows: the variable \textit{A-certificate} assumes value 1 if the student gets an ``A-certificate" at the end of the school year (which is the most favorable outcome) and 0 if not.} Since we do not have data on standardized test scores for Flemish students, \textit{A-certificate} is a good, available proxy of student performance. Every year, each student performs a final test and gets a ranking from ``A" to ``C". Students that get an ``A" can progress school without any restriction, while the students that get either ``B" or ``C" can progress school but only in specific programs or have some grade retention. {\color{blue} Furthermore, we additionally analyse -- in the Supplementary Material -- the results that would be obtained using the variable \textit{progress school} as outcome. This variable assumes value 1 if the student progresses to the following year without any grade retention and 0 if not (this variable is a complement of school retention).} Both these outcome variables are proxies for different levels of students' performance: a positive \textit{A-certificate} proxies for a higher level of performance than a positive \textit{progress school}. In principle, the target of a policy-maker could be to have the highest possible share of students getting ``A-certificates" and the lowest share of students not progressing through school.

\subsection{Identification Strategy}\label{subsec:identification_strategy}

To evaluate the impact of the policy on students' performance we apply the BCF-IV within a regression discontinuity design \citep{trochim1984research, hahn2001identification}. Regression Discontinuity Design (RDD) is a method that aims at evaluating the causal effects of interventions in settings where the assignment to the treatment is determined (at least partly) by the values of an observed covariate lying on either sides of a threshold point. The idea is that subjects just above and below this threshold are very similar and one can assume a quasi-randomization around the threshold \citep{mealli2012evaluating}. RDDs are categorized in sharp RDDs and fuzzy RDDs.

In sharp RDDs, the central assumption is that, around the threshold, there is a sharp discontinuity (from 0 to 1) in the probability of being treated. This is due to the fact that in sharp RDDs there is no room for imperfect compliance. In many real world scenarios, however, thresholds are not strictly implemented, as in the case of our application.
To deal with these situations, one can use fuzzy RDDs, which are applicable when around the threshold the probability of being actually treated changes discontinuously, but not sharply from 0 to 1 (i.e., the jump in the probability of being treated is less than 1). 

In our application of the fuzzy RDD technique, we exploit two cutoffs around the 10\% share of disadvantaged students in the first stage of secondary education. The students in schools just below the threshold are assigned to the control group ($Z_i=0$), while the students in schools just above the threshold are assigned to the treated group ($Z_i=1$).
The bandwidth around the threshold (from which one obtains the two cutoffs) is determined using the data-driven methodology proposed by \cite{calonico2014robust} and implemented in the \texttt{rdrobust package} in \texttt{R} \citep{calonico2015rdrobust}.  This methodology is used as a stand-alone bandwidth selector, and inference on the treatment effect is performed, on the inference subsample, using the estimators introduced in Section \ref{subsec:estimation}.  
Following the indication of \cite{imbens2008regression}, we focus on the outcome equation to select the bandwidth and then we use the same bandwidth for the estimation of the treatment equation. Moreover, as pointed out by \cite{lee2010regression}, the usage of the same bandwidth guarantees the validity of the 2SLS estimators used to estimate the causal effects. 

The selected bandwidth around the threshold is 3.5\% for the outcome variables \textit{A-certificate}. To further validate this bandwidth, selected using the method of \cite{calonico2014robust}, we run additional analyses implementing the Bayesian methods proposed by \cite{li2015evaluating} and by \cite{mattei2016regression}. In particular, we run a series of robustness checks for the selection of the bandwidth around the threshold implementing a hierarchical Bayesian model for assessing the balance of the covariates between the groups of observations assigned to the treatment and the ones assigned to the control. For the selected bandwidth, the probability of the pre-assignment variables being well-balanced is high for the subpopulations defined by values of the cutoff strictly lower than 3.5\%. Indeed, these probabilities are larger than or close to 0.8, indicating that the covariates are balanced in the two groups.

Moreover, to guarantee an equal representation to all the schools, and to avoid biases related to the over-representation of biggest schools' students, we sample 50 pupils from each school.  In turn, this leads to a higher balance among the averages between the observations assigned to the treatment and the observations assigned to the control, as shown in panel (a) of Figure C.1 in the Supplementary Material.

In the Supplementary Material, we run a series of tests to show that the RDD (Regression Discontinuity Design) is valid for this application. Moreover, as a robustness check we sample a higher number of students according to the size of the smallest school (62 pupils) from every school. We show the balance in the samples of units assigned to the treatment and to the control in the second scenario.

\subsection{Results} \label{results}

This Section assesses the effects of the additional funding on students' performances and highlights the main drivers of the heterogeneity in causal effects. {\color{blue}These analyses are made for the outcome variable \textit{A-certificate}. 
In the Supplementary Material, we report the results for a \textit{progress school}. It is important to highlight that the results for both the outcomes, considered separately, in terms of effects and heterogeneity drivers, remain roughly the same when we (i) widen the sample of units included in the analysis, and (ii) perform the heterogeneity discovery on the ITT  (results are reported in the Supplementary Material)}.

The variable \textit{A-certificate} serves as a proxy for positive performance. In our sample, the students that got an ``A-certificate" are the 91.73\% of the total population. In Figure \ref{fig:BCF-IV-certificate-50}, the heterogeneous Complier Average Causal Effects (CACE) estimated using the proposed model are depicted. The darker the shade of blue in the node the higher the causal effect. 

Although positive, the overall effect of the additional funding is not significant.  This finding is in line with the recent literature on school spending and students' performance in a cross-country scenario \citep{hanushek2016handbook, hanushek2017school} and in the Flanders, in particular \citep{dewitte2018disadvantaged}.  However, it is compelling to observe the main drivers of the heterogeneity in the causal effect. {\color{blue} It is worth highlighting that we did not implement any leaves trimming using p-values corrections as the aim of this application was to discover an interpretable representation of potential heterogeneity in the effects, even in the absence of significant results.}

The first driver in the heterogeneity of the effects is the variable \textit{principal age}: for students in schools with younger principals, the effects of funding are larger. These results, even if not significant, show that the treatment effects are higher for students that are in schools with principals younger than 55 years old. 
The second driver of heterogeneity is the seniority of the teachers: students in schools with younger principals and with less senior teachers -- namely, teachers with less than 11 years of experience --  have an increase in their performance. The conditional effect is 0.051, meaning that being treated leads to an increase of 5.1\% in the probability of getting the highers possible grade. On the opposite side, students in schools with older principals and with more senior teachers -- namely, teachers with more than 11 years of experience --  have a decrease in their performance. The conditional effect is -0.039, meaning that being treated leads to a decrease of 3.9\% in the probability of getting the highest possible grade.

Both these heterogeneity drivers, namely, the seniority and age of the teachers and principals, are particularly appreciable, as there are evidences in the education literature that connect teachers' seniority \citep{rice2010impact, harris2011teacher},  teachers' age \citep{holmlund2008gender} to their teaching performance, and in turn teaching performance to students' positive achievements \citep{goldhaber2010using}. Moreover, there is a compelling evidence in the literature regarding the role of principals in driving higher students' achievements \citep{eberts1988student, gentilucci2007principals}, however this is, to the extent of our knowledge, the first research that highlights the role of principal's age and seniority as drivers of treatment effect variations. Interestingly, another source of treatment variation is student's gender. Female students in schools with older principals, depict a positive effect while male students a negative effect. Yet, the effects for both subgroups are not significant.

This evidence can be interpreted in the following way: the additional funding has a negative, but not statistically significant, effect in the performance of students in the overall population, but it increases its effect in a notable way for those students in schools with less senior teachers and younger principals. These results are in line with the evidence that additional school funding does not boost the performance of the overall population of students \citep{hanushek2016handbook, hanushek2017school, dewitte2018disadvantaged} and with the literature that connects students' achievements with principals \citep{eberts1988student, gentilucci2007principals} and teachers performance \citep{holmlund2008gender, rice2010impact, harris2011teacher, goldhaber2010using}. It is important to highlight that even if we find some evidence of treatment effects variation connected to teachers' seniority, principal age, and students gender the conditional causal effects are not significantly different from zero. {\color{blue} These results are robust when the outcome variable considered is \textit{progress school} (see Supplementary Material).}

\begin{figure}[t!]  
\centering
\begin{tikzpicture}[level distance=100pt, sibling distance=15pt, edge from parent path={(\tikzparentnode) -- (\tikzchildnode)}]
\tikzset{every tree node/.style={align=center}}
\Tree [.\node[fill=non-photoblue,circle,draw]{CACE\\ -0.004\\ 100\%};
\edge node[auto=right,pos=.6]{Principal Age $<7$};[. \node[fill=non-photoblue, circle,draw]{CACE\\-0.001\\38\%};
\edge node[auto=right,pos=.6]{Teacher Senior. $< 4$}; \node[fill=tuftsblue,circle,draw]{CACE\\0.051\\10\%};
\edge node[auto=left,pos=.6]{Teacher Senior. $\geq 4$}; \node[fill=bubbles, circle,draw]{CACE\\-0.039\\52\%}; ]
\edge node[auto=left,pos=.6]{Principal Age $\geq 7$};[. \node[fill=bubbles,circle,draw]{CACE\\-0.028\\62\%};
\edge node[auto=right,pos=.6]{Gender = ``Female"}; \node[fill=ceruleanblue,circle,draw]{CACE\\0.186\\20\%};
\edge node[auto=left,pos=.6]{Gender = ``Male"}; \node[fill=whitish,circle,draw]{CACE\\-0.237\\18\%}; ]]
\end{tikzpicture}
\caption{\footnotesize Visualization of the heterogeneous Complier Average Causal Effects (CACE) of additional funding on \textit{A-certificate} estimated using the proposed BCF-IV model. The overall learning sample size is 4,300. The tree is a summarizing classification tree fit to posterior point estimates of individual treatment effects. \\
The significance level is * for a significance level of 0.1, ** for a significance level of 0.05 and *** for a significance level of 0.01.}
\label{fig:BCF-IV-certificate-50}
\end{figure}

\section{Conclusion and Discussion} \label{conclusion}

This paper developed a novel Bayesian machine learning technique, BCF-IV, to draw causal inference in scenarios with imperfect compliance. By investigating the heterogeneity in the causal effects, the technique expedites targeted policies. We manifested that the BCF-IV technique outperforms other machine learning techniques tailored for causal inference in precisely estimating the causal effects and converges to an optimal large sample performance in identifying the subgroups with heterogeneous effects.  Moreover, using Monte-Carlo simulations, we showed that the competitive advantages of using BCF-IV, as compared to GRF or HCT-IV, are substantial.

In our application, we evaluated the effects of additional funding on students' performances. While the overall effects are negative but not significant, there are significant differences among different sub-populations of students. Indeed, for students in schools with less senior and younger principals (principals younger than 55 years old and with less than 30 years of experience) the effects of the policy are greater. We want to highlight that age and seniority of principals as treatment variation drivers are robust to different definitions of the outcome variable (see results in Section \ref{application}), to variations in the algorithm used (BCF-IV vs BCF-ITT), in the size and definition of the learning sample and, in turn, in the balance between the group of treated and control units (see the Supplementary Material).

On one hand, as an underlying mechanism, the need for additional funds can be higher in schools with younger and less senior principals, who are more often observed in the most disadvantaged schools. This phenomenon arises as senior  principals select themselves out of the most disadvantaged schools and more into advantaged schools, thereby creating relatively more vacancies in disadvantaged schools. Therefore, on average, younger and less senior principals lack a real choice but to start working in the most disadvantaged schools. Moreover, we can think of the motivation for principals to decrease as they grow older and this, in turn, have an impact on their performance, and their ability to effectively allocate the additional funds. To the best of our knowledge, this is the first study that investigates the effects of age and seniority of principals on enhancing the effectiveness of school funding on students' performance. The investigation of the true causal channel is beyond the goals of this paper and is left to further investigation where more granular teachers' and principals' characteristics are available. 

These results are relevant to the policy as they furnish the instruments to policy-makers to enhance the effects of additional funding on students' performance. Indeed, on one side policy-makers could target just students in school with positive, effects reducing the overall costs of the policy and using the savings to experiment more effective policies in the other schools. On the other side, policy-makers could analyze the reason of lack of the effectiveness of funding in schools with certain characteristics and implement policies to boost the effects of future funding. Furthermore, the added value of our algorithm is that it could enable policy-makers to target just those units that benefit the most from the treatment and it provides an insight on possible inefficiencies in the allocation and/or usage of funding. From our analysis it seems that there is room for policies that support less senior principals since students in their schools show higher returns in terms of performance from additional funding.

The extension of these methods to other fields of economic investigation and the development of novel machine learning algorithms for targeted policies and welfare maximization can form the future scope of further research.  In particular, the development of an algorithm that could deal with welfare maximization in the context of multiple outcomes is of interest. {\color{blue} As a further extension, it may be worth exploring hierarchical testing procedures to control for familywise error rate \cite[see, e.g.,][]{marcus1976closed} or for false discovery rate \cite[see, e.g.,][]{yekutieli2008hierarchical} at each node of the tree discovered by BCF-IV and not just in the terminal nodes. Such algorithms have been explored in the context of matching procedures by \cite{johnson2019detecting} and \cite{lee2021discovering}.} Moreover, the ``usual" Bayesian way for estimating CACE is via a data augmentation scheme, \cite[e.g., imputing compliance status and estimating impacts among estimated compliers, {\color{blue} as in}][]{imbens2015causal}. In our algorithm we do not implement such a methodology, however, as a further line of research, it could be extended including a data augmentation scheme.


\setlength{\bibsep}{0.0pt}
\bibliography{sample}

@article{holmlund2008gender,
  title={Is the gender gap in school performance affected by the sex of the teacher?},
  author={Holmlund, Helena and Sund, Krister},
  journal={Labour Economics},
  volume={15},
  number={1},
  pages={37--53},
  year={2008},
  publisher={Elsevier}
}

@article{goldhaber2010using,
  title={Using performance on the job to inform teacher tenure decisions},
  author={Goldhaber, Dan and Hansen, Michael},
  journal={American Economic Review},
  volume={100},
  number={2},
  pages={250--55},
  year={2010}
}

@article{Psacharopoulos2018,
  title={Returns to investment in education: a decennial review of the global literature},
  author={Psacharopoulos, George and Patrinos, Harry Anthony},
  journal={Education Economics},
  volume={26},
  number={5},
  pages={445--458},
  year={2018}
}

@article{wang2018instrumental,
  title={An instrumental variable tree approach for detecting heterogeneous treatment effects in observational studies},
  author={Wang, Guihua and Li, Jun and Hopp, Wallace J},
  journal={Available at SSRN 3045327},
  year={2018}
}

@article{lee2010regression,
  title={Regression discontinuity designs in economics},
  author={Lee, David S and Lemieux, Thomas},
  journal={Journal of Economic Literature},
  volume={48},
  number={2},
  pages={281--355},
  year={2010}
}

@article{friedman1984classification,
  title={Classification and regression trees},
  author={Breiman, Leo and Friedman, Jerome H and Olshen, Richard A and Stone, Charles J},
  journal={Belmont, CA: Wadsworth \& Brooks},
  year={1984}
}

@book{breiman1984classification,
  title={Classification and regression trees},
  author={Breiman, Leo},
  year={1984},
  publisher={Routledge},
  address = {New York, New York}
}

@book{wooldridge2002,
  title={Econometric analysis of cross section and panel data},
  author={Wooldridge, Jeffrey M.},
  year={2002},
  publisher={MIT Press},
  address = {Cambridge, Massachusetts}
}

@article{mullainathan2017machine,
  title={Machine learning: an applied econometric approach},
  author={Mullainathan, Sendhil and Spiess, Jann},
  journal={Journal of Economic Perspectives},
  volume={31},
  number={2},
  pages={87--106},
  year={2017}
}

@article{athey2019generalized,
  title={Generalized random forests},
  author={Athey, Susan and Tibshirani, Julie and Wager, Stefan and others},
  journal={The Annals of Statistics},
  volume={47},
  number={2},
  pages={1148--1178},
  year={2019},
  publisher={Institute of Mathematical Statistics}
}

@article{athey2016recursive,
  title={Recursive partitioning for heterogeneous causal effects},
  author={Athey, Susan and Imbens, Guido},
  journal={Proceedings of the National Academy of Sciences},
  volume={113},
  number={27},
  pages={7353--7360},
  year={2016},
  publisher={National Acad Sciences}
}

@article{hahn2020bayesian,
author = "Hahn, P. Richard and Murray, Jared S. and Carvalho, Carlos M.",
doi = "10.1214/19-BA1195",
fjournal = "Bayesian Analysis",
journal = "Bayesian Anal.",
month = "09",
number = "3",
pages = "965--1056",
publisher = "International Society for Bayesian Analysis",
title = "Bayesian Regression Tree Models for Causal Inference: Regularization, Confounding, and Heterogeneous Effects (with Discussion)",
volume = "15",
year = "2020"
}

@article{mealli2012evaluating,
  title={Evaluating the effects of university grants by using regression discontinuity designs},
  author={Mealli, Fabrizia and Rampichini, Carla},
  journal={Journal of the Royal Statistical Society: Series A (Statistics in Society)},
  volume={175},
  number={3},
  pages={775--798},
  year={2012},
  publisher={Wiley Online Library}
}

@article{bargagli2018estimating,
  title={Estimating Heterogeneous Causal Effects in the Presence of Irregular Assignment Mechanisms},
  author={Bargagli Stoffi, Falco J and Gnecco, Giorgio},
  pages = {1--10},
  journal={In Proceedings of the 5th IEEE Conference in Data Science and Advanced Analytics},
  year={2018}
}

@article{chipman2010bart,
  title={BART: Bayesian additive regression trees},
  author={Chipman, Hugh A and George, Edward I and McCulloch, Robert E and others},
  journal={The Annals of Applied Statistics},
  volume={4},
  number={1},
  pages={266--298},
  year={2010},
  publisher={Institute of Mathematical Statistics}
}

@article{angrist1996identification,
  title={Identification of causal effects using instrumental variables},
  author={Angrist, Joshua D and Imbens, Guido W and Rubin, Donald B},
  journal={Journal of the American Statistical Association},
  volume={91},
  number={434},
  pages={444--455},
  year={1996},
  publisher={Taylor \& Francis}
}

@article{mccrary2008manipulation,
  title={Manipulation of the running variable in the regression discontinuity design: A density test},
  author={McCrary, Justin},
  journal={Journal of Econometrics},
  volume={142},
  number={2},
  pages={698--714},
  year={2008},
  publisher={Elsevier}
}

@article{hahn2019atlantic,
  title={Atlantic causal inference conference (ACIC) data analysis challenge 2017},
  author={Hahn, P Richard and Dorie, Vincent and Murray, Jared S},
  journal={arXiv preprint arXiv:1905.09515},
  year={2019}
}

@article{dewitte2018disadvantaged,
  title={Impact evaluation in a multi-input multi-output setting: Evidence on the effect of additional resources for schools},
  author={D’Inverno, Giovanna and Smet, Mike and De Witte, Kristof},
  journal={European Journal of Operational Research},
  volume={290},
  number={3},
  pages={1111--1124},
  year={2021},
  publisher={Elsevier}
}

@article{rubin1974estimating,
  title={Estimating causal effects of treatments in randomized and nonrandomized studies.},
  author={Rubin, Donald B},
  journal={Journal of Educational Psychology},
  volume={66},
  number={5},
  pages={688},
  year={1974},
  publisher={American Psychological Association}
}

@article{rubin1978bayesian,
  title={Bayesian inference for causal effects: The role of randomization},
  author={Rubin, Donald B},
  journal={The Annals of Statistics},
  pages={34--58},
  year={1978},
  publisher={JSTOR}
}

@book{imbens2015causal,
  title={Causal inference in statistics, social, and biomedical sciences},
  author={Imbens, Guido W and Rubin, Donald B},
  year={2015},
  publisher={Cambridge University Press}
}

@book{angrist2008mostly,
  title={Mostly harmless econometrics: An empiricist's companion},
  author={Angrist, Joshua D and Pischke, J{\"o}rn-Steffen},
  year={2008},
  publisher={Princeton University Press}
}

@article{breiman2001random,
  title={Random forests},
  author={Breiman, Leo},
  journal={Machine Learning},
  volume={45},
  number={1},
  pages={5--32},
  year={2001},
  publisher={Springer}
}

@article{hill2011bayesian,
  title={Bayesian nonparametric modeling for causal inference},
  author={Hill, Jennifer L},
  journal={Journal of Computational and Graphical Statistics},
  volume={20},
  number={1},
  pages={217--240},
  year={2011},
  publisher={Taylor \& Francis}
}

@article{kapelner2013bartmachine,
  title={BartMachine: Machine learning with Bayesian additive regression trees},
  author={Kapelner, Adam and Bleich, Justin},
  journal={arXiv preprint, arXiv:1312.2171},
  year={2013}
}

@article{hahn2018regularization,
  title={Regularization and confounding in linear regression for treatment effect estimation},
  author={Hahn, P Richard and Carvalho, Carlos M and Puelz, David and He, Jingyu and others},
  journal={Bayesian Analysis},
  volume={13},
  number={1},
  pages={163--182},
  year={2018},
  publisher={International Society for Bayesian Analysis}
}

@article{wendling2018comparing,
  title={Comparing methods for estimation of heterogeneous treatment effects using observational data from health care databases},
  author={Wendling, T and Jung, K and Callahan, A and Schuler, A and Shah, NH and Gallego, B},
  year={2018},
  volume = {37},
  number = {23},
  pages = {3309--3324},
  journal={Statistics in Medicine},
  publisher={Wiley Online Library}
}

@article{su2012facilitating,
  title={Facilitating score and causal inference trees for large observational studies},
  author={Su, Xiaogang and Kang, Joseph and Fan, Juanjuan and Levine, Richard A and Yan, Xin},
  journal={Journal of Machine Learning Research},
  volume={13},
  number={Oct},
  pages={2955--2994},
  year={2012}
}

@unpublished{oecd2017,
  title={Educational Opportunity for All:
Overcoming Inequality throughout the Life Course},
  author={OECD},
  year={2017}
}

@article{calonico2015rdrobust,
  title={rdrobust: An r package for robust nonparametric inference in regression-discontinuity designs},
  author={Calonico, Sebastian and Cattaneo, Matias D and Titiunik, Rocio},
  journal={R Journal},
  volume={7},
  number={1},
  pages={38--51},
  year={2015},
  publisher={Citeseer}
}

@article{hahn2001identification,
  title={Identification and estimation of treatment effects with a regression-discontinuity design},
  author={Hahn, Jinyong and Todd, Petra and Van der Klaauw, Wilbert},
  journal={Econometrica},
  volume={69},
  number={1},
  pages={201--209},
  year={2001},
  publisher={Wiley Online Library}
}

@book{trochim1984research,
  title={Research design for program evaluation: The regression-discontinuity approach},
  author={Trochim, William MK},
  volume={6},
  year={1984},
  publisher={SAGE Publications, Inc}
}

@book{hanushek2016handbook,
  title={Handbook of the Economics of Education},
  author={Hanushek, Eric A and Machin, Stephen J and Woessmann, Ludger},
  year={2016},
  publisher={Elsevier}
}

@article{hanushek2003failure,
  title={The failure of input-based schooling policies},
  author={Hanushek, Eric A},
  journal={The Economic Journal},
  volume={113},
  number={485},
  pages={F64--F98},
  year={2003},
  publisher={Oxford University Press Oxford, UK}
}

@article{coleman1966equality,
  title={Equality of educational opportunity},
  author={Coleman, James S},
  journal={Washington DC: US Government Printing Office},
  pages={1--32},
  year={1966}
}

@article{jackson2015effects,
  title={The effects of school spending on educational and economic outcomes: Evidence from school finance reforms},
  author={Jackson, C Kirabo and Johnson, Rucker C and Persico, Claudia},
  journal={The Quarterly Journal of Economics},
  volume={131},
  number={1},
  pages={157--218},
  year={2015},
  publisher={Oxford University Press}
}

@techreport{jackson2018does,
  title={Does School Spending Matter? The New Literature on an Old Question},
  author={Jackson, C Kirabo},
  year={2018},
  institution={National Bureau of Economic Research}
}

@incollection{hanushek2017school,
  title={School resources and student achievement: A review of cross-country economic research},
  author={Hanushek, Eric A and Woessmann, Ludger},
  booktitle={Cognitive abilities and educational outcomes},
  pages={149--171},
  year={2017},
  publisher={Springer}
}

@article{van2007super,
  title={Super learner},
  author={Van der Laan, Mark J and Polley, Eric C and Hubbard, Alan E},
  journal={Statistical Applications in Genetics and Molecular Biology},
  volume={6},
  number={1},
  year={2007},
  publisher={De Gruyter}
}

@inproceedings{hartford2017deep,
  title={Deep IV: A flexible approach for counterfactual prediction},
  author={Hartford, Jason and Lewis, Greg and Leyton-Brown, Kevin and Taddy, Matt},
  booktitle={International Conference on Machine Learning},
  pages={1414--1423},
  year={2017},
  organization={PMLR}
}

@article{Dominici2021From,
journal = {Harvard Data Science Review},
doi = {10.1162/99608f92.8102afed},
publisher = {},
title = {From Controlled to Undisciplined Data: Estimating Causal Effects in the Era of Data Science Using a Potential Outcome Framework},
author = {Dominici, Francesca and Bargagli-Stoffi, Falco J. and Mealli, Fabrizia},
date = {2021-08-31},
year = {2021},
month = {8},
day = {31},
}

@article{lechner2019modified,
  title={Modified Causal Forests for Estimating Heterogeneous Causal Effects},
  author={Lechner, Michael},
  year={2019},
  journal={CEPR Discussion Paper No. DP13430}
}

@article{green2012modeling,
  title={Modeling heterogeneous treatment effects in survey experiments with Bayesian additive regression trees},
  author={Green, Donald P and Kern, Holger L},
  journal={Public Opinion Quarterly},
  volume={76},
  number={3},
  pages={491--511},
  year={2012},
  publisher={Oxford University Press UK}
}

@article{foster2011subgroup,
  title={Subgroup identification from randomized clinical trial data},
  author={Foster, Jared C and Taylor, Jeremy MG and Ruberg, Stephen J},
  journal={Statistics in Medicine},
  volume={30},
  number={24},
  pages={2867--2880},
  year={2011},
  publisher={Wiley Online Library}
}

@article{mattei2016regression,
  title={Regression discontinuity designs as local randomized experiments},
  author={Mattei, Alessandra and Mealli, Fabrizia},
  journal={Observational Studies},
  volume={66},
  pages={156--173},
  year={2016}
}

@article{li2015evaluating,
  title={Evaluating the causal effect of university grants on student dropout: evidence from a regression discontinuity design using principal stratification},
  author={Li, Fan and Mattei, Alessandra and Mealli, Fabrizia},
  journal={The Annals of Applied Statistics},
  pages={1906--1931},
  year={2015},
  publisher={JSTOR}
}

@article{starling2018functional,
  title={Functional response regression with funBART: an analysis of patient-specific stillbirth risk},
  author={Starling, Jennifer E and Murray, Jared S and Carvalho, Carlos M and Bukowski, Radek and Scott, James G},
  journal={arXiv preprint, arXiv:1805.07656},
  year={2018}
}

@article{wager2018estimation,
  title={Estimation and inference of heterogeneous treatment effects using random forests},
  author={Wager, Stefan and Athey, Susan},
  journal={Journal of the American Statistical Association},
  volume={113},
  number={523},
  pages={1228--1242},
  year={2018},
  publisher={Taylor \& Francis}
}

@article{bargagli2020causal,
  title={Causal tree with instrumental variable: an extension of the causal tree framework to irregular assignment mechanisms},
  author={Bargagli-Stoffi, Falco J and Gnecco, Giorgio},
  journal={International Journal of Data Science and Analytics},
  volume={9},
  number={3},
  pages={315--337},
  year={2020},
  publisher={Springer}
}

@book{wooldridge2015introductory,
  title={Introductory econometrics: A modern approach},
  author={Wooldridge, Jeffrey M},
  year={2015},
  publisher={Nelson Education}
}

@article{imbens1997estimating,
  title={Estimating outcome distributions for compliers in instrumental variables models},
  author={Imbens, Guido W and Rubin, Donald B},
  journal={The Review of Economic Studies},
  volume={64},
  number={4},
  pages={555--574},
  year={1997},
  publisher={Wiley-Blackwell}
}

@article{murray2017log,
  title={Log-linear Bayesian additive regression trees for categorical and count responses},
  author={Murray, Jared S},
  journal={arXiv preprint, arXiv:1701.01503},
  year={2017}
}

@article{linero2018bayesian1,
  title={Bayesian regression tree ensembles that adapt to smoothness and sparsity},
  author={Linero, Antonio R and Yang, Yun},
  journal={Journal of the Royal Statistical Society: Series B (Statistical Methodology)},
  volume={80},
  number={5},
  pages={1087--1110},
  year={2018},
  publisher={Wiley Online Library}
}

@article{hernandez2018bayesian,
  title={Bayesian additive regression trees using Bayesian model averaging},
  author={Hern{\'a}ndez, Belinda and Raftery, Adrian E and Pennington, Stephen R and Parnell, Andrew C},
  journal={Statistics and Computing},
  volume={28},
  number={4},
  pages={869--890},
  year={2018},
  publisher={Springer}
}

@article{linero2018bayesian2,
  title={Bayesian regression trees for high-dimensional prediction and variable selection},
  author={Linero, Antonio R},
  journal={Journal of the American Statistical Association},
  volume={113},
  number={522},
  pages={626--636},
  year={2018},
  publisher={Taylor \& Francis}
}

@article{logan2019decision,
  title={Decision making and uncertainty quantification for individualized treatments using Bayesian Additive Regression Trees},
  author={Logan, Brent R and Sparapani, Rodney and McCulloch, Robert E and Laud, Purushottam W},
  journal={Statistical Methods in Medical Research},
  volume={28},
  number={4},
  pages={1079--1093},
  year={2019},
  publisher={SAGE Publications Sage UK: London, England}
}

@article{nethery2019estimating,
  title={Estimating population average causal effects in the presence of non-overlap: The effect of natural gas compressor station exposure on cancer mortality},
  author={Nethery, Rachel C and Mealli, Fabrizia and Dominici, Francesca},
  journal={The Annals of Applied Statistics},
  volume={13},
  number={2},
  pages={1242},
  year={2019},
  publisher={NIH Public Access}
}

@article{ding2019decomposing,
  title={Decomposing treatment effect variation},
  author={Ding, Peng and Feller, Avi and Miratrix, Luke},
  journal={Journal of the American Statistical Association},
  volume={114},
  number={525},
  pages={304--317},
  year={2019},
  publisher={Taylor \& Francis}
}

@article{harris2011teacher,
  title={Teacher training, teacher quality and student achievement},
  author={Harris, Douglas N and Sass, Tim R},
  journal={Journal of Public Economics},
  volume={95},
  number={7-8},
  pages={798--812},
  year={2011},
  publisher={Elsevier}
}

@article{rice2010impact,
  title={The Impact of Teacher Experience: Examining the Evidence and Policy Implications},
  author={Rice, Jennifer King},
  journal={National Center for Analysis of Longitudinal Data in Education Research},
  year={2010},
  publisher={ERIC}
}

@article{eberts1988student,
  title={Student achievement in public schools: Do principals make a difference?},
  author={Eberts, Randall W and Stone, Joe A},
  journal={Economics of Education Review},
  volume={7},
  number={3},
  pages={291--299},
  year={1988},
  publisher={Elsevier}
}

@article{gentilucci2007principals,
  title={Principals' influence on academic achievement: The student perspective},
  author={Gentilucci, James L and Muto, Cindy C},
  journal={NASSP bulletin},
  volume={91},
  number={3},
  pages={219--236},
  year={2007},
  publisher={Sage Publications Sage CA: Los Angeles, CA}
}

@article{johnson2019detecting,
  title={Detecting Heterogeneous Treatment Effect with Instrumental Variables},
  author={Johnson, Michael and Cao, Jiongyi and Kang, Hyunseung},
  journal={arXiv preprint arXiv:1908.03652},
  year={2019}
}

@article{thistlethwaite1960regression,
  title={Regression-discontinuity analysis: An alternative to the ex post facto experiment.},
  author={Thistlethwaite, Donald L and Campbell, Donald T},
  journal={Journal of Educational Psychology},
  volume={51},
  number={6},
  pages={309},
  year={1960},
  publisher={American Psychological Association}
}

@article{cook2008waiting,
  title={“Waiting for life to arrive”: a history of the regression-discontinuity design in psychology, statistics and economics},
  author={Cook, Thomas D},
  journal={Journal of Econometrics},
  volume={142},
  number={2},
  pages={636--654},
  year={2008},
  publisher={Elsevier}
}

@article{lee2020causal,
  title={Causal rule ensemble: Interpretable inference of heterogeneous treatment effects},
  author={Lee, Kwonsang and Bargagli-Stoffi, Falco J and Dominici, Francesca},
  journal={arXiv preprint arXiv:2009.09036},
  year={2020}
}

@article{dominici2020controlled,
  title={From controlled to undisciplined data: estimating causal effects in the era of data science using a potential outcome framework},
  author={Dominici, Francesca and Bargagli-Stoffi, Falco Joannes and Mealli, Fabrizia},
  journal={Working Paper},
  year={2020}
}

@article{knaus2020double,
  title={Double machine learning based program evaluation under unconfoundedness},
  author={Knaus, Michael C},
  journal={arXiv preprint arXiv:2003.03191},
  year={2020}
}

@inproceedings{zhang2017sensitivity,
  title={A Sensitivity Analysis of (and Practitioners’ Guide to) Convolutional Neural Networks for Sentence Classification},
  author={Zhang, Ye and Wallace, Byron C},
  booktitle={Proceedings of the Eighth International Joint Conference on Natural Language Processing (Volume 1: Long Papers)},
  pages={253--263},
  year={2017}
}

@article{novak2018sensitivity,
  title={Sensitivity and generalization in neural networks: an empirical study},
  author={Novak, Roman and Bahri, Yasaman and Abolafia, Daniel A and Pennington, Jeffrey and Sohl-Dickstein, Jascha},
  journal={arXiv preprint arXiv:1802.08760},
  year={2018}
}

@article{prosperi2020causal,
  title={Causal inference and counterfactual prediction in machine learning for actionable healthcare},
  author={Prosperi, Mattia and Guo, Yi and Sperrin, Matt and Koopman, James S and Min, Jae S and He, Xing and Rich, Shannan and Wang, Mo and Buchan, Iain E and Bian, Jiang},
  journal={Nature Machine Intelligence},
  volume={2},
  number={7},
  pages={369--375},
  year={2020},
  publisher={Nature Publishing Group}
}

@article{van2006targeted,
  title={Targeted maximum likelihood learning},
  author={Van Der Laan, Mark J and Rubin, Daniel},
  journal={The International Journal of Biostatistics},
  volume={2},
  number={1},
  year={2006},
  publisher={De Gruyter}
}

@article{imai2013estimating,
  title={Estimating treatment effect heterogeneity in randomized program evaluation},
  author={Imai, Kosuke and Ratkovic, Marc and others},
  journal={The Annals of Applied Statistics},
  volume={7},
  number={1},
  pages={443--470},
  year={2013},
  publisher={Institute of Mathematical Statistics}
}

@article{kunzel2019metalearners,
  title={Metalearners for estimating heterogeneous treatment effects using machine learning},
  author={K{\"u}nzel, S{\"o}ren R and Sekhon, Jasjeet S and Bickel, Peter J and Yu, Bin},
  journal={Proceedings of the National Academy of Sciences},
  volume={116},
  number={10},
  pages={4156--4165},
  year={2019},
  publisher={National Academy of Sciences}
}

@article{wang2017bayesian,
  title={A bayesian framework for learning rule sets for interpretable classification},
  author={Wang, Tong and Rudin, Cynthia and Doshi-Velez, Finale and Liu, Yimin and Klampfl, Erica and MacNeille, Perry},
  journal={The Journal of Machine Learning Research},
  volume={18},
  number={1},
  pages={2357--2393},
  year={2017},
  publisher={JMLR. org}
}

@article{semenova2020debiased,
  title={Debiased machine learning of conditional average treatment effects and other causal functions},
  author={Semenova, Vira and Chernozhukov, Victor},
  journal={The Econometrics Journal},
  year={2020},
  number={forthcoming}
}

@article{fan2020estimation,
  title={Estimation of conditional average treatment effects with high-dimensional data},
  author={Fan, Qingliang and Hsu, Yu-Chin and Lieli, Robert P and Zhang, Yichong},
  journal={Journal of Business \& Economic Statistics},
  pages={1--15},
  year={2020},
  publisher={Taylor \& Francis}
}

@article{starling2019targeted,
  title={Targeted smooth bayesian causal forests: An analysis of heterogeneous treatment effects for simultaneous versus interval medical abortion regimens over gestation},
  author={Starling, Jennifer E and Murray, Jared S and Lohr, Patricia A and Aiken, Abigail RA and Carvalho, Carlos M and Scott, James G},
  journal={arXiv preprint arXiv:1905.09405},
  year={2019}
}

@article{zimmert2019nonparametric,
  title={Nonparametric estimation of causal heterogeneity under high-dimensional confounding},
  author={Zimmert, Michael and Lechner, Michael},
  journal={arXiv preprint arXiv:1908.08779},
  year={2019}
}

@article{kennedy2020optimal,
  title={Optimal doubly robust estimation of heterogeneous causal effects},
  author={Kennedy, Edward H},
  journal={arXiv preprint arXiv:2004.14497},
  year={2020}
}

@article{athey2019machine,
  title={Machine learning methods that economists should know about},
  author={Athey, Susan and Imbens, Guido W},
  journal={Annual Review of Economics},
  volume={11},
  pages={685--725},
  year={2019},
  publisher={Annual Reviews}
}

@article{angrist2001instrumental,
  title={Instrumental variables and the search for identification: From supply and demand to natural experiments},
  author={Angrist, Joshua D and Krueger, Alan B},
  journal={Journal of Economic perspectives},
  volume={15},
  number={4},
  pages={69--85},
  year={2001}
}

@article{rubin1986comment,
  title={Comment: Which ifs have causal answers},
  author={Rubin, Donald B},
  journal={Journal of the American Statistical Association},
  volume={81},
  number={396},
  pages={961--962},
  year={1986},
  publisher={Taylor \& Francis Group}
}

@article{woody2020estimating,
  title={Estimating heterogeneous effects of continuous exposures using Bayesian tree ensembles: revisiting the impact of abortion rates on crime},
  author={Woody, Spencer and Carvalho, Carlos M and Hahn, P Richard and Murray, Jared S},
  journal={arXiv preprint arXiv:2007.09845},
  year={2020}
}

@Article{friedman2001greedy,
  author    = {Friedman, Jerome H},
  title     = {Greedy function approximation: a gradient boosting machine},
  journal   = {The Annals of Statistics},
  year      = {2001},
  volume    = {29},
  number    = {9},
  pages     = {1189--1232},
  publisher = {JSTOR},
}

@Article{cook2004subgroup,
  author    = {Cook, David I and Gebski, Val J. and Keech, Anthony C.},
  title     = {Subgroup analysis in clinical trials},
  journal   = {Medical Journal of Australia},
  year      = {2004},
  volume    = {180},
  number    = {6},
  pages     = {289--291},
  publisher = {Australasian Medical Publishing Company Proprietary, Ltd.},
}

@article{stone1974cross,
  title={Cross-validatory choice and assessment of statistical predictions},
  author={Stone, Mervyn},
  journal={Journal of the Royal Statistical Society: Series B (Methodological)},
  volume={36},
  number={2},
  pages={111--133},
  year={1974},
  publisher={Wiley Online Library}
}

@article{cox1975note,
  title={A note on data-splitting for the evaluation of significance levels},
  author={Cox, David R},
  journal={Biometrika},
  volume={62},
  number={2},
  pages={441--444},
  year={1975},
  publisher={Oxford University Press}
}

@article{hsu2015strong,
  title={Strong control of the familywise error rate in observational studies that discover effect modification by exploratory methods},
  author={Hsu, Jesse Y and Zubizarreta, Jos{\'e} R and Small, Dylan S and Rosenbaum, Paul R},
  journal={Biometrika},
  volume={102},
  number={4},
  pages={767--782},
  year={2015},
  publisher={Oxford University Press}
}

@book{zhang2010recursive,
  title={Recursive partitioning and applications},
  author={Zhang, Heping and Singer, Burton H},
  year={2010},
  publisher={Springer Science \& Business Media}
}

@article{strobl2009introduction,
  title={An introduction to recursive partitioning: rationale, application, and characteristics of classification and regression trees, bagging, and random forests.},
  author={Strobl, Carolin and Malley, James and Tutz, Gerhard},
  journal={Psychological Methods},
  volume={14},
  number={4},
  pages={323},
  year={2009},
  publisher={American Psychological Association}
}

@InProceedings{kim2016examples,
  author    = {Kim, Been and Khanna, Rajiv and Koyejo, Oluwasanmi O.},
  title     = {Examples are not enough, learn to criticize! criticism for interpretability},
  booktitle = {Advances in Neural Information Processing Systems},
  year      = {2016},
  pages     = {2280--2288},
}

@Article{miller2018explanation,
  author    = {Miller, Tim},
  title     = {Explanation in artificial intelligence: Insights from the social sciences},
  journal   = {Artificial Intelligence},
  year      = {2019},
  volume    = {267},
  pages     = {1--38},
  publisher = {Elsevier},
}

@article{therneau2015package,
  title={Package ‘rpart’ package version 4.1-15},
  author={Therneau, Terry and Atkinson, Beth and Ripley, Brian and Ripley, Maintainer Brian},
  journal={Available online: https://cran.r-project.org/web/packages/rpart/rpart.pdf},
  year={2015}
}

@article{dorie2020package,
  title={Package ‘bartCause’},
  author={Dorie, Vincent and Hill, Jennifer and Dorie, Maintainer Vincent},
  year={2020}
}

@article{dorie2019automated,
  title={Automated versus do-it-yourself methods for causal inference: Lessons learned from a data analysis competition},
  author={Dorie, Vincent and Hill, Jennifer and Shalit, Uri and Scott, Marc and Cervone, Dan},
  journal={Statistical Science},
  volume={34},
  number={1},
  pages={43--68},
  year={2019},
  publisher={Institute of Mathematical Statistics}
}

@article{calonico2014robust,
  title={Robust nonparametric confidence intervals for regression-discontinuity designs},
  author={Calonico, Sebastian and Cattaneo, Matias D and Titiunik, Rocio},
  journal={Econometrica},
  volume={82},
  number={6},
  pages={2295--2326},
  year={2014},
  publisher={Wiley Online Library}
}

@article{imbens2008regression,
  title={Regression discontinuity designs: A guide to practice},
  author={Imbens, Guido W and Lemieux, Thomas},
  journal={Journal of Econometrics},
  volume={142},
  number={2},
  pages={615--635},
  year={2008},
  publisher={Elsevier}
}

@article{nelson1990distribution,
  title={The distribution of the instrumental variables estimator and its t-ratio when the instrument is a poor one},
  author={Nelson, Charles R and Startz, Richard},
  journal={Journal of Business},
  pages={S125--S140},
  year={1990},
  publisher={JSTOR}
}

@article{nelson1990some,
  title={Some further results on the exact small sample properties of the instrumental variable estimator},
  author={Nelson, Charles R and Stratz, Richard},
  journal={Econometrica},
  volume={58},
  number={4},
  pages={967--976},
  year={1990},
  publisher={Wiley-Blackwell}
}

@article{balke1997bounds,
  title={Bounds on treatment effects from studies with imperfect compliance},
  author={Balke, Alexander and Pearl, Judea},
  journal={Journal of the American Statistical Association},
  volume={92},
  number={439},
  pages={1171--1176},
  year={1997},
  publisher={Taylor \& Francis}
}

@article{andrews2019weak,
  title={Weak instruments in instrumental variables regression: Theory and practice},
  author={Andrews, Isaiah and Stock, James H and Sun, Liyang},
  journal={Annual Review of Economics},
  volume={11},
  pages={727--753},
  year={2019},
  publisher={Annual Reviews}
}

@article{holm1979simple,
  title={A simple sequentially rejective multiple test procedure},
  author={Holm, Sture},
  journal={Scandinavian journal of statistics},
  pages={65--70},
  year={1979},
  publisher={JSTOR}
}

@article{hochberg1988sharper,
  title={A sharper Bonferroni procedure for multiple tests of significance},
  author={Hochberg, Yosef},
  journal={Biometrika},
  volume={75},
  number={4},
  pages={800--802},
  year={1988},
  publisher={Oxford University Press}
}

@article{hommel1988stagewise,
  title={A stagewise rejective multiple test procedure based on a modified Bonferroni test},
  author={Hommel, Gerhard},
  journal={Biometrika},
  volume={75},
  number={2},
  pages={383--386},
  year={1988},
  publisher={Oxford University Press}
}

@article{benjamini1995controlling,
  title={Controlling the false discovery rate: a practical and powerful approach to multiple testing},
  author={Benjamini, Yoav and Hochberg, Yosef},
  journal={Journal of the Royal statistical society: series B (Methodological)},
  volume={57},
  number={1},
  pages={289--300},
  year={1995},
  publisher={Wiley Online Library}
}

@article{benjamini2001control,
  title={The control of the false discovery rate in multiple testing under dependency},
  author={Benjamini, Yoav and Yekutieli, Daniel},
  journal={Annals of statistics},
  pages={1165--1188},
  year={2001},
  publisher={JSTOR}
}

@article{sarkar1997simes,
  title={The Simes method for multiple hypothesis testing with positively dependent test statistics},
  author={Sarkar, Sanat K and Chang, Chung-Kuei},
  journal={Journal of the American Statistical Association},
  volume={92},
  number={440},
  pages={1601--1608},
  year={1997},
  publisher={Taylor \& Francis}
}

@article{lee2021discovering,
  title={Discovering Heterogeneous Exposure Effects Using Randomization Inference in Air Pollution Studies},
  author={Lee, Kwonsang and Small, Dylan S and Dominici, Francesca},
  journal={Journal of the American Statistical Association},
  pages={1--12},
  year={2021},
  publisher={Taylor \& Francis}
}

@article{marcus1976closed,
  title={On closed testing procedures with special reference to ordered analysis of variance},
  author={Marcus, Ruth and Eric, Peritz and Gabriel, K Ruben},
  journal={Biometrika},
  volume={63},
  number={3},
  pages={655--660},
  year={1976},
  publisher={Oxford University Press}
}

@article{yekutieli2008hierarchical,
  title={Hierarchical false discovery rate--controlling methodology},
  author={Yekutieli, Daniel},
  journal={Journal of the American Statistical Association},
  volume={103},
  number={481},
  pages={309--316},
  year={2008},
  publisher={Taylor \& Francis}
}

\pagebreak

\newpage
\appendix
\counterwithin{figure}{section}

\captionsetup{labelformat=AppendixTables}
\setcounter{table}{0}

\numberwithin{equation}{section}
\makeatletter 
\newcommand{\section@cntformat}{Appendix \thesection:\ }

\begin{center}
 {\bf \huge Supplementary Material}
\end{center}
 \pagenumbering{arabic}
    \setcounter{page}{1}

\section{Discussion on the Instrumental Variable Approach in the Empirical Application} \label{appendix-sec1}

\subsection{Assumptions}

In a typical IV scenario one can express the treatment received as a function of the treatment assigned: $W_i(Z_i)$. This leads to distinguish four sub-populations of units ($G_i$) \citep{angrist1996identification, imbens2015causal}: (i) those that comply with the assignment (\textit{compliers}: $G_i=C: W_i(Z_i=0)=0$ and  $W_i(Z_i = 1)=1$); (ii) those that never comply with the assignment (\textit{defiers}: $G_i=D: W_i(Z_i=0)=1$ and $W_i(Z_i = 1)=0$); (iii) those that even if not assigned to the treatment always take it (\textit{always-takers}: $G_i=AT: W_i(Z_i=0)=1, W_i(Z_i=1)=1$); (iv) those that even if assigned to the treatment never take it (\textit{never-takers}: $G_i=NT: W_i(Z_i=0)=0, W_i(Z_i=1)=0$). 
In such a scenario what ``one directly gets from the data" is the so-called Intention-To-Treat ($ITT_Y$):
    \begin{equation}\label{super_IV}
    	ITT_Y=\mathbb{E}[Y_i\mid Z_i=1] - \mathbb{E}[Y_i\mid Z_i=0],
    \end{equation}
which is defined as the effect of the intention to treat a unit on the outcome of the same unit. (\ref{super_IV}) can be written as the weighted average of the intention-to-treat effects across the four sub-populations of compliers, defiers, always-takers and never-takers:
    \begin{equation}
     ITT_Y = \pi_C ITT_{Y,C} + \pi_D ITT_{Y,D} + \pi_{NT} ITT_{Y,NT} + \pi_{AT} ITT_{Y,AT},
    \end{equation}
where $ITT_{Y,G}$ is the effect of the treatment assignment on units of type $G$ and $\pi_G$ is the proportion of units of type $G$.
\par $ITT_Y$ does not represent the effect of the treatment itself but just the effect of the assignment to the treatment. If we want to draw proper causal inference in such a scenario we need to invoke the four classical IV assumptions \citep{angrist1996identification}:
\begin{enumerate} 
        \item \textit{exclusion restriction}: 
    	$Y_i(0) = Y_i(1),  \ \ {\rm for } \ G_i \in \{AT,NT\}$ where, for each sub-population and $z \in \{0,1\}$, the shortened notation $Y_i(z)$ is used to denote $Y_i(z, W_i(z))$
    	\item \textit{monotonicity}: $W_i (1) \geq W_i (0) \rightarrow \pi_D=0$;
    	\item \textit{existence of compliers}: $	P(W_{i}(0)<W_{i}(1))>0 \rightarrow \pi_C\neq 0$;
    	\item \textit{unconfoundedness of the instrument}:\\ 
    	$Z_i \independent (Y_i(0,0), Y_i(0,1), Y_i(1,0),Y_i(1,1 ),W_i(0), W_i(1))$.
    \end{enumerate}
    
In our application, these four assumptions are assumed to hold. Let us look at them in detail. The exclusion restriction is assumed to hold since we can reasonably rule out a direct effect of being eligible (around the threshold) on the performance of students. The effect, in this case, can be reasonably assumed to go through the actual reception of additional funding. Monotonicity holds by design: since we are in a one-sided non-compliance scenario there is no possibility for those who are not assigned to the treatment to defy and get the treatment. The same can be said about the existence of compliers. Since the sub-populations of always-takers and defiers can be ruled out by design, this leads to the fact that units receiving the treatment are compliers. Unconfoundedness of the instrument can also reasonably be assumed to hold since observations around the exogenous threshold are as good as if they were randomized to the assigned-to-the-treatment group and the assigned-to-the control group. This holds true especially since we do not observe any manipulation around the threshold and sorting of the units into the treated group. 

{\color{ao(english)}\subsection{Method-of-Moments Estimator}\label{subsec:MoM}

The conditional CACE can be estimated in a generic sub-sample (i.e., for each $\bX_i \in {\mathbb{X}}_j$, where ${\mathbb{X}}_j$ is a generic node of the discovered tree, like a non-terminal node or a leaf) as:
\begin{equation}\label{eq:conditionalCACEestimate}
    \hat{\tau}^{cace}(\bX_i) = \frac{\reallywidehat{ITT}_{Y}(\bX_i)}{\hat{\pi}_C(\bX_i)},
\end{equation}
where $\hat{\pi}_{C}(\bX_i)$ is estimated as:
\begin{equation}
\hat{\pi}_{C}(\bX_i)=\frac{1}{N_{1,j}}\sum_{l: X_l \in \mathbb{X}_j}  W_l   Z_l -\frac{1}{N_{0,j}}\sum_{l: X_l \in \mathbb{X}_j} W_l  (1-Z_l),
\end{equation} 
and $\reallywidehat{ITT}_{Y}(\bX_i)$ as:
    \begin{equation} \label{IV_propensity}
    \reallywidehat{ITT}_Y(\bX_i)=\frac{1}{N_{1,j}}\sum_{l: X_l \in \mathbb{X}_j}  Y_l^{obs}   {Z_l } - \frac{1}{N_{0,j}}\sum_{l: X_l \in \mathbb{X}_j}  Y_l^{obs}   (1-Z_l),
 \end{equation}
where $N_{k,j}$ (where $k \in \{0,1\})$ is the number of observations with $Z_l=k$ in the sub-sample of observations with  $X_l \in \mathbb{X}_j$: where $N_{1,j}=\sum_{l: X_l \in \mathbb{X}_j} Z_l$ and $N_{0,j}=\sum_{l: X_l \in \mathbb{X}_j}(1- Z_l)$. It is worth highlighting that, since the supervised machine learning technique is used in the discovery phase and not in the estimation phase, the estimators that are proposed here could be used in a more ``traditional way", in settings where the subgroups are defined ex-ante by the researcher.

\subsection{Theoretical Properties}

The 2SLS estimator associated with (\ref{IV_{X_j}_reg})-(\ref{IV_reg_red}) satisfies the next properties. They can be proved likewise in the application of 2SLS to the population case \citet{imbens2015causal}.

\noindent  \textbf{Theorem 1: Consistency of the Conditional 2SLS Estimator.} 
\noindent {\em Let $\mathbb{E}(Z^2_{i,\mathbb{X}_j}) \neq 0$ (Assumption 1), $\mathbb{E}(Z_{i,\mathbb{X}_j} \epsilon_{i,\mathbb{X}_j}) = 0$ (Assumption 2) and $\pi_{C,\mathbb{X}_j} \neq 0$ (Assumption 3) hold. Then
\begin{equation} 
  \hat{\tau}^{2SLS}_{\mathbb{X}_j} - {\tau}_{\mathbb{X}_j}  \:\:\overset{p}{\to}\:\: 0 \:\:\: \text{as} \:\:\: N_{\mathbb{X}_j} \rightarrow \infty,
\end{equation}
where $\overset{p}{\to}$ denotes convergence in probability, and $N_{\mathbb{X}_j}$ is the number of observations within the node $\mathbb{X}_j$.}

It should be noted that Assumption 3 is not necessarily guaranteed even if the overall instrument is strong.  However, this assumption is standard in treatment effects variation papers such as the contribution of \citet{ding2019decomposing}. \vspace{0.25cm} 

\noindent \textbf{Theorem 2: Asymptotic Normality of the Conditional 2SLS Estimator.} 

\noindent {\em Let Assumptions 1, 2, and 3 hold. Let also $\mathbb{E}(Z^2_{i,\mathbb{X}_j} \epsilon^2_{i,\mathbb{X}_j})$ be finite (Assumption 4). Then
\begin{equation}
  \sqrt{N_{\mathbb{X}_j}}\bigl(\hat{\tau}^{2SLS}_{\mathbb{X}_j} - {\tau}_{\mathbb{X}_j}\bigl) \:\:\overset{d}{\to}\:\: \mathcal{N}\bigl(0, N_{\mathbb{X}_j}  avar(\hat{\tau}^{2SLS}_{\mathbb{X}_j})\bigl) \:\:\: \text{as} \:\:\: N_{\mathbb{X}_j} \rightarrow \infty,
\end{equation}
where $\overset{d}{\to}$ denotes convergence in distribution, $\mathcal{N}$ stands for normal distribution, and $avar(\hat{\tau}^{2SLS}_{\mathbb{X}_j})$ is the asymptotic variance of the 2SLS estimator that can be approximated as in Chapter 15 of \cite{wooldridge2015introductory}.}

The proofs of the two Theorems above directly follow from their unconditional versions. For further details on these proofs we refer to \cite[Section 5.2]{wooldridge2002}. In this case, for the convergence of our estimator to ${\tau}_{\mathbb{X}_j}$ and its normality to hold approximately we need to have a sufficient number of observations within every node. Hence, we suggest to perform our algorithm on sufficiently large datasets and to trim those nodes where the number of observations is not large enough. Moreover, as argued by \cite{lee2021discovering}, smaller trees guarantee higher levels of interpretability of each discovered subgroup and higher statistical power.}

\section{Additional Monte Carlo Simulations}
\label{appendix-sec3}

{\color{ao(english)}

\subsection{Goodness-of-fit Measures for Simulations}\label{appendix:goodness_of_fit}

The goodness-of-fit measures adopted for the simulations are reported as follows:
\begin{enumerate}
\item average number of truly discovered heterogeneous subgroups corresponding to the leaves of the generated CART;
\item   Monte Carlo estimated bias for the heterogeneous subgroups:
\begin{eqnarray*}
\text{Bias}_m(\mathcal{I}^{inf})&=& \frac{1}{N^{inf}}\sum_{i=1}^{N^{inf}}\sum_{l=1}^{L}\Big(\tau^{cace}_{i}(\ell_l)-\widehat{\tau}^{cace}_{i}(\ell_l, \Pi_m, \mathcal{I}^{inf})\Big),\\
\text{Bias}(\mathcal{I}^{inf})&=&\frac{1}{M}\sum_{m=1}^M \text{Bias}_m(\mathcal{I}^{inf})
\end{eqnarray*}
where $\Pi_m$ is the partition selected in simulation $m$, $L$ is the number of subgroups with heterogeneous effects (i.e., two for the case of \textit{strong heterogeneity} and four for the case of \textit{slight heterogeneity}), and $N^{inf}$ is the number of observations in the inference sample;
\item  Monte Carlo estimated MSE for the heterogeneous subgroups:
\begin{eqnarray*}
\text{MSE}_m(\mathcal{I}^{inf}) &=& \frac{1}{N^{inf}}\sum_{i=1}^{N^{inf}}\sum_{l=1}^{L}\Big(\tau^{cace}_{i}(\ell_l)-\widehat{\tau}^{cace}_{i}(\ell_l, \Pi_m, \mathcal{I}^{inf})\Big)^2,\\
\text{MSE}(\mathcal{I}^{inf})&=&\frac{1}{M}\sum_{m=1}^M \text{MSE}_m(\mathcal{I}^{inf});
\end{eqnarray*}
\item Monte Carlo coverage, computed as the average proportion of units for which the estimated $95\%$ confidence interval of the causal effect in the assigned leaf  includes the true value, for the heterogeneous subgroups:
\begin{eqnarray*}
\text{C}_m(\mathcal{I}^{inf})&=&\frac{1}{N^{inf}}\sum_{i=1}^{N^{inf}}\sum_{l=1}^{L}
\Big(\tau^{cace}_{i}(\ell_l) \in \widehat{\text{CI}}_{95}\Big(\widehat{\tau}^{cace}_{i}(\ell_l, \Pi_m, \mathcal{I}^{inf})\Big) \Big),\\
\text{C}(\mathcal{I}^{inf})&=&\frac{1}{M}\sum_{m=1}^M \text{C}_m(\mathcal{I}^{inf}).
\end{eqnarray*}
\end{enumerate}

\subsection{Mild and Weak Instruments with Varying Effect Size for the Heterogeneous Causal Effects}\label{subsec:sim2}

In real world applications, it can often be the case that the proportion of units complying with the treatment assignment is not high. This can lead to the well-known issue of weak instrument \citep{nelson1990distribution, nelson1990some, andrews2019weak}. Hence, it is critical to assess the performance of the proposed algorithms in scenarios in which the proportion of compliers decreases from 75\% to 50\% (what we call \textit{mild instrument} scenario) and then to 25\% (what we call \textit{weak instrument} scenario). In both cases, we keep a fixed sample size (1,000 units), in a setting with \textit{strong heterogeneity}, while we let the effect sizes to vary between 0 and 2.

Figure \ref{fig:rules_weak} shows the results for the TPR in the case of a \textit{mild instrument} and a \textit{weak instrument}. Again, both BCF-IV and BCF-ITT perform very similarly and they both outperform HCT-IV. In the case of a mild instrument both BCF-IV and BCF-ITT are able to correctly discover all the heterogeneous subgroups, while in the case of a weak instrument the discovery rate decreases, hinting at the fact that we might need larger sample and/or effect sizes to correctly discover all the subgroups. {\color{blue} The FPR is approaching zero in both scenarios. }

With respect to the estimations precision, the results for \textit{mild instrument} and a \textit{weak instrument} scenarios are depicted in Tables \ref{tab:sims_0.50_1000} and \ref{tab:sims_0.25_1000}, respectively. As expected, in both scenarios there is an increase in the Monte Carlo estimated MSE and bias with respect to the scenario with a strong instrument in Table \ref{tab:sims_0.75_1000} for BCF-IV and GRF. However, BCF-IV is still outperforming GRF and its Monte Carlo coverages are consistent with the 95\% coverage. Moreover, the decrease in the estimation performance of BCF-IV is less steep than the one of GRF as we move from a mild to a weak instrument. 

\begin{figure}[H]
    \centering
    \includegraphics[width=1\linewidth]{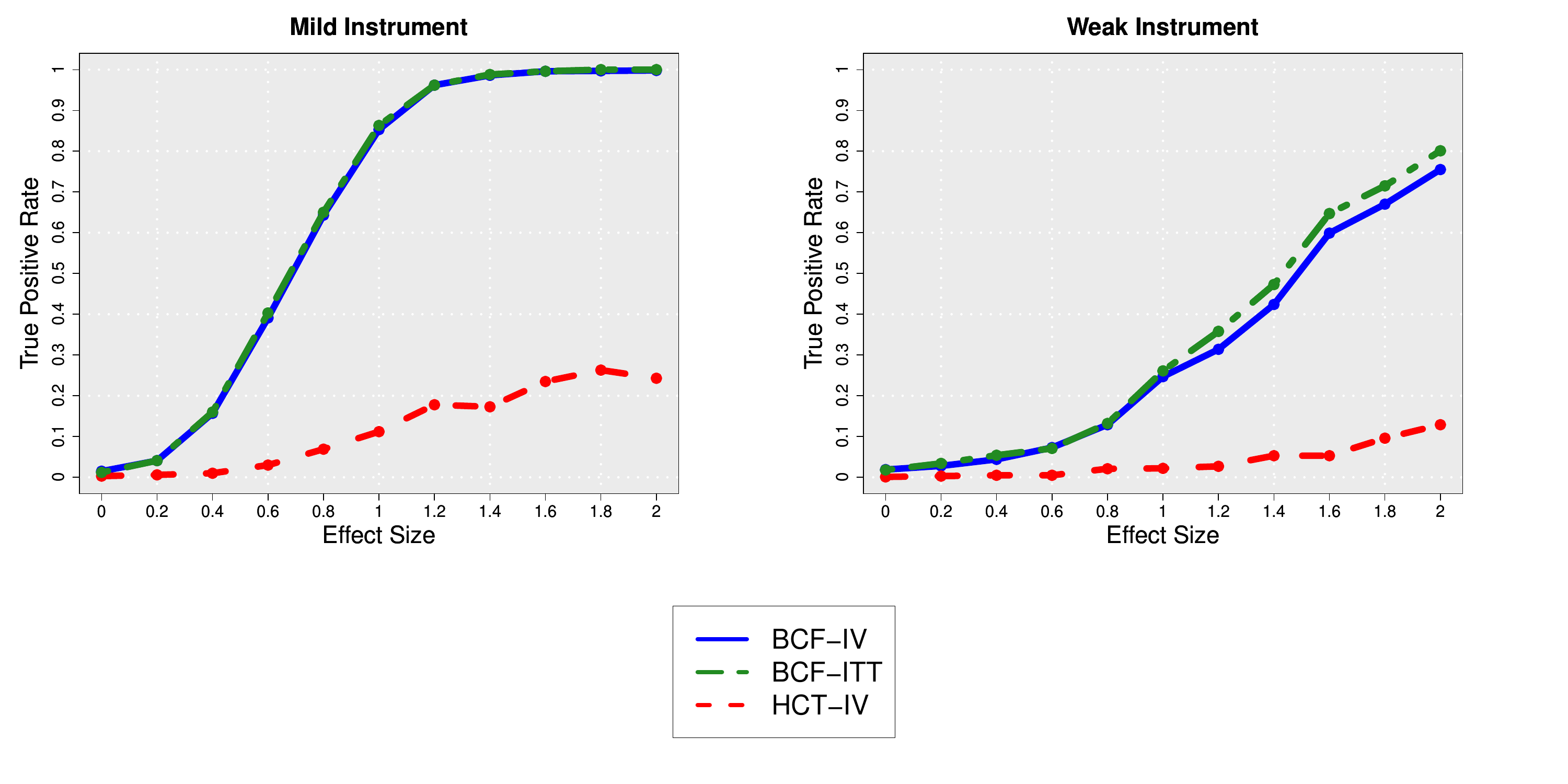}
    \caption{TPR in a \textit{mild instrument} scenario on the left, and a \textit{weak instrument} scenario on the right.}
    \label{fig:rules_weak}
\end{figure}

\begin{figure}[H]
    \centering
    \includegraphics[width=1\linewidth]{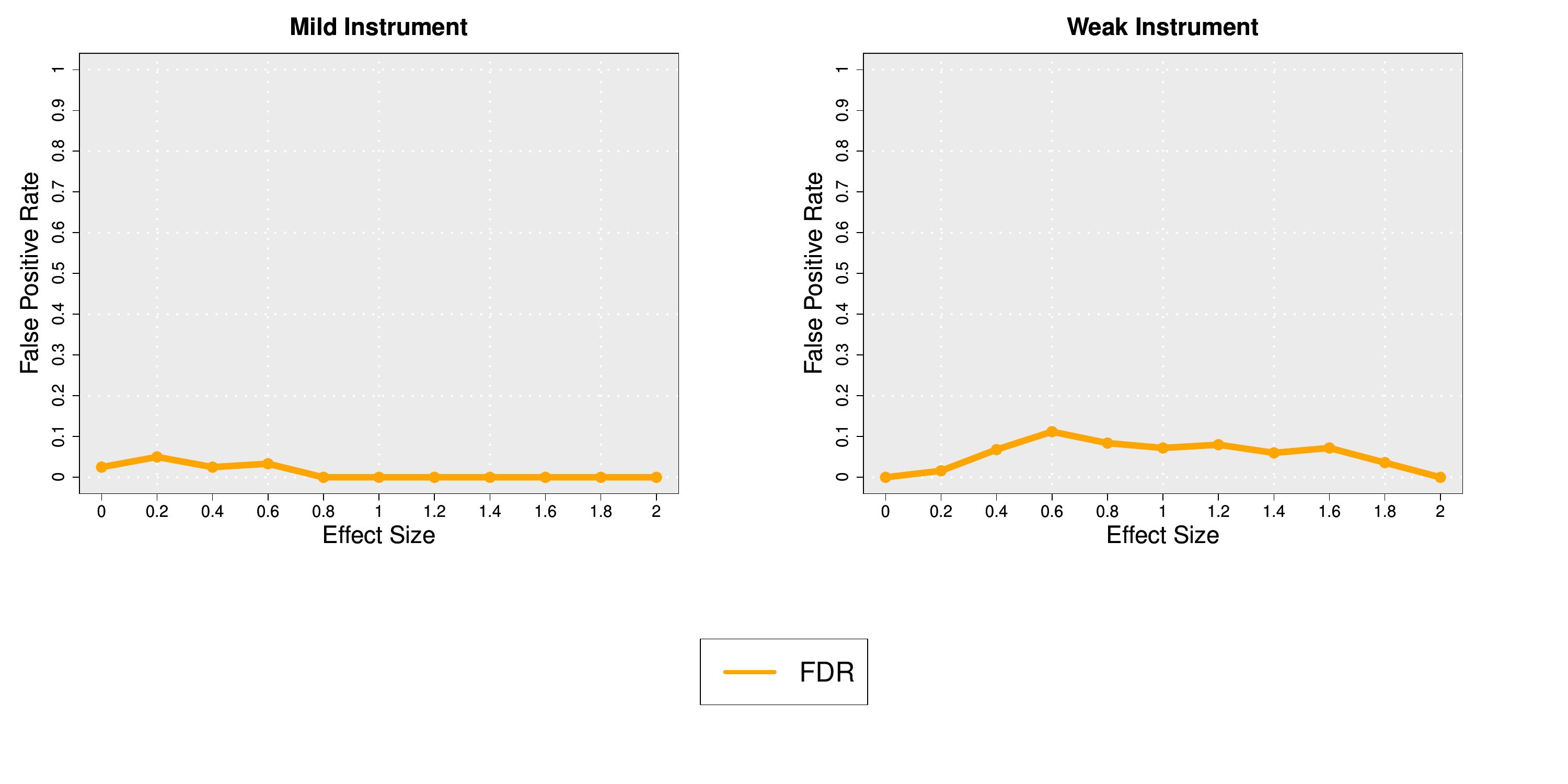}
    \caption{FPR in a \textit{mild instrument} scenario on the left, and a \textit{weak instrument} scenario on the right.}
    \label{fig:FPR_weak}
\end{figure}

\begin{table}[H]
\centering \scriptsize
\begin{tabular}{ccccccc}
\textit{Effect Size} & MSE($\hat{\tau}^{cace}_{l_1}$) & Bias($\hat{\tau}^{cace}_{l_1}$) & Coverage($\hat{\tau}^{cace}_{l_1}$) & \multicolumn{1}{l}{MSE($\hat{\tau}^{cace}_{l_2}$)} & \multicolumn{1}{l}{Bias($\hat{\tau}^{cace}_{l_2}$)} & \multicolumn{1}{l}{Coverage($\hat{\tau}^{cace}_{l_2}$)} \\ \toprule
\multicolumn{1}{l}{} & \multicolumn{6}{c}{BCF-IV}\\ \toprule
0                    & 0.071                   & 0.210                    & 0.948                        & 0.071                                       & 0.211                                        & 0.956                                            \\
0.1                  & 0.046                   & 0.071                    & 0.966                        & 0.038                                       & 0.051                                        & 0.984                                            \\
0.2                  & 0.057                   & 0.026                    & 0.964                        & 0.054                                       & 0.006                                        & 0.976                                            \\
0.3                  & 0.066                   & 0.001                    & 0.960                        & 0.060                                       & 0.026                                        & 0.968                                            \\
0.4                  & 0.063                   & 0.006                    & 0.964                        & 0.060                                       & 0.009                                        & 0.970                                            \\
0.5                  & 0.067                   & -0.011                   & 0.950                        & 0.068                                       & -0.002                                       & 0.954                                            \\
0.6                  & 0.068                   & -0.038                   & 0.960                        & 0.065                                       & -0.016                                       & 0.950                                            \\
0.7                  & 0.067                   & -0.006                   & 0.956                        & 0.062                                       & 0.005                                        & 0.954                                            \\
0.8                  & 0.068                   & 0.011                    & 0.960                        & 0.077                                       & 0.002                                        & 0.940                                            \\
0.9                  & 0.067                   & -0.004                   & 0.958                        & 0.061                                       & -0.006                                       & 0.972                                            \\
1                    & 0.070                   & -0.008                   & 0.942                        & 0.071                                       & -0.005                                       & 0.956                                            \\ \midrule
\multicolumn{1}{l}{} & \multicolumn{6}{c}{GRF}                                                                                                                                                                                                           \\ \toprule 
0                    & 0.047                   & 0.171                    & 0.996                        & 0.045                                       & 0.168                                        & 0.998                                            \\
0.1                  & 0.019                   & -0.020                   & 1.000                        & 0.018                                       & -0.038                                       & 1.000                                            \\
0.2                  & 0.061                   & -0.190                   & 1.000                        & 0.060                                       & -0.199                                       & 1.000                                            \\
0.3                  & 0.155                   & -0.349                   & 0.958                        & 0.148                                       & -0.335                                       & 0.974                                            \\
0.4                  & 0.263                   & -0.459                   & 0.778                        & 0.259                                       & -0.462                                       & 0.792                                            \\
0.5                  & 0.361                   & -0.542                   & 0.696                        & 0.362                                       & -0.540                                       & 0.698                                            \\
0.6                  & 0.479                   & -0.630                   & 0.608                        & 0.454                                       & -0.610                                       & 0.638                                            \\
0.7                  & 0.533                   & -0.655                   & 0.654                        & 0.516                                       & -0.646                                       & 0.656                                            \\
0.8                  & 0.579                   & -0.675                   & 0.634                        & 0.593                                       & -0.671                                       & 0.644                                            \\
0.9                  & 0.599                   & -0.680                   & 0.692                        & 0.603                                       & -0.688                                       & 0.676                                            \\
1                    & 0.639                   & -0.695                   & 0.684                        & 0.637                                       & -0.691                                       & 0.700                                            \\ \bottomrule
\end{tabular}%
\caption{Simulation results for 1,000 data points in a \textit{mild instrument} scenario.}
\label{tab:sims_0.50_1000}
\end{table}

\begin{table}[H]
\centering \scriptsize
\begin{tabular}{ccccccc}
\textit{Effect Size} & MSE($\hat{\tau}^{cace}_{l_1}$) & Bias($\hat{\tau}^{cace}_{l_1}$) & Coverage($\hat{\tau}^{cace}_{l_1}$) & \multicolumn{1}{l}{MSE($\hat{\tau}^{cace}_{l_2}$)} & \multicolumn{1}{l}{Bias($\hat{\tau}^{cace}_{l_2}$)} & \multicolumn{1}{l}{Coverage($\hat{\tau}^{cace}_{l_2}$)} \\ \toprule
\multicolumn{1}{l}{} & \multicolumn{6}{c}{BCF-IV}\\ \toprule
0                    & 0.281                   & 0.421                    & 0.976                        & 0.264                                       & 0.399                                        & 0.974                                            \\
0.1                  & 0.216                   & 0.255                    & 0.982                        & 0.185                                       & 0.244                                        & 0.986                                            \\
0.2                  & 0.178                   & 0.121                    & 0.982                        & 0.208                                       & 0.170                                        & 0.976                                            \\
0.3                  & 0.216                   & 0.065                    & 0.982                        & 0.216                                       & 0.109                                        & 0.978                                            \\
0.4                  & 0.221                   & 0.018                    & 0.984                        & 0.251                                       & 0.051                                        & 0.976                                            \\
0.5                  & 0.257                   & -0.062                   & 0.978                        & 0.236                                       & 0.033                                        & 0.974                                            \\
0.6                  & 0.250                   & 0.005                    & 0.970                        & 0.275                                       & 0.022                                        & 0.964                                            \\
0.7                  & 0.310                   & 0.016                    & 0.964                        & 0.297                                       & -0.036                                       & 0.972                                            \\
0.8                  & 0.286                   & 0.021                    & 0.962                        & 0.282                                       & 0.005                                        & 0.960                                            \\
0.9                  & 0.340                   & 0.018                    & 0.950                        & 0.285                                       & -0.018                                       & 0.970                                            \\
1                    & 0.263                   & -0.006                   & 0.964                        & 0.314                                       & 0.008                                        & 0.968                                            \\ \midrule
\multicolumn{1}{l}{} & \multicolumn{6}{c}{GRF}                                                                                                                                                                                                           \\ \toprule
0                    & 0.180                   & 0.337                    & 1.000                        & 0.159                                       & 0.319                                        & 1.000                                            \\
0.1                  & 0.090                   & 0.131                    & 0.998                        & 0.083                                       & 0.138                                        & 1.000                                            \\
0.2                  & 0.073                   & -0.054                   & 1.000                        & 0.077                                       & -0.043                                       & 1.000                                            \\
0.3                  & 0.142                   & -0.206                   & 1.000                        & 0.132                                       & -0.214                                       & 1.000                                            \\
0.4                  & 0.243                   & -0.396                   & 1.000                        & 0.250                                       & -0.397                                       & 1.000                                            \\
0.5                  & 0.439                   & -0.560                   & 0.996                        & 0.384                                       & -0.525                                       & 0.996                                            \\
0.6                  & 0.621                   & -0.703                   & 0.942                        & 0.591                                       & -0.670                                       & 0.970                                            \\
0.7                  & 0.800                   & -0.799                   & 0.904                        & 0.816                                       & -0.812                                       & 0.912                                            \\
0.8                  & 1.017                   & -0.915                   & 0.820                        & 1.069                                       & -0.937                                       & 0.820                                            \\
0.9                  & 1.272                   & -1.019                   & 0.736                        & 1.302                                       & -1.037                                       & 0.748                                            \\
1                    & 1.539                   & -1.135                   & 0.698                        & 1.562                                       & -1.141                                       & 0.692                                            \\ \bottomrule
\end{tabular}%
\caption{Simulation results for 1,000 data points in a \textit{weak instrument} scenario.}
\label{tab:sims_0.25_1000}
\end{table}

\subsection{Weak Instruments with  Varying Effect Size for the Heterogeneous Causal Effects and Varying Heterogeneity in Compliance Rates}\label{subsec:sim3}

Thus far, we assumed constant compliance rates across the various subgroups. However, in real-world applications it can be the case that compliance is varying based on the characteristics of units or individuals. For instance, in the empirical application reported later in Section \ref{application}, it may be the case that eligible schools are more likely to comply and get funded based on some, say, characteristics of the principal and the teaching staff.

{\color{blue} To assess the effectiveness of the BCF-IV and BCF-ITT algorithm to deal with varying heterogeneity in compliance rates we designed two simulations scenarios keeping a fixed sample size of 1,000. In the first scenario, we assume the same heterogeneity structure of the strong heterogeneity case, fixing $k$ to one, and introducing a variation in the compliance rates. Hence, we have that for $\ell_1$ and $\ell_2$ the effect size $k$ is fixed to one -- hence, there is always heterogeneity in the effects -- while the compliance rates are varying. The compliance rates are constant on all the subgroups with no heterogeneity in the effects. Regarding the variation in the compliance rate, we have that $\pi_{\ell_1} \in [0.25, 0.5]$ and $\pi_{\ell_2} \in [0.5, 0.75]$, where we start from a constant compliance rate in the overall population and then we move away from it -- i.e., in the first iteration we have $\pi_{\ell_1}=\pi_{\ell_2}=0.5$, in the second one $\pi_{\ell_1}=0.475$ and $\pi_{\ell_2}=0.525$, and so on, until we get to the last iteration where $\pi_{\ell_1}=0.25$ and $\pi_{\ell_2}=0.75$. The maximal difference in the compliance rates among the subgroups is 0.5, while the overall compliance rate is always 0.5. In the second scenario, we define in the same way the variation in the compliance rates, but we keep the effect of the ITT constant and equal to one. This means that we start from a situation in which both ITT and CACE are homogeneous -- hence, there is no heterogeneity in the effects -- and then, by introducing increasingly bigger variations in the proportion of compliers in the different subgroups we generate heterogeneous CACE while keeping the ITT constant. For simplicity, in this second scenario, we define $\ell_{1}=\{\mathbf\bX_i: X_{i1}=0\}$ and $\ell_{2}=\{\mathbf\bX_i: X_{i1}=1\}$, and the number of covariates $p = 5$. The first scenario is designed to assess the performance of the method in situations where both the conditional CACE, $\tau^{cace}(\bX_i$), and the conditional proportion of compliers, $\pi_C(\bX_i)$, are heterogeneous across various subgroups. The second scenario is designed to mimic those settings where the conditional ITT, $ITT(\bX_i)$, is constant while $\pi_C(\bX_i)$ is heterogeneous leading to heterogeneity in $\tau^{cace}(\bX_i$) purely driven by the heterogeneity in $\pi_C(\bX_i)$. In this second scenario, CACE heterogeneity could be masked by ITT homogeneity.}

Figure \ref{fig:rules_samples} shows the TPR as a function of the distance in the proportion of compliers in the two subgroups defined by $\ell_1$ and $\ell_2$ (e.g., this difference is 0 in the case that both the subgroups have a compliance rate of 0.5, while this difference equals 0.5 in the case where $\pi_{\ell_1}=0.25$ and $\pi_{\ell_2}=0.75$). In the first scenario, the TPR is similar for BCF-IV and BCF-ITT and both these techniques are outperforming HCT-IV. The big difference comes in the second scenario, where BCF-IV strongly outperforms BCF-ITT and HCT-IV. In fact, as the heterogeneity in CACE is introduced by the increasing variations in the proportion of compliers in the two subgroups $\ell_1$ and $\ell_2$, BCF-IV is able to perform a correct identification of the heterogeneous subgroups, while BCF-ITT and HCT-IV are not. This hints at the higher ability of this technique to discover effect variation in scenarios where the intention-to-treat appears to be constant while the heterogeneity is driven by a varying proportion of compliers in the subgroups. This is due to the fact that BCF-IV is designed to discover and estimate the heterogeneity in CACE while both BCF-ITT and HCT-IV are designed to discover the heterogeneity in the ITT and then estimate the conditional CACE for the discovered subgroups. {\color{blue} Please note that, in both the simulations scenarios introduced, there is a slight decline in the TPR as the difference in compliance rates approaches its maximum. This is due to the fact that, in this case, for the subgroup $\ell_1$ the conditional proportion of compliers approaches $\pi_{\ell_1}=0.25$ hence leading to a partially weak IV scenario. As shown in the simulations in the previous Section of the Supplementary Material, in the setting with $\pi_C(X_i)=0.25$, it becomes harder to discover the heterogeneity in the effect as the algorithm faces potential weak instrument issues. Moreover, the FPR in this scenario is larger for small differences in the compliance rates. This is due to the fact that, when there is no difference in the compliance rates, both the ITT and the proportion of compliers are constant, leading to a zero variance in the CACE estimator. In turn, for even small variations in the effect detected by the tree, the BCF-IV algorithm is not able to discard this spurious heterogeneity due to the variance approaching zero. However, as the heterogeneity in the proportion of compliers grows, the algorithm is able to reach a zero FPR. We want to highlight that the scenario of constant ITT and constant proportion of compliers is highly implausible in real-world applications. Moreover, in such cases, once could discard the spurious heterogeneity by trimming the node whose effect size is too close to the population's effect. We leave this effect-size based trimming procedure as scope for future research.}

Table \ref{tab:compliance_scenario_1} shows the results for the estimation again, as a function of the difference in the compliance rates in the two subgroups. As we can see, the precision of BCF-IV and GRF deteriorates for $\ell_1$ as the difference in the compliance rates decreases, while it improves for $\ell_2$ as the difference in the compliance rates increases. Again, we can observe from the table that BCF-IV has a better performance in term of Monte Carlo estimated MSE, estimated bias and coverage than GRF for all the various differences in the compliance rates.

\begin{figure}[H]
    \centering
    \includegraphics[width=1\linewidth]{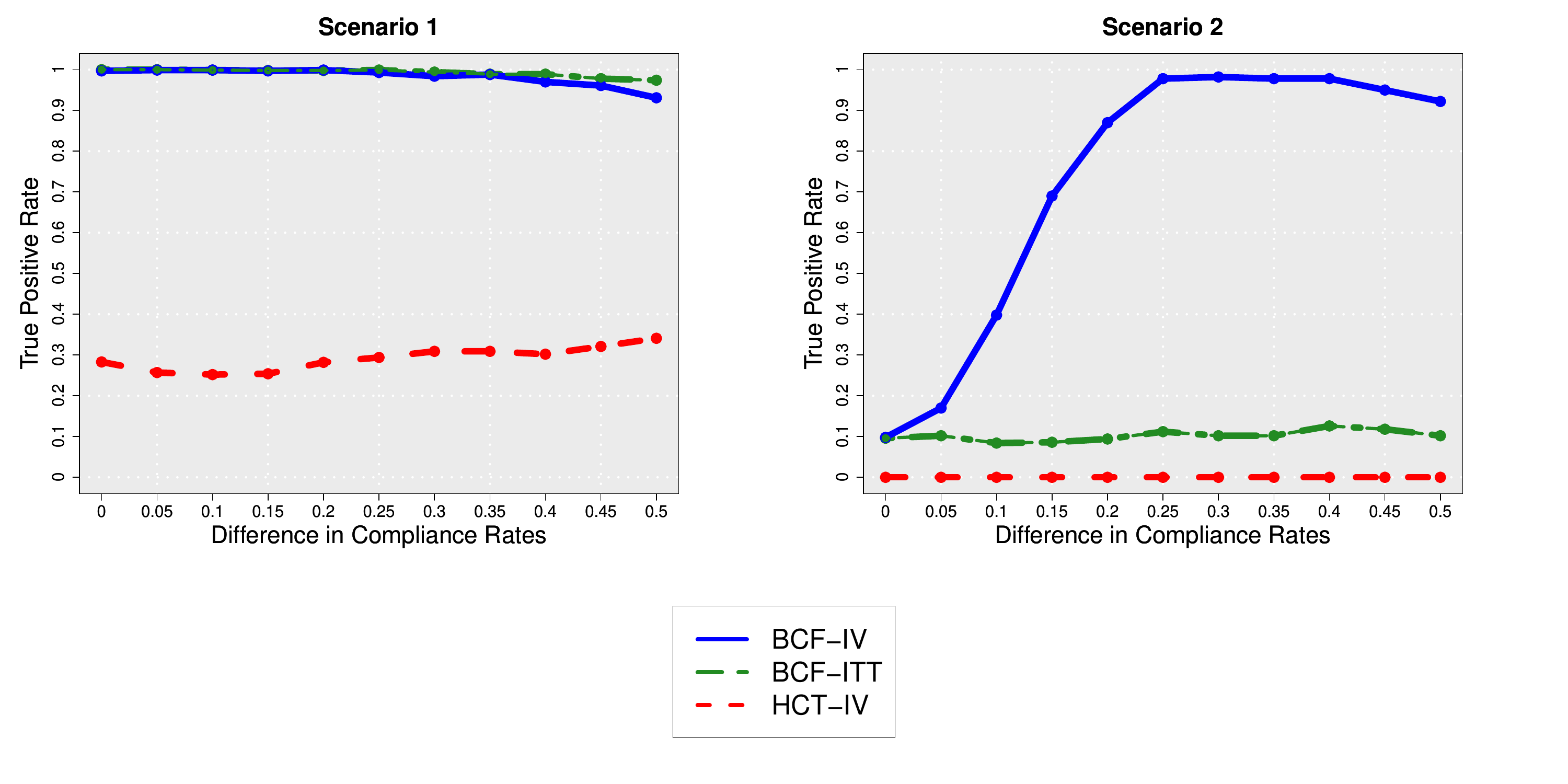}
    \caption{TPR with varying compliance rates among the subpopulations.}
    \label{fig:rules_samples}
\end{figure}

\begin{figure}[H]
    \centering
    \includegraphics[width=1\linewidth]{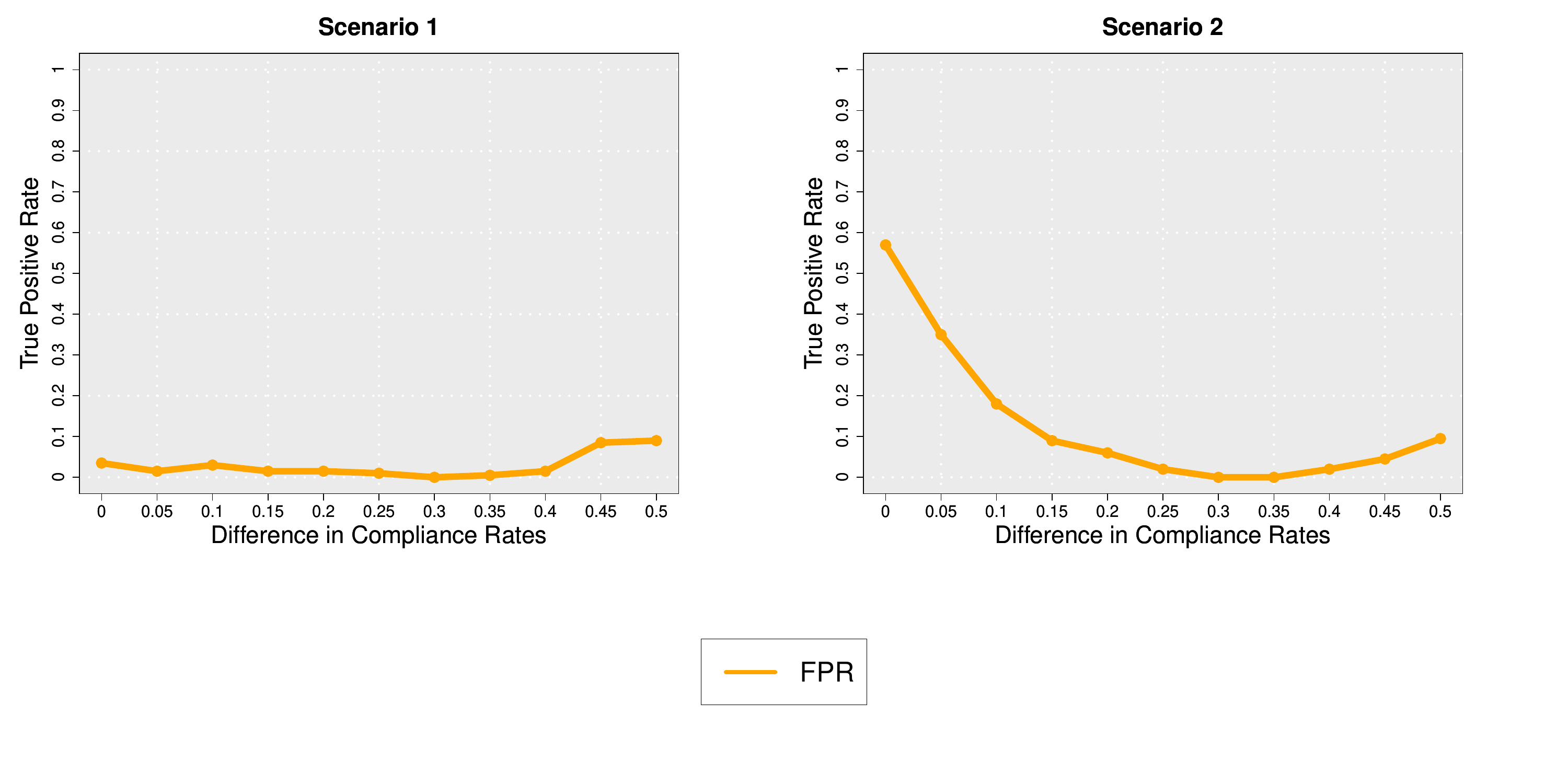}
    \caption{FPR with varying compliance rates among the subpopulations.}
    \label{fig:rules_fpr_pw}
\end{figure}

\begin{table}[H] 
\centering \footnotesize
\begin{tabular}{ccccccccc}
\textit{$\pi_{\ell_1}$} & MSE($\hat{\tau}^{cace}_{l_1}$) & Bias($\hat{\tau}^{cace}_{l_1}$) & Coverage($\hat{\tau}^{cace}_{l_1}$) & &
\textit{$\pi_{\ell_2}$} & \multicolumn{1}{l}{MSE($\hat{\tau}^{cace}_{l_2}$)} & \multicolumn{1}{l}{Bias($\hat{\tau}^{cace}_{l_2}$)} & \multicolumn{1}{l}{Coverage($\hat{\tau}^{cace}_{l_2}$)} \\ \toprule
\multicolumn{1}{l}{} & \multicolumn{8}{c}{BCF-IV}\\  \toprule
0.500   & 0.017 & 0.010  & 0.952 & & 0.500 & 0.016 & 0.011  & 0.954 \\
0.475 & 0.018 & 0.003  & 0.964 & & 0.525 & 0.014 & 0.000  & 0.958 \\
0.450 & 0.021 & 0.000  & 0.950 & & 0.550 & 0.013 & -0.007 & 0.962 \\
0.425 & 0.022 & 0.000  & 0.958 & & 0.575 & 0.012 & 0.001  & 0.966 \\
0.400 & 0.030 & 0.007  & 0.944 & & 0.600 & 0.012 & 0.000  & 0.938 \\
0.375 & 0.031 & 0.017  & 0.960 & & 0.625 & 0.011 & -0.001 & 0.954 \\
0.350 & 0.035 & -0.009 & 0.966 & & 0.650 & 0.010 & -0.001 & 0.952 \\
0.325 & 0.047 & 0.005  & 0.950 & & 0.675 & 0.009 & 0.001  & 0.942 \\
0.300 & 0.053 & 0.008  & 0.952 & & 0.700 & 0.008 & -0.005 & 0.950 \\
0.275 & 0.059 & 0.024  & 0.950 & & 0.725 & 0.008 & 0.004  & 0.952 \\
0.250   & 0.082 & 0.014  & 0.958 & & 0.750 & 0.008 & 0.007  & 0.936 \\ \midrule
\multicolumn{1}{l}{} & \multicolumn{8}{c}{GRF} \\  \toprule
0.500   & 0.151 & -0.333 & 0.714 & & 0.500 & 0.158 & -0.341 & 0.698 \\
0.475 & 0.171 & -0.362 & 0.684 & & 0.525 & 0.152 & -0.342 & 0.688 \\
0.450 & 0.196 & -0.385 & 0.660 & & 0.550 & 0.140 & -0.327 & 0.692 \\
0.425 & 0.211 & -0.403 & 0.650 & & 0.575 & 0.120 & -0.303 & 0.752 \\
0.400 & 0.243 & -0.439 & 0.636 & & 0.600 & 0.115 & -0.293 & 0.724 \\
0.375 & 0.255 & -0.446 & 0.648 & & 0.625 & 0.104 & -0.278 & 0.720 \\
0.350 & 0.296 & -0.487 & 0.598 & & 0.650 & 0.097 & -0.268 & 0.722 \\
0.325 & 0.322 & -0.512 & 0.580 & & 0.675 & 0.093 & -0.263 & 0.712 \\
0.300 & 0.344 & -0.531 & 0.604 & & 0.700 & 0.092 & -0.260 & 0.700 \\
0.275 & 0.366 & -0.554 & 0.566 & & 0.725 & 0.075 & -0.232 & 0.770 \\
0.250   & 0.420 & -0.596 & 0.528 & & 0.750 & 0.070 & -0.223 & 0.748 \\ \bottomrule
\end{tabular}%
\caption{Simulation results for 1,000 data points and varying the compliance rates.}
\label{tab:compliance_scenario_1}
\end{table}
}

\subsection{Robustness Checks for Monte Carlo Simulations}

We introduce some changes in the synthetic models used to test the fits of BCF-IV and BCF-ITT (as compared to GRF  and HCT-IV). The model from which we start is the simplest model introduced in Section \ref{simul} with 1,000 observations, strong heterogeneity, and a fixed compliance rate of 0.75. As in the Monte Carlo simulations in the main text, the results are obtained by aggregating over 500 rounds of simulations.  In order to make the results for the Monte Carlo simulations robust, we introduce 3 different modifications in this model: (i) confounding in the generation of the IV; (ii) covariance in the covariates matrix; and (iii) misspecification in the propensity score. 

{\color{blue} The first variation could potentially hinder both the discovery of the heterogeneous subgroups, but also the estimation of the causal effects.  Confounding is introduced by partially modifying the model for the generation of the treatment assignment and the model for the generation of the potential outcomes in Section \ref{simul}. We assume the confounding to be a linear function of $X_{i3}$ and $X_{i4}$. In particular, the model for the treatment becomes $Z_i  \sim \textit{Binom}(\pi_i)$ where $\pi_i = \text{logit}(-1 + X_{i3} - X_{i4})$, while the model for the potential outcome becomes $Y_i(0) \sim N(X_{i3} + X_{i4}, 1)$. A similar simulation model was introduced in \cite{lee2020causal} to assess the performance of the heterogeneity discovery in situations where the confounders are different from the effect modifiers (namely, variables that drive the heterogeneity in the effects). This scenario can be particularly challenging as the non-overlap between these variables may reduce the algorithm's capacity to disentangle between confounders and effects modifiers. The second modification could potentially have an effect on the detection of the heterogeneous subgroups.  In fact, as the correlation between the covariates increases (in this particular case the correlation in the covariance matrix of the covariates is set to be 0.25), it could lead to potential problems in disentangling between the true drivers of effect variation and spurious effect variation. In particular, we introduce correlation in the data generating process for the covariates matrix $\bX$ by assuming a positive correlation between all the covariates of 0.25. The third and last modification is to plug-in, in the first stage of the BCF-IV algorithm used for the discovery of the effects in \eqref{cace_bcf}, a misspecified propensity score. Again, this modification could affect both discovery and estimation of the conditional effects. Misspecification of the propensity score is introduced through a 10\% bias in the vector of the estimated propensity score $\hat{\pi}(x)$ plugged in the BCF-IV.}

\begin{figure}[H]
    \centering
    \includegraphics[width=1\linewidth]{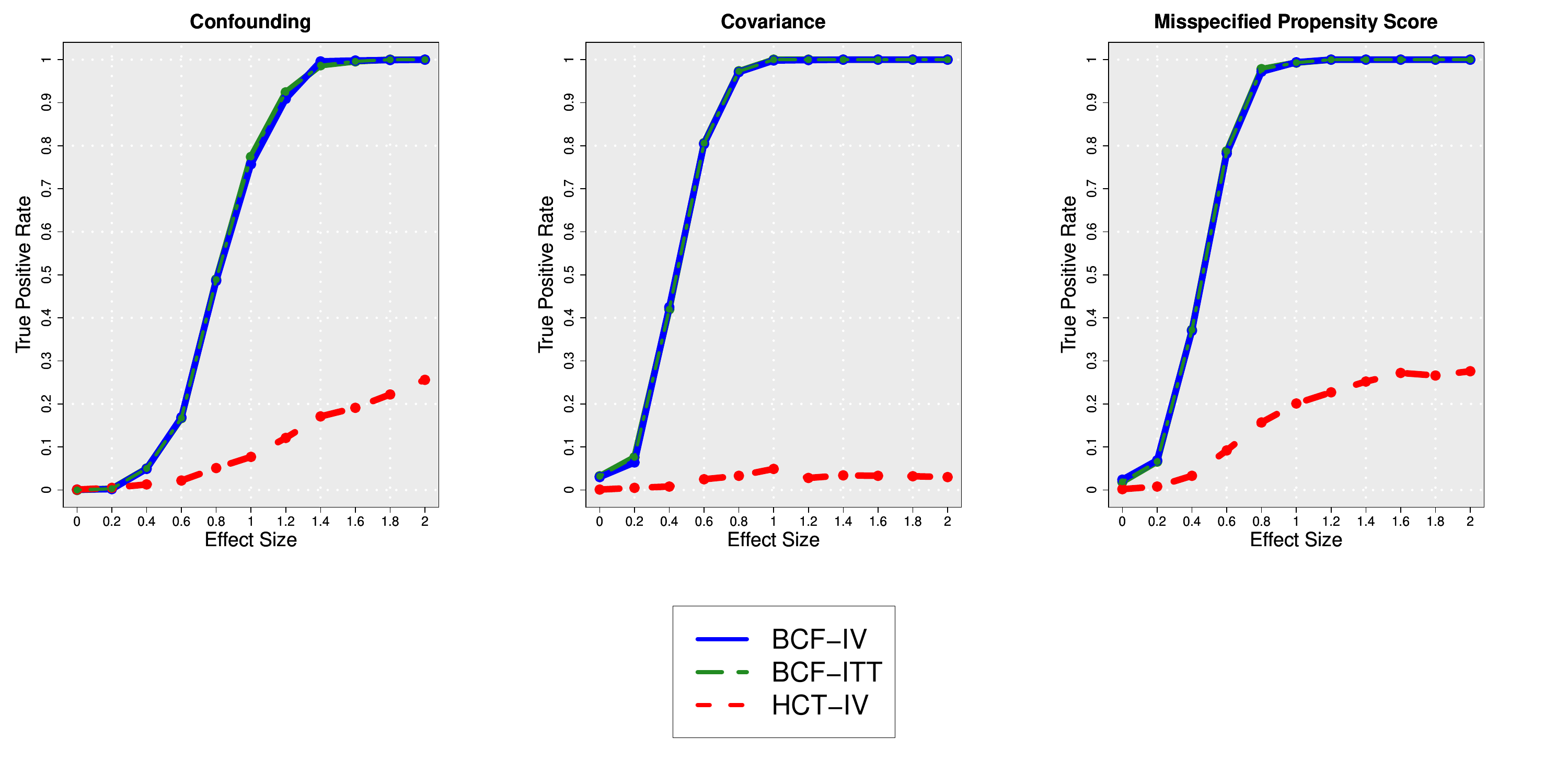}
    \caption{TPR with confounding (left panel), covariance (middle panel) and misspecified propensity score (right panel).}
    \label{fig:rules_appendix}
\end{figure}

Figure \ref{fig:rules_appendix} depicts the results for the discovery step in all the three scenarios. This figure should be directly compared with the left panel of Figure \ref{fig:rules_mtx1}. As one can easily observe the performance of BCF-IV does not deteriorate, while the one of HCT-IV deteriorates especially in the case of correlated covariates. 
{\color{blue} The FDR is approaching zero in all the scenarios.} With respect to the estimation set, results are reported in Tables \ref{tab:sims_conf_1000}, \ref{tab:sims_covariance_1000}, and \ref{tab:sims_missps_1000}. Again, we do not observe any steep deterioration in the performance of the BCF-IV algorithm, that is still able to outperform the results obtained from GRF.

\begin{figure}[H]
    \centering
    \includegraphics[width=1\linewidth]{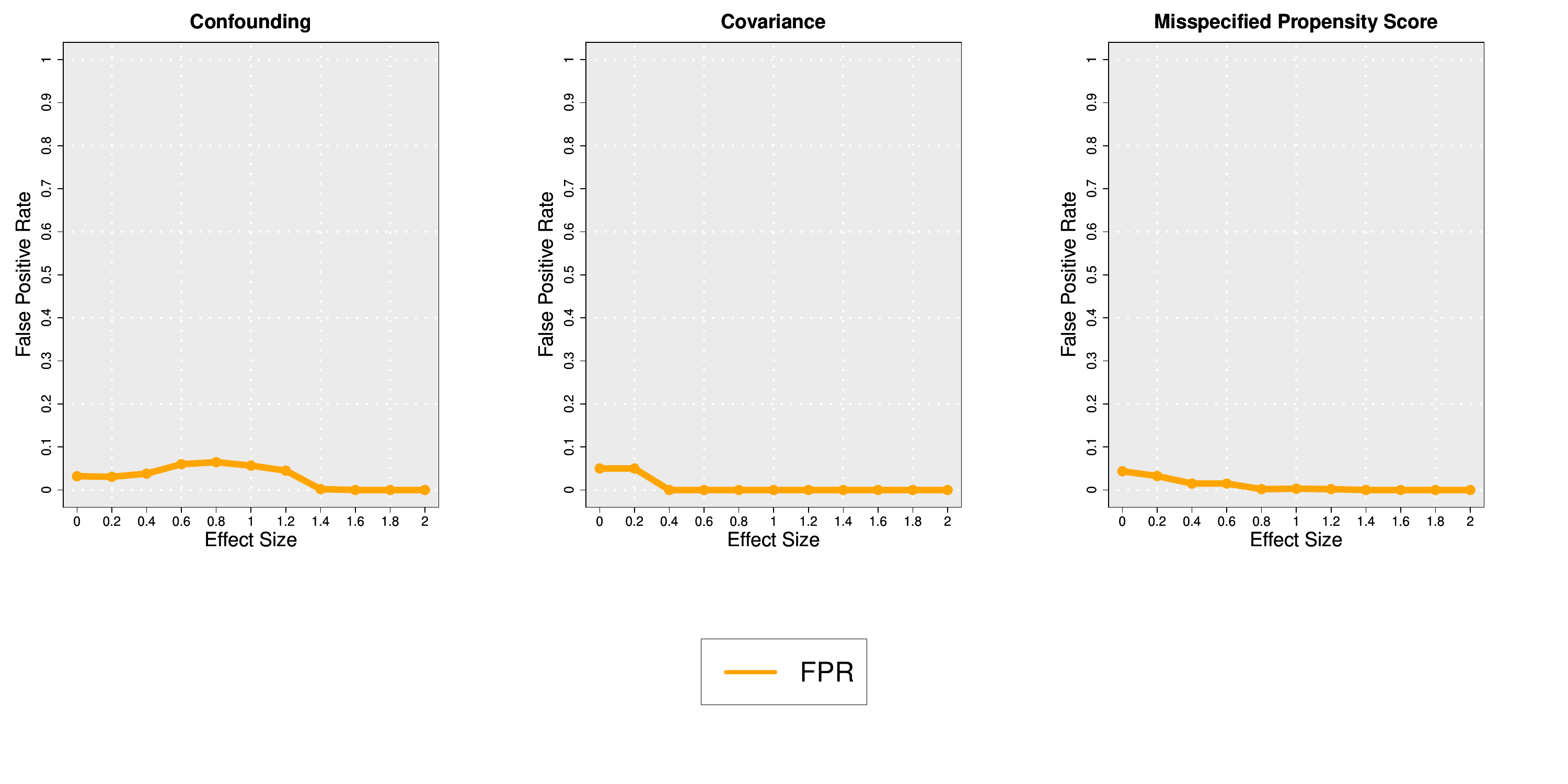}
    \caption{FPR with confounding (left panel), covariance (middle panel) and misspecified propensity score (right panel).}
    \label{fig:rules_appendix_fpr}
\end{figure}

\begin{table}[H]
\centering \scriptsize
\begin{tabular}{ccccccc}                                          
\textit{Effect Size} & MSE($\hat{\tau}^{cace}_{l_1}$) & Bias($\hat{\tau}^{cace}_{l_1}$) & Coverage($\hat{\tau}^{cace}_{l_1}$) & \multicolumn{1}{l}{MSE($\hat{\tau}^{cace}_{l_2}$)} & \multicolumn{1}{l}{Bias($\hat{\tau}^{cace}_{l_2}$)} & \multicolumn{1}{l}{Coverage($\hat{\tau}^{cace}_{l_2}$)} \\ \toprule
\multicolumn{1}{l}{} & \multicolumn{6}{c}{BCF-IV}\\ \toprule
0                    & 0.047                   & 0.175                    & 0.940                        & 0.042                                       & 0.163                                        & 0.956                                            \\
0.1                  & 0.027                   & 0.029                    & 0.978                        & 0.029                                       & 0.040                                        & 0.970                                            \\
0.2                  & 0.043                   & 0.016                    & 0.968                        & 0.039                                       & 0.011                                        & 0.966                                            \\
0.3                  & 0.043                   & 0.010                    & 0.948                        & 0.042                                       & -0.011                                       & 0.946                                            \\
0.4                  & 0.043                   & -0.003                   & 0.950                        & 0.041                                       & -0.001                                       & 0.970                                            \\
0.5                  & 0.049                   & 0.007                    & 0.944                        & 0.043                                       & 0.003                                        & 0.958                                            \\
0.6                  & 0.044                   & -0.007                   & 0.952                        & 0.048                                       & 0.003                                        & 0.940                                            \\
0.7                  & 0.043                   & -0.002                   & 0.960                        & 0.046                                       & -0.001                                       & 0.944                                            \\
0.8                  & 0.046                   & -0.009                   & 0.942                        & 0.042                                       & -0.015                                       & 0.962                                            \\
0.9                  & 0.049                   & -0.008                   & 0.938                        & 0.047                                       & 0.007                                        & 0.950                                            \\
1                    & 0.042                   & 0.004                    & 0.952                        & 0.041                                       & 0.016                                        & 0.946                                            \\ \midrule
\multicolumn{1}{l}{} & \multicolumn{6}{c}{GRF}                                                                                                                                                                                                           \\ \toprule
0                    & 0.017                   & 0.104                    & 0.996                        & 0.016                                       & 0.101                                        & 0.996                                            \\
0.1                  & 0.014                   & -0.077                   & 1.000                        & 0.015                                       & -0.076                                       & 1.000                                            \\
0.2                  & 0.064                   & -0.213                   & 0.968                        & 0.066                                       & -0.220                                       & 0.962                                            \\
0.3                  & 0.127                   & -0.317                   & 0.776                        & 0.142                                       & -0.339                                       & 0.726                                            \\
0.4                  & 0.188                   & -0.387                   & 0.686                        & 0.189                                       & -0.388                                       & 0.674                                            \\
0.5                  & 0.223                   & -0.411                   & 0.700                        & 0.227                                       & -0.426                                       & 0.684                                            \\
0.6                  & 0.255                   & -0.442                   & 0.668                        & 0.249                                       & -0.428                                       & 0.694                                            \\
0.7                  & 0.242                   & -0.421                   & 0.746                        & 0.246                                       & -0.416                                       & 0.742                                            \\
0.8                  & 0.257                   & -0.407                   & 0.758                        & 0.260                                       & -0.426                                       & 0.766                                            \\
0.9                  & 0.245                   & -0.393                   & 0.780                        & 0.219                                       & -0.365                                       & 0.836                                            \\
1                    & 0.207                   & -0.341                   & 0.860                        & 0.181                                       & -0.309                                       & 0.878                                            \\ \bottomrule
\end{tabular}%
\caption{Results for 1,000 data points and confounding.}
\label{tab:sims_conf_1000}
\end{table}

\begin{table}[H]
\centering \scriptsize
\begin{tabular}{ccccccc}  
\textit{Effect Size} & MSE($\hat{\tau}^{cace}_{l_1}$) & Bias($\hat{\tau}^{cace}_{l_1}$) & Coverage($\hat{\tau}^{cace}_{l_1}$) & \multicolumn{1}{l}{MSE($\hat{\tau}^{cace}_{l_2}$)} & \multicolumn{1}{l}{Bias($\hat{\tau}^{cace}_{l_2}$)} & \multicolumn{1}{l}{Coverage($\hat{\tau}^{cace}_{l_2}$)} \\ \toprule
\multicolumn{1}{l}{} & \multicolumn{6}{c}{BCF-IV}\\ \toprule
0                    & 0.025                   & 0.126                    & 0.954                        & 0.026                                       & 0.129                                        & 0.948                                            \\
0.1                  & 0.019                   & 0.013                    & 0.978                        & 0.021                                       & 0.025                                        & 0.966                                            \\
0.2                  & 0.026                   & -0.005                   & 0.954                        & 0.025                                       & -0.003                                       & 0.960                                            \\
0.3                  & 0.027                   & 0.002                    & 0.928                        & 0.025                                       & -0.003                                       & 0.954                                            \\
0.4                  & 0.027                   & 0.004                    & 0.930                        & 0.027                                       & -0.003                                       & 0.936                                            \\
0.5                  & 0.025                   & 0.002                    & 0.946                        & 0.024                                       & 0.012                                        & 0.950                                            \\
0.6                  & 0.025                   & -0.010                   & 0.966                        & 0.027                                       & -0.004                                       & 0.944                                            \\
0.7                  & 0.026                   & 0.001                    & 0.934                        & 0.025                                       & 0.005                                        & 0.952                                            \\
0.8                  & 0.023                   & 0.012                    & 0.946                        & 0.023                                       & -0.003                                       & 0.950                                            \\
0.9                  & 0.026                   & 0.002                    & 0.940                        & 0.023                                       & -0.005                                       & 0.960                                            \\
1                    & 0.026                   & -0.010                   & 0.964                        & 0.026                                       & -0.011                                       & 0.942                                            \\ \midrule
\multicolumn{1}{l}{} & \multicolumn{6}{c}{GRF}                                                                                                                                                                                                           \\ \toprule
0                    & 0.017                   & 0.103                    & 0.992                        & 0.019                                       & 0.110                                        & 0.998                                            \\
0.1                  & 0.014                   & -0.066                   & 1.000                        & 0.015                                       & -0.072                                       & 1.000                                            \\
0.2                  & 0.058                   & -0.197                   & 0.970                        & 0.062                                       & -0.203                                       & 0.974                                            \\
0.3                  & 0.109                   & -0.283                   & 0.818                        & 0.112                                       & -0.288                                       & 0.794                                            \\
0.4                  & 0.146                   & -0.327                   & 0.744                        & 0.147                                       & -0.332                                       & 0.784                                            \\
0.5                  & 0.166                   & -0.347                   & 0.774                        & 0.158                                       & -0.335                                       & 0.784                                            \\
0.6                  & 0.177                   & -0.359                   & 0.788                        & 0.186                                       & -0.362                                       & 0.744                                            \\
0.7                  & 0.179                   & -0.344                   & 0.810                        & 0.177                                       & -0.346                                       & 0.776                                            \\
0.8                  & 0.163                   & -0.315                   & 0.818                        & 0.170                                       & -0.334                                       & 0.804                                            \\
0.9                  & 0.165                   & -0.314                   & 0.836                        & 0.172                                       & -0.329                                       & 0.842                                            \\
1                    & 0.159                   & -0.308                   & 0.840                        & 0.167                                       & -0.314                                       & 0.852                                            \\ \bottomrule
\end{tabular}%
\caption{Results for 1,000 data points and covariance in the covariates' matrix.}
\label{tab:sims_covariance_1000}
\end{table}

\begin{table}[H]
\centering \scriptsize
\begin{tabular}{ccccccc}  
\textit{Effect Size} & MSE($\hat{\tau}^{cace}_{l_1}$) & Bias($\hat{\tau}^{cace}_{l_1}$) & Coverage($\hat{\tau}^{cace}_{l_1}$) & \multicolumn{1}{l}{MSE($\hat{\tau}^{cace}_{l_2}$)} & \multicolumn{1}{l}{Bias($\hat{\tau}^{cace}_{l_2}$)} & \multicolumn{1}{l}{Coverage($\hat{\tau}^{cace}_{l_2}$)} \\ \toprule
\multicolumn{1}{l}{} & \multicolumn{6}{c}{BCF-IV}\\   \toprule
0   & 0.030 & 0.138  & 0.946 & 0.030 & 0.136  & 0.944 \\
0.1 & 0.020 & 0.011  & 0.978 & 0.022 & 0.018  & 0.976 \\
0.2 & 0.028 & -0.004 & 0.950 & 0.026 & 0.003  & 0.962 \\
0.3 & 0.029 & -0.004 & 0.950 & 0.031 & -0.016 & 0.938 \\
0.4 & 0.029 & 0.001  & 0.946 & 0.029 & 0.002  & 0.958 \\
0.5 & 0.029 & -0.003 & 0.966 & 0.028 & 0.012  & 0.946 \\
0.6 & 0.029 & -0.001 & 0.958 & 0.028 & 0.015  & 0.958 \\
0.7 & 0.030 & 0.000  & 0.942 & 0.024 & -0.001 & 0.970 \\
0.8 & 0.031 & 0.003  & 0.948 & 0.035 & 0.011  & 0.926 \\
0.9 & 0.031 & 0.003  & 0.940 & 0.028 & 0.013  & 0.950 \\
1   & 0.028 & -0.002 & 0.960 & 0.030 & -0.003 & 0.944 \\ \midrule
\multicolumn{1}{l}{} & \multicolumn{6}{c}{GRF}      \\ \toprule                    
0   & 0.016 & 0.100  & 0.998 & 0.018 & 0.106  & 0.998 \\
0.1 & 0.015 & -0.080 & 1.000 & 0.015 & -0.079 & 1.000 \\
0.2 & 0.066 & -0.224 & 0.958 & 0.067 & -0.227 & 0.960 \\
0.3 & 0.134 & -0.328 & 0.724 & 0.143 & -0.342 & 0.698 \\
0.4 & 0.190 & -0.388 & 0.686 & 0.189 & -0.391 & 0.688 \\
0.5 & 0.230 & -0.419 & 0.668 & 0.238 & -0.425 & 0.628 \\
0.6 & 0.245 & -0.423 & 0.696 & 0.225 & -0.410 & 0.716 \\
0.7 & 0.235 & -0.396 & 0.750 & 0.225 & -0.400 & 0.778 \\
0.8 & 0.234 & -0.385 & 0.796 & 0.223 & -0.365 & 0.812 \\
0.9 & 0.199 & -0.343 & 0.842 & 0.188 & -0.326 & 0.858 \\
1   & 0.198 & -0.334 & 0.866 & 0.209 & -0.347 & 0.846  \\ \bottomrule                   \end{tabular}%
\caption{Results for 1,000 data points and misspecified propensity score.}
\label{tab:sims_missps_1000}
\end{table}

\section{RDD Checks} \label{RDD}

\begin{figure}[H]
    \centering
    \subfloat[Balance improvement obtained with sampling. ``Initial" refers to the initial sample, while ``Sampled" refers to the bootstrapped sample.]{\includegraphics[width=0.45\textwidth]{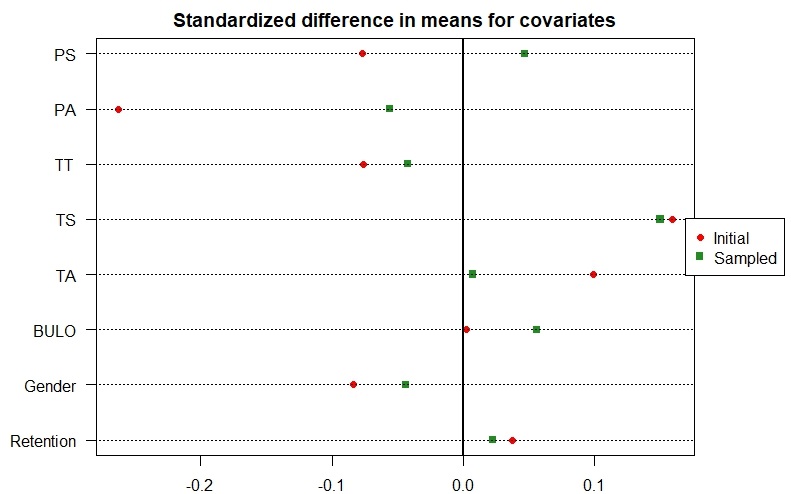}} \quad
    \subfloat[Standardized difference in means (SDM) and 95\% confidence interval around the threshold with a bandwidth of 3.5\%. ]{\includegraphics[width=0.45\textwidth]{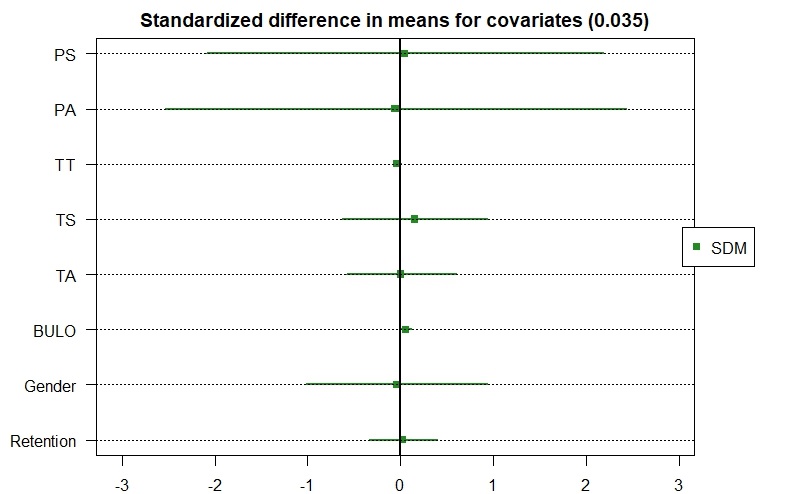}}
    \caption{\footnotesize The label ``PS" refers to Principal Seniority, the label ``PA" to Principal Age, the label ``TS" to Teacher Seniority, the label ``TA" to Teacher Age, the label ``TT" to Teacher Training and the label ``BULO" refers to students with special needs in primary education.}
    \label{fig:2}
\end{figure}

\begin{figure}[H]
    \centering
    \includegraphics[width=0.6\linewidth]{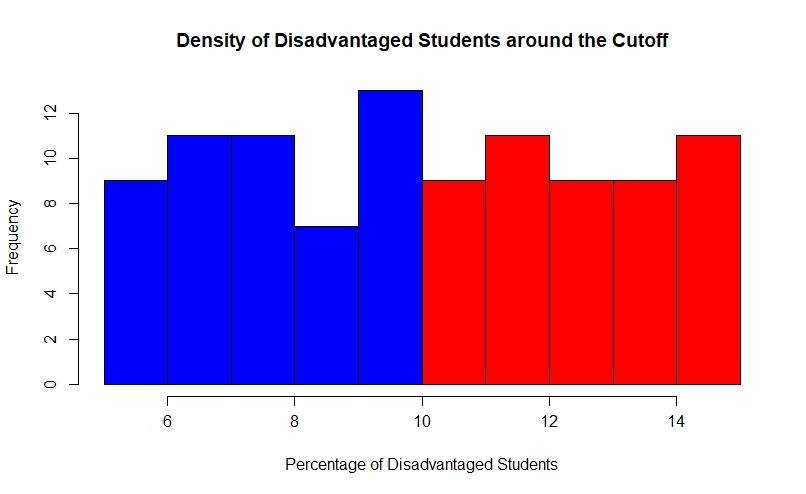}
    \caption{\footnotesize Frequency distribution of disadvantaged students around the threshold (10\%). In red the density of the disadvantaged students in the units assigned to the treatment and in blue the density for the units assigned to the control. The densities are aggregated at school level.}
    \label{fig:3}
\end{figure}

\begin{figure}[H]
    \centering
    \includegraphics[width=0.8\linewidth]{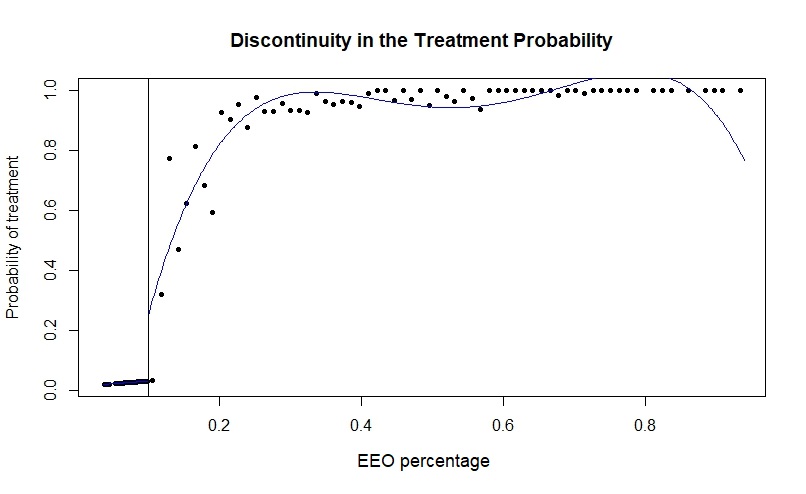}
    \caption{\footnotesize Probability of treatment given the share of disadvantaged students (EEO percentage) in the first stage of secondary education (threshold 10\%).}
    \label{fig:4}
\end{figure}

In order to check whether or not the RDD (Regression Discontinuity Design) setting is valid, we implement the following checks \citep{lee2010regression}: (i) we check the balance in the sample of units assigned to the treatment just above and below the threshold (this is done to check if the randomization holds); (ii) we examine if there are manipulations in the distribution of schools with respect to the share of disadvantaged students around the threshold, (iii) we employ a formal manipulation test, the McCrary test \citep{mccrary2008manipulation}, to discover potential sorting around the threshold; (iv) we check if there is a discontinuity in the probability of being assigned to the treatment around the threshold. The checks depicted in this Subsection are made on the sample of 50 students introduced in Subsection \ref{subsec:identification_strategy}. Table \ref{table_2} shows that the averages of the control variables are not statistically different for the group of units assigned to the treatment and assigned to the control around the threshold, with the exception of \textit{teacher seniority}.  Thus, there is evidence that more senior teachers self-select in schools with lower disadvantaged students. However, as {\color{blue} shown} in Section \ref{results}, this variable does not surface in any model as a driver of significant heterogeneity in the estimated causal effects. This is due to the fact that our model is robust to \textit{spurious} heterogeneity coming from unbalances in the samples, as shown by \cite{hahn2020bayesian} in randomized and regular assignment mechanisms' scenarios. Moreover, panel (b) of Figure \ref{fig:2} shows the standardized difference in the means for these two groups with the relative standardized confidence intervals. The McCrary manipulation test implemented in \cite{calonico2015rdrobust} through a Local-Polynomial Density Estimation leads to the rejection of the null hypothesis of the threshold manipulation. The McCrary test leads to a T-value of -0.7497 corresponding to a p-value of  0.4534. The test is performed aggregating the student data at school level. Both these results and the plot of the distribution of schools with respect to the share of disadvantaged students around the threshold in Figure \ref{fig:3} indicate that there is no evidence of manipulation. Finally, Figure \ref{fig:4} shows a clear discontinuity in the probability of being assigned to the treatment around the threshold. 
\par However, as we pointed out in Section \ref{application}, schools that are assigned to the treatment actually \textit{receive} the treatment if they satisfy an additional condition of a minimum of six teaching hours. This leads to a fuzzy-regression discontinuity design where the jump in the probability of being assigned to the treatment around the threshold is not sharp. This scenario is characterized by imperfect compliance.
\par Students can be sorted, with respect to their compliance status, into two types: (i) students in schools above the threshold with more than six teaching hours or students in schools below the threshold (\textit{compliers}: $W_i(Z_i=1)=1$ or $W_i(Z_i=0)=0$); (ii) students in schools above the threshold but with less than six teaching hours (\textit{never-takers}: $W_i(Z_i=1)=0$). This is a so-called case of one-sided-non-compliance, in which we do not observe any \textit{always-takers} since for those that are sorted out of the assignment to the treatment ($Z_i=0$) there is no possibility to access the treatment.

The assignment to the treatment variable (i.e., studying in a school just below or above the threshold) is a relevant IV in our scenario (namely, the correlation between $Z_i$ and $W_i$ is roughly 0.62). Moreover, we can reasonably assume both the exogeneity condition and the exclusion restriction to hold in this situation. On one side, since the randomization of the instrument holds there is no reason not to assume conditional independence between the instrument and the unobservables. On the other side, the exclusion restriction seems to hold as well since we can believe that being just below or above the threshold does not affect the performance of students in any way other than through the additional funding.
In this imperfect compliance setting, the causal effect of the additional funding on the students' performance can be assessed through the Complier Average Causal Effect in (\ref{cace}). Moreover, using our novel BCF-IV algorithm we can estimate the Conditional Complier Average Causal Effect, (\ref{ITT_C}), to assess the heterogeneity in the causal effects.

\captionsetup[subfigure]{labelformat=simple, labelsep=colon}

\begin{table}[H]
\centering
\begin{tabular}{llllllll}
             & \multicolumn{2}{c}{Above Threshold} & \multicolumn{2}{c}{Below Threshold} & \multicolumn{2}{c}{Full Sample} & p-value \\ \toprule
Retention & 0.036 & (0.187) & 0.037 & (0.189) & 0.037 & (0.188) & 0.913 \\ 
  Gender & 0.492 & (0.500) & 0.471 & (0.499) & 0.482 & (0.500) & 0.155 \\ 
  Special Needs & 0.000 & (0.000) & 0.002 & (0.044) & 0.001 & (0.030) & 0.045 \\ 
  Teacher Age & 4.022 & (0.333) & 4.024 & (0.269) & 4.023 & (0.304) & 0.814 \\ 
  Teacher Seniority & 3.867 & (0.452) & 3.927 & (0.342) & 3.895 & (0.404) & 0.000 \\ 
  Teacher Training & 0.982 & (0.025) & 0.981 & (0.026) & 0.982 & (0.026) & 0.169 \\
  Principal Age & 6.022 & (1.308) & 5.951 & (1.229) & 5.988 & (1.271) & 0.067 \\ 
  Principal Seniority & 5.778 & (1.228) & 5.829 & (0.935) & 5.802 & (1.098) & 0.120 \\ 
  \midrule
Observations & \multicolumn{2}{c}{2250}         & \multicolumn{2}{c}{2050}         & \multicolumn{2}{c}{4300}        &         \\ \bottomrule
\end{tabular}
\caption{\footnotesize Results for 3.5\% discontinuity sample with bootstrapped samples of size 50. Standard deviations are in parentheses and the p-value corresponds to a t-test for the difference between the means in the group above and below the threshold.}
\label{table_2}
\end{table}

\section{Robustness Checks for Policy Evaluation}
\label{appendix-robustness-application}

{\color{ao(english)}
\subsection{Progress in School}\label{sec:progress}

The second outcome variable, \textit{progress school}, assumes value 1 if the student progresses to the following year without any grade retention and 0 if not: roughly 98\% of the students in the sample manage to progress in school in the first two years of secondary education. 
For the students unable to progress in school, this variable is used as a proxy of negative achievements. The bandwidth around the threshold -- selected and validated in the same manner as for the outcome reported in the main text -- is 3.7\%.

It is relevant to understand if additional funding was effective in driving students away from negative performance. Figure \ref{fig:BCF-IV-progress-50} depicts the heterogeneous conditional CACEs: the darker the shade of green in the node, the higher the causal effect. 

The additional funding has a slightly negative, but statistically insignificant, impact on the chance of progress in school for the overall students in the sample (again, this is in line with what was found by \cite{dewitte2018disadvantaged} at the school level). Nevertheless, rather than focusing on the overall average effect, it is more interesting to explore the heterogeneous effects.

The first driver of heterogeneity is the gender of the student. In this case, the funding seems to be effective, even if not significant for the male students, while it is not-effective, and has even a slightly negative effect for the female students.
The first driver of heterogeneity is the seniority level of principals: the treatment has an increased effect for students in schools with less senior principals (less than 30 years of experience). In particular, for male students in treated schools with principals with less than 30 years of experience there is an increase of 1.6\% in the probability of progressing through school. On the the opposite side,  for female students with more senior teachers there is a decrease of 5\% in the probability of progressing through school.

Clearly, these characteristics could possibly correlate with unobservables, such as the effectiveness of principals (which may decrease as principals grow older). In any case, this finding opens up new fields for further investigation, in line with the newly established role of machine learning in the economic literature as a ``theory-driving/theory-testing" tool \citep{mullainathan2017machine}.

\begin{figure}[t!]  
\centering
\begin{tikzpicture}[level distance=100pt, sibling distance=5pt, edge from parent path={(\tikzparentnode) -- (\tikzchildnode)}]
\tikzset{every tree node/.style={align=center}}
\Tree [.\node[fill=pistachio,circle,draw]{CACE\\ -0.005\\ 100\%};
\edge node[auto=right,pos=.6]{Gender = ``Female"};[. \node[fill=teagreen, circle,draw]{CACE\\-0.021\\51\%};
\edge node[auto=right,pos=.6]{\small Principal Senior. $< 7$}; \node[fill=teagreen,circle,draw]{CACE\\-0.016\\35\%};
\edge node[auto=left,pos=.6]{\small Principal Senior. $\geq 7$}; \node[fill=tea,circle,draw]{CACE\\-0.050\\16\%}; ]
\edge node[auto=left,pos=.6]{Gender = ``Male"};[. \node[fill=olivedrab(web)(olivedrab3),circle,draw]{CACE\\0.012\\49\%};
\edge node[auto=right,pos=.6]{\small Principal Senior. $< 7$}; \node[fill=olivedrab(web),circle,draw]{CACE\\0.016\\34\%};
\edge node[auto=left,pos=.6]{\small Principal Senior. $\geq 7$}; \node[fill=olivedrab(web)(olivedrab3), circle,draw]{CACE\\0.014\\15\%}; ]]
\end{tikzpicture}
\caption{\footnotesize Visualization of the heterogeneous Complier Average Causal Effects (CACE) of additional funding on \textit{Progress School} estimated using the proposed BCF-IV model. The overall sample (discovery plus inference subsamples) size is 4,450. The tree is a summarizing classification tree fit to posterior point estimates of individual treatment effects. The significance level is * for a significance level of 0.1, ** for a significance level of 0.05 and *** for a significance level of 0.01.}
\label{fig:BCF-IV-progress-50}
\end{figure}
}

\subsection{Sampling Variations}

This Section tests the robustness of our models to sampling variations. The sampling variations introduced come from the following two sources: (i) a wider bandwidth around the threshold (changing from 3.5\% to 3.7\%); (ii) an expansion in the number of sampled units (from 50 up to the lowest number of students per school, which is 62). Moreover, an algorithm which detects the heterogeneity in the ITT effects is applied (BCF-ITT). To understand if the balance and the results are robust, we manifest the balance in the averages in the samples of units assigned to the treatment and assigned to the control (Tables \ref{table_3}, \ref{table_4}, \ref{table_5}), the results of the causal effects when we increase the number of units sampled (Figures \ref{fig:BCF-IV-certificate-62} and \ref{fig:BCF-IV-progress-62}) and the results for the BCF-ITT algorithm (Figures \ref{fig:BCF-ITT-certificate-62} and \ref{fig:BCF-ITT-progress-62}).

In all the different samples the school level characteristics remain widely balanced (with the exception of teacher seniority). This could be due to the fact that less senior principals select themselves in schools with a lower percentage of disadvantaged students. \textit{Primary retention} and \textit{Gender} seem to be slightly unbalanced when we widen the bandwidth, this however holds true just in the case where we sample through bootstrap 50 units (\textit{Gender} in this case gets back to a good balance).

With respect to the results of the BCF-IV algorithm, when we increase the number of sampled units the main differences between the results for the sample of 50 students for the \textit{A-certificate} outcome (Figure \ref{fig:BCF-IV-certificate-50}) and the ones for the sample of 62 students (Figure \ref{fig:BCF-IV-certificate-62}) are the following: (i) the first split is performed on the age of the principal, however this time the chosen split category is being younger or older than 60 years old (in Figure \ref{fig:BCF-IV-certificate-50} it was 55 years old); (ii) the split on the seniority level of the teacher disappears. With respect to the \textit{Progress School} outcome (Figure \ref{fig:BCF-IV-progress-50} vs Figure \ref{fig:BCF-IV-progress-62}) the main differences are: (i) the first split is performed on the seniority of principals (that was detected as an important variable also in Figure \ref{fig:BCF-IV-progress-50}); (ii) the less senior principals seem to boost the effect for students in schools consistently with the results from the analysis on the sample of 50 students; (iii) for the subgroup of students in schools with principals with less than 30 years of experience and without primary retention the additional funding leads to a significant increase of 2.6\% in the probability of progressing to the next year of school; and (iv) for the subgroups of (a) students in schools with principals with seniority of 30 years of more, (b) female students in schools with principals with seniority of 30 years of more, (c) retained students in schools with principals with seniority of less than 30 years, the additional funding leads to a significant decrease of 8.1\%, 7.4\% and 9.4\% in the probability of progressing through school, respectively. The results for both trees hint, consistently with the main analysis, at an important role of principal's age and seniority. Moreover, the significance of the results for the heterogeneous effects in the case of the \textit{Progress School} outcome hint at the fact that non-significance of the treatment effect is probably due to the smaller number of observations in the nodes in the main analysis.

\begin{table}[H]
\centering
\begin{tabular}{llllllll}
             & \multicolumn{2}{c}{Above Threshold} & \multicolumn{2}{c}{Below Threshold} & \multicolumn{2}{c}{Full Sample} & p-value \\ \toprule
Retention    & 0.039          & (0.194)         & 0.035          & (0.184)         & 0.037         & (0.189)         & 0.418   \\
Gender       & 0.471          & (0.499)         & 0.493          & (0.500)         & 0.482         & (0.499)         & 0.110   \\
Special Needs         & 0.001          & (0.039)         & 0.000          & (0.000)         & 0.001         & (0.027)         & 0.045   \\
Teacher Age           & 4.024          & (0.269)         & 4.002          & (0.333)         & 4.023         & (0.304)         & 0.793   \\
Teacher Seniority           & 3.926          & (0.341)         & 3.867          & (0.452)         & 3.895         & (0.404)         & 0.000   \\
Teacher Training & 0.982 & (0.025) & 0.981 & (0.026) & 0.982 & (0.026) & 0.126 \\ 
Principal Age           & 5.951          & (1.228)         & 6.002          & (1.308)         & 5.988         & (1.271)         & 0.041   \\
Principal Seniority           & 5.829          & (0.934)         & 5.777          & (1.227)         & 5.802         & (1.097)         & 0.083   \\ \midrule
Observations & \multicolumn{2}{c}{2790}         & \multicolumn{2}{c}{2542}         & \multicolumn{2}{c}{5332}        &         \\ \bottomrule
\end{tabular}
\caption{\footnotesize Results for 3.5\% discontinuity sample with bootstrapped samples of size 62. Standard deviations are in parentheses and the p-value corresponds to a t-test for the difference between the means in the group above and below the threshold.}
\label{table_3}
\end{table}

\begin{table}[H]
\centering
\begin{tabular}{llllllll}
             & \multicolumn{2}{c}{Above Threshold} & \multicolumn{2}{c}{Below Threshold} & \multicolumn{2}{c}{Full Sample} & p-value \\ \toprule
Retention & 0.030 & (0.170) & 0.042 & (0.201) & 0.036 & (0.186) & 0.025 \\ 
  Gender & 0.497 & (0.500) & 0.461 & (0.499) & 0.479 & (0.500) & 0.015 \\ 
  Special Needs & 0.000 & (0.021) & 0.001 & (0.037) & 0.001 & (0.030) & 0.309 \\ 
  Teacher Age & 4.022 & (0.333) & 4.023 & (0.260) & 4.022 & (0.299) & 0.955 \\ 
  Teacher Seniority & 3.867 & (0.452) & 3.932 & (0.330) & 3.899 & (0.398) & 0.000 \\
  Teacher Training & 0.982 & (0.025) & 0.983 & (0.026) & 0.983 & (0.026) & 0.805 \\ 
  Principal Age & 6.022 & (1.308) & 6.000 & (1.206) & 6.011 & (1.259) & 0.556 \\ 
  Principal Seniority & 5.778 & (1.228) & 5.818 & (0.912) & 5.798 & (1.083) & 0.212 \\  
  \midrule
Observations & \multicolumn{2}{c}{2250}         & \multicolumn{2}{c}{2200}         & \multicolumn{2}{c}{4450}        &         \\ \bottomrule
\end{tabular}
\caption{\footnotesize Results for 3.7\% discontinuity sample with bootstrapped samples of size 50. Standard deviations are in parentheses and the p-value corresponds to a t-test for the difference between the means in the group above and below the threshold.}
\label{table_4}
\end{table}

 \begin{table}[H]
\centering
\begin{tabular}{llllllll}
             & \multicolumn{2}{c}{Above Threshold} & \multicolumn{2}{c}{Below Threshold} & \multicolumn{2}{c}{Full Sample} & p-value \\ \toprule
Retention & 0.029 & (0.168) & 0.040 & (0.196) & 0.034 & (0.182) & 0.026 \\ 
  Gender & 0.490 & (0.500) & 0.464 & (0.499) & 0.477 & (0.500) & 0.058 \\ 
  Special Needs & 0.000 & (0.019) & 0.001 & (0.038) & 0.001 & (0.030) & 0.174 \\ 
  Teacher Age & 4.022 & (0.333) & 4.023 & (0.260) & 4.022 & (0.299) & 0.950 \\ 
  Teacher Seniority & 3.867 & (0.452) & 3.932 & (0.330) & 3.899 & (0.398) & 0.000 \\
  Teacher Training & 0.982 & (0.025) & 0.983 & (0.026) & 0.983 & (0.026) & 0.784 \\ 
  Principal Age & 6.022 & (1.308) & 6.000 & (1.206) & 6.011 & (1.259) & 0.512 \\ 
  Principal Seniority & 5.778 & (1.227) & 5.818 & (0.912) & 5.798 & (1.083) & 0.165 \\
\midrule
Observations & \multicolumn{2}{c}{2790}         & \multicolumn{2}{c}{2728}         & \multicolumn{2}{c}{5518}        &         \\ \bottomrule
\end{tabular}
\caption{\footnotesize Results for 3.7\% discontinuity sample with bootstrapped samples of size 62 (the smallest school in the sample). Standard deviations are in parentheses and the p-value corresponds to a t-test for the difference between the means in the group above and below the threshold.}
\label{table_5}
\end{table}

\begin{figure}[H]  
\centering
\begin{tikzpicture}[level distance=100pt, sibling distance=30pt, edge from parent path={(\tikzparentnode) -- (\tikzchildnode)}]
\tikzset{every tree node/.style={align=center}}
\Tree [.\node[fill=non-photoblue,circle,draw]{CACE\\ -0.019\\ 100\%};
\edge node[auto=right,pos=.6]{Principal Age $< 8$};[. \node[fill=non-photoblue, circle,draw]{CACE\\-0.010\\81\%};]
\edge node[auto=left,pos=.6]{Principal Age $\geq 8$};[. \node[fill=bubbles,circle,draw]{CACE\\-0.062\\9\%};
\edge node[auto=right,pos=.6]{Gender = ``Female"}; \node[fill=whitish,circle,draw]{CACE\\-0.152\\5\%};
\edge node[auto=left,pos=.6]{Gender = ``Male"}; \node[fill=non-photoblue, circle,draw]{CACE\\-0.010\\4\%}; ]]
\end{tikzpicture}
\caption{\footnotesize Visualization of the heterogeneous Complier Average Causal Effects (CACE) of additional funding on \textit{A-certificate} estimated using the proposed BCF-IV model. The overall sample (discovery plus inference subsamples) size is 5,332. The tree is a summarizing classification tree fit to posterior point estimates of individual treatment effects. The significance level is * for a significance level of 0.1, ** for a significance level of 0.05 and *** for a significance level of 0.01.}
\label{fig:BCF-IV-certificate-62}
\end{figure}

\begin{figure}[H]  
\centering
\begin{tikzpicture}[level distance=100pt, sibling distance=5pt, edge from parent path={(\tikzparentnode) -- (\tikzchildnode)}]
\tikzset{every tree node/.style={align=center}}
\Tree [.\node[fill=pistachio,circle,draw]{CACE\\ -0.003\\ 100\%};
\edge node[auto=right,pos=.6]{Principal Seniority$<7$};[. \node[fill=olivedrab(web)(olivedrab3), circle,draw]{CACE\\0.014\\69\%};
\edge node[auto=right,pos=.6]{\small Primary Retention $=0$}; \node[fill=olivedrab(web),circle,draw]{CACE\\0.026**\\65\%};
\edge node[auto=left,pos=.6]{\small Primary Retention $=1$}; \node[fill=tea,circle,draw]{CACE\\-0.094*\\4\%}; ]
\edge node[auto=left,pos=.6]{Principal Seniority $\geq 7$};[. \node[fill=teagreen,circle,draw]{CACE\\-0.081**\\31\%};
\edge node[auto=right,pos=.6]{Gender = ``Female"}; \node[fill=teagreen,circle,draw]{CACE\\-0.074*\\16\%};
\edge node[auto=left,pos=.6]{Gender = ``Male"}; \node[fill=tea, circle,draw]{CACE\\-0.100\\15\%}; ]]
\end{tikzpicture}
\caption{\footnotesize Visualization of the heterogeneous Complier Average Causal Effects (CACE) of additional funding on \textit{Progress School} estimated using the proposed BCF-IV model. The overall sample (discovery plus inference subsamples) size is 5,518. The tree is a summarizing classification tree fit to posterior point estimates of individual treatment effects. The significance level is * for a significance level of 0.1, ** for a significance level of 0.05 and *** for a significance level of 0.01.}
\label{fig:BCF-IV-progress-62}
\end{figure}

\subsection{ITT Analyses}\label{sec:itt}

Figures \ref{fig:BCF-ITT-certificate-62} and \ref{fig:BCF-ITT-progress-62} depict the results obtained from the BCF-ITT. Figure \ref{fig:BCF-ITT-certificate-62} depicts the results for the ITT for the \textit{A-certificate}, while Figure \ref{fig:BCF-ITT-progress-62} depicts the results for the ITT for the \textit{Progress School}. The results again are robust with the ones of the main analyses in highlighting the role of principals age and seniority as drivers of the heterogeneity in the effects.

\begin{figure}[H]  
\centering
\begin{tikzpicture}[level distance=80pt, sibling distance=5pt, edge from parent path={(\tikzparentnode) -- (\tikzchildnode)}]
\tikzset{every tree node/.style={align=center}}
\Tree [.\node[fill=bubbles,circle,draw]{CACE\\ -0.029\\ 100\%};
\edge node[auto=right,pos=.6]{Principal Age $<6$};[. \node[fill=non-photoblue, circle,draw]{CACE\\0.004\\30\%};
\edge node[auto=right,pos=.6]{\small Teacher Senior. $< 4$}; \node[fill=ceruleanblue, circle,draw]{CACE\\0.019\\8\%};
\edge node[auto=left,pos=.6]{\small Teacher Senior. $\geq 4$}; \node[fill=bubbles,circle,draw]{CACE\\-0.026\\22\%}; ]
\edge node[auto=left,pos=.6]{Principal Age $\geq 6$};[. \node[fill=whitish,circle,draw]{CACE\\-0.048\\70\%};
\edge node[auto=right,pos=.6]{\small Principal Age $< 8$}; \node[fill=bubbles,circle,draw]{CACE\\-0.044\\61\%};
\edge node[auto=left,pos=.6]{\small Principal Age $\geq 8$}; \node[fill=bubbles,circle,draw]{CACE\\-0.042\\9\%}; ]]
\end{tikzpicture}
\caption{\footnotesize Visualization of the heterogeneous Complier Average Causal Effects (CACE) of additional funding on \textit{A-certificate} estimated using the proposed BCF-ITT model. The overall sample (discovery plus inference subsamples) size is 5,332. The tree is a summarizing classification tree fit to posterior point estimates of individual treatment effects. The significance level is * for a significance level of 0.1, ** for a significance level of 0.05 and *** for a significance level of 0.01.}
\label{fig:BCF-ITT-certificate-62}
\end{figure}

\begin{figure}[H]  
\centering
\begin{tikzpicture}[level distance=100pt, sibling distance=15pt, edge from parent path={(\tikzparentnode) -- (\tikzchildnode)}]
\tikzset{every tree node/.style={align=center}}
\Tree [.\node[fill=pistachio,circle,draw]{CACE\\ -0.008\\ 100\%};
\edge node[auto=right,pos=.6]{Teacher Seniority$<4$};[. \node[fill=pistachio, circle,draw]{CACE\\0.006\\14\%};
\edge node[auto=right,pos=.6]{Gender = ``Female"}; \node[fill=pistachio,circle,draw]{CACE\\-0.009\\7\%};
\edge node[auto=left,pos=.6]{Gender = ``Male"};  \node[fill=olivedrab(web),circle,draw]{CACE\\0.025\\7\%}; ]
\edge node[auto=left,pos=.6]{Teacher Seniority $\geq 4$};[. \node[fill=teagreen,circle,draw]{CACE\\-0.014\\86\%};
\edge node[auto=right,pos=.6]{Principal Senior. $< 6$}; \node[fill=teagreen,circle,draw]{CACE\\-0.013\\26\%};
\edge node[auto=left,pos=.6]{Principal Senior. $\geq 6$}; \node[fill=teagreen, circle,draw]{CACE\\-0.015\\60\%}; ]]
\end{tikzpicture}
\caption{\footnotesize Visualization of the heterogeneous Complier Average Causal Effects (CACE) of additional funding on \textit{Progress School} estimated using the proposed BCF-ITT model. The overall sample (discovery plus inference subsamples) size is 5,518. The tree is a summarizing classification tree fit to posterior point estimates of individual treatment effects. The significance level is * for a significance level of 0.1, ** for a significance level of 0.05 and *** for a significance level of 0.01.}
\label{fig:BCF-ITT-progress-62}
\end{figure}

\end{document}